\documentclass[10pt,preprint]{aastex}
\usepackage{amssymb}
\usepackage{textcomp}

\begin{document}

\shortauthors{Engel et al.}

\title{Arp\,220: extinction \& merger-induced star formation}

\author{H. Engel, R. I. Davies, R. Genzel, L. J. Tacconi, E. Sturm}
\affil{Max-Planck-Institut f\"ur extraterrestrische Physik, Postfach 1312,
  85741 Garching, Germany}

\and

\author{D. Downes}
\affil{Institut de Radio Astronomie Millim\'etrique, Domaine Universitaire, 38406 St.~Martin d'H\`eres, France}


\begin{abstract}
We analyse new spatially resolved integral field spectroscopic H- and K-band data at a
resolution of 0.3\arcsec\ (100\,pc) and re-analyse interferometric CO(2-1) line
observations of the prototypical merging system \object{Arp\,220}.
We find that the majority of the K-band luminosity is due to a 10\,Myr old starburst, with a significant contribution from an underlying $\gtrsim$\,1\,Gyr old stellar population, and a small contribution from stars $\lesssim$\,8\,Myr old. The \cite{calzetti00} reddening law provides the best fit to photometric datapoints spanning 0.45\,$\mu$m -- 2.12\,$\mu$m. Furthermore, estimates of the bolometric luminosity from \textit{IRAS} fluxes in conjunction with our stellar population analysis indicate that we observe less than 10\% of the emitted K-band light.
The stellar and CO(2-1) kinematic centre of the western nucleus coincides with the compact hot dust emission, indicating that the latter marks the centre of the gravitational potential.
In the eastern nucleus, the CO(2-1) data are well matched by a model in which the gas orbits around the peak of the dust emission. This, and the similarity of the K-band tracer kinematics, shows that despite the irregular morphology, the eastern nucleus is also a kinematically coherent structure. Comparison of the extinction map and EW$_{CO}$ and EW$_{Br\gamma}$ maps indicates that the lower half of the eastern nucleus is significantly more extincted than the upper half, suggesting that the lower half is buried in the larger-scale gas disk.
\end{abstract}


\maketitle

\section{Introduction}
\label{sec:intro}

Local Ultraluminous Infrared Galaxies (ULIRGs, with $L_{IR}\geq\,10^{12}L_\odot$, \citealt{sandersmirabel96}) have received considerable attention since their discovery by \textit{IRAS} \citep{neugebauer84}. They are almost exclusively major mergers \citep{sanders88a,sandersmirabel96,veilleux02,jogee04}, and widely held to constitute a key epoch in the transformation of gas-rich spiral galaxies to elliptical galaxies \citep[e.g.][]{hopkins06}. Although not making a significant contribution to the cosmic energy budget at current times, their relative importance increases towards higher redshifts \citep[e.g.][]{caputi07,magnelli09,magnelli10}. Thus the study of local ULIRGs is important for several reasons; to properly understand the major merger process believed to trigger quasar activity and result in elliptical galaxies; and as local analogues of the most luminous high-z galaxies \citep[e.g.][]{engel10b}.

Arp\,220 ($L_{IR}$\,=\,1.4\,$\times$\,10$^{12}L_\odot$, \citealt{soif87}) is the closest ULIRG representative (D\,=\,73\,Mpc; z\,=\,0.018, 1\arcsec=\,352\,pc) and thus a unique laboratory for the investigation of this important class of galaxies. As such, it has been extensively studied and is widely regarded as the prototypical ULIRG.
It is an advanced merger; tidal tails and distortions are observed at optical wavelengths and in HI emission \citep{arp66,josephwright85,hibbard00}. There is evidence for a galactic-scale outflow from H$\alpha$ and soft X-ray observations \citep{armus90,heckman96,mcdowell03}. In the centre, two nuclei are discernible with a separation of $\sim$\,0.4\,kpc in the near-infrared \citep{scoville98}, millimeter \citep{scoville97,downes98,sakamoto99,downes07} and radio \citep{mundell01} regimes. There also is a more extended (kpc-scale), rotating gas disk \citep{downes98,sakamoto99,sakamoto09,mundell01,wiedner02}.
Arp\,220 is very gas-rich; \cite{scoville97} measure a molecular gas mass of $\sim$\,9\,$\times$\,10$^9M_\odot$, concentrated in the central 750\,pc -- implying an astonishing molecular gas surface density of 5\,$\times$\,10$^4M_\odot$\,pc$^{-2}$. 
It therefore comes as no surprise that obscuration is severe for this system; estimates for $A_V$ range from 50 to 1000 \citep{sturm96,downes98}, and even at near-infrared wavelengths obscuring dust lanes are visible \citep{scoville98}. Observations of CO(3-2) and the 860\,$\mu$m continuum \citep{sakamoto08} and CO(6-5) and the 435\,$\mu$m continuum \citep{matsushita09} indicate substantial dust optical depths even at submillimetre wavelengths.
These extreme levels of extinction have hindered a definitive determination of the power source of Arp\,220; 
a significant contribution to its prodigious luminosity comes from a starburst \citep{sturm96,lutz96,lutz98,genzel98}, but there is also significant (and possibly dominant) dust emission from heavily obscured central star formation or perhaps AGN \citep{spoon04}.
The presence of a deeply dust-enshrouded major nuclear power source is confirmed by millimetre observations; most recently \cite{sakamoto08} estimate the bolometric luminosity of the 50--80\,pc core within the western nucleus to be at least 2\,$\times$\,10$^{11}L_\odot$ and possibly as much as $\sim$\,10$^{12}L_\odot$.
The high obscuration has prevented confirmation or rejection of a possible active nucleus through X-ray observations \citep{iwasawa01,iwasawa05,clements02}; the obscuring column density is of order $N_H\sim$\,10$^{25}$cm$^{-2}$ \citep{iwasawa01,sakamoto08}. \cite{downes07} interpret the very compact dust continuum emission in the western nucleus as being heated by a black hole accretion disk, whereas \cite{sakamoto08} contend that the starbursting west nuclear disk must have in its center a dust enshrouded AGN or a very young starburst equivalent to hundreds of super star clusters.

Since Arp\,220 is the ULIRG most frequently used as a starburst galaxy template, it is vital to understand its power source and star formation history.
A photometric analysis by \cite{wilson06} of \textit{HST~UBVI} observations showed that the central cluster population divides into two groups; one with ages $\lesssim$\,10\,Myr, and one with ages $\sim$\,300\,Myr. However, it must be kept in mind that due to photometric uncertainties, the age/reddening degeneracy limits the accuracy of such work. \cite{rod08} use optical spectroscopic observations of the extended diffuse light along three slit positions for a stellar population analysis; they find the optical spectrum to be dominated by an intermediate-age stellar population (0.5--0.9\,Gyr), and a young stellar population ($<$\,100\,Myr) which has an increasing contribution towards the central part of the galaxy. Besides the extinction limitations afflicting all optical-wavelength observations of dusty galaxies, it must be cautioned that only one of their slits covers the central region, from which the vast majority of the luminosity is emitted, and hence that their results may be biased towards the star formation history of the sub-dominant outer regions. \cite{parra07} (see also \citealt{rov05,lonsdale06}) investigate radio supernovae and supernova remnants in the nuclear region of Arp\,22, finding that they are indicative either of a radically different stellar initial mass function, or a very short, intense burst of star formation $\sim$\,3\,Myr ago. 

Here, we use adaptive optics near-infrared integral field spectroscopy data and interferometric mm CO(2-1) and continuum observations, to investigate the star formation history, the extinction, and the nature and orientation of the two nuclei.

In \S\ref{sec:obs}, we introduce the data, and data analysis procedures. In \S\ref{sec:sfh} we investigate the star formation history in the SINFONI field of view, and in \S\ref{sec:extinction} look at the extinction in the near-infrared regime. We then focus on the western (\S\ref{sec:WN:analysis}) and eastern (\S\ref{sec:EN:analysis}) nucleus. We put this together in a coherent picture of the nuclear region (\S\ref{sec:setup}), before summarising and concluding in \S\ref{sec:summ}. 


\section{Observations and Data Processing}
\label{sec:obs}

\subsection{SINFONI Data}

\subsubsection{Observations and Reduction}
\label{sec:obs:red}

\begin{figure}
\begin{center}
\includegraphics[width=0.91\textwidth]{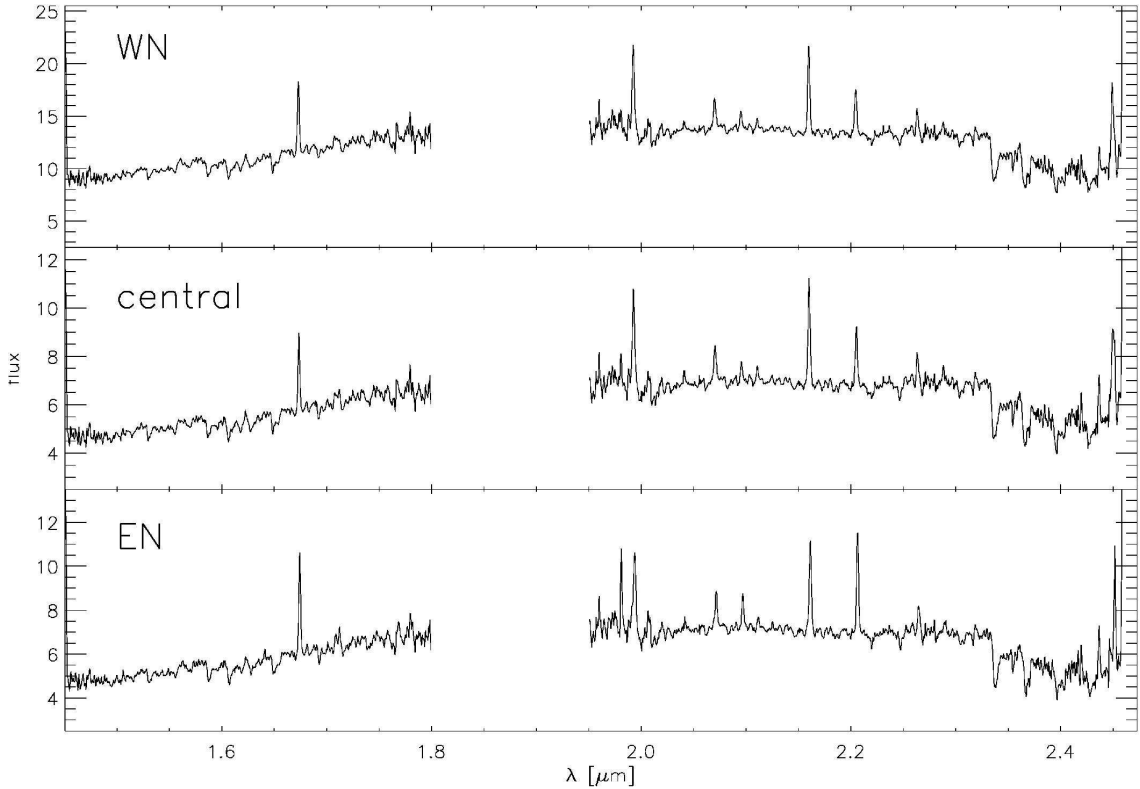} 
\caption{Spectra extracted in 0.25\arcsec\ radius apertures on the western nucleus (WN), eastern nucleus (EN), and in between the two nuclei (central).\label{fig:spec}}
\end{center}
\end{figure}

Observations of Arp\,220 were performed on the nights of March 7th and April 18th to 21st 2007 on Cerro Paranal, Chile, at the Very Large Telescope (VLT) with SINFONI. SINFONI is a near-infrared integral field spectrometer \citep{eis03} which includes a curvature-based adaptive optics system \citep{bon04} and can operate with the VLT's laser guide star facility \citep{bon06,rab04}. The laser guide star was used for these observations, without tip-tilt correction due to lack of a suitable tip-tilt star. SINFONI's 0.05\arcsec$\times$\,0.10\arcsec\ pixel scale was used, giving a 3.2\arcsec$\times$3.2\arcsec\ field of view.
The science data and standard star observations were reduced using the SPRED software package \citep{abu06}. Telluric correction and flux calibration of the reconstructed data cubes were performed using these standard star frames.
Fig.~\ref{fig:spec} shows spectra from three different positions of the Arp\,220 nuclear region.


\subsubsection{PSF Estimation}
\label{sec:obs:psf}

Several avenues exist to estimate the point spread function (PSF hereafter) of adaptive optics data. Here we employ the method outlined by \cite{dav08} and used e.g.~by \cite{mue06} and \cite{engel10a}, which proceeds by comparing the data for which the PSF is not known, denoted $I_{low}$ here, to higher-resolution data with known PSF ($I_{high}$). Since an observed image is the convolution of the instrumental PSF with the intrinsic on-sky image, $I_{obs} = I_{intr} \otimes PSF$, the PSF of the AO data can be estimated by finding the broadening function $B$ which, applied to the higher-resolution image, yields the best match to the AO image; $I_{low} = I_{high} \otimes B$. Since $I_{intr}$ is the same for both $I_{low}$ and $I_{high}$, the PSF of $I_{low}$ can be estimated by convolving the PSF of the high resolution image with the broadening function $B$; $PSF_{low} = PSF_{high} \otimes B$.

\begin{figure}
\begin{center}
\includegraphics[width=0.31\textwidth]{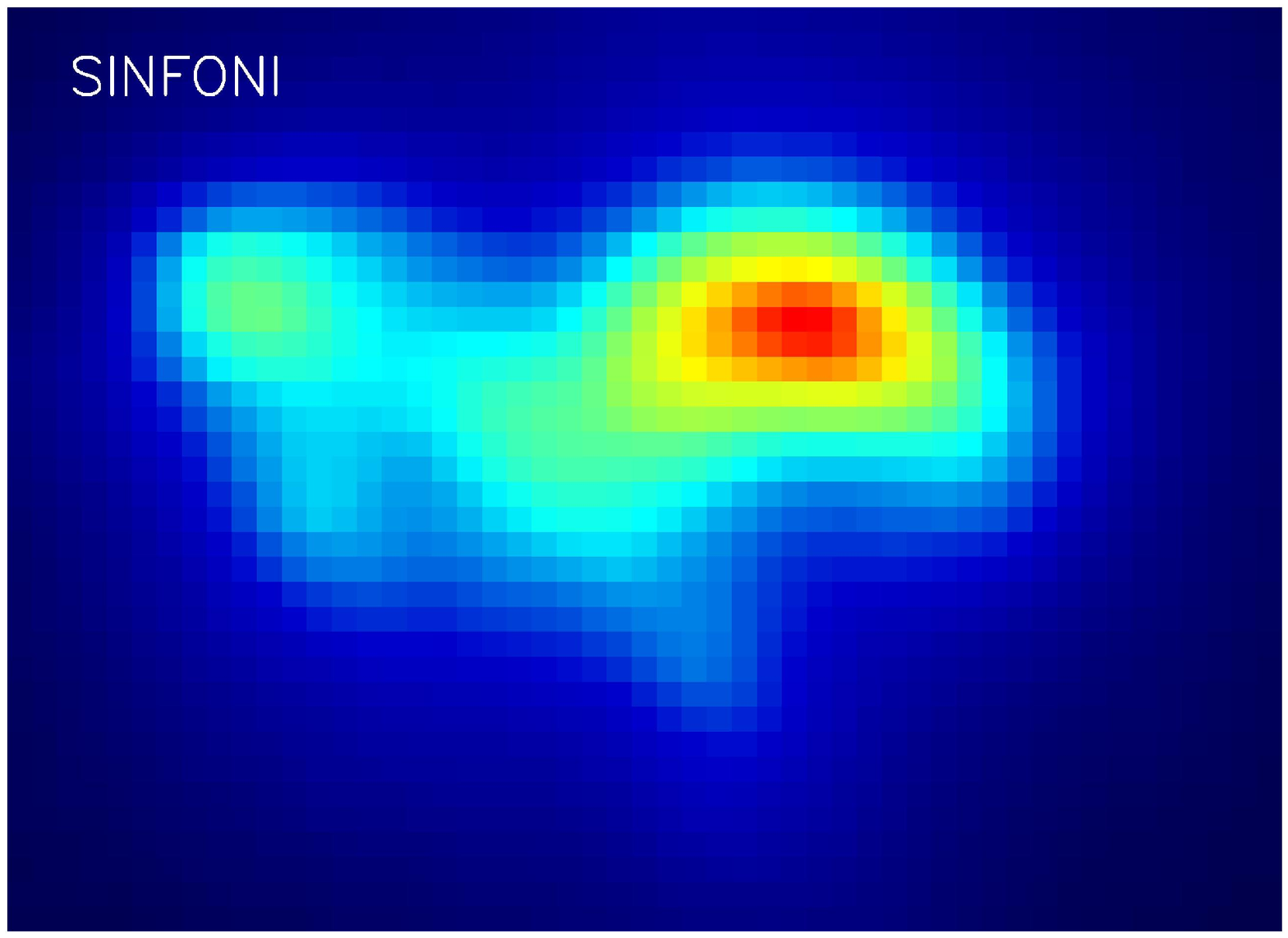} 
\includegraphics[width=0.31\textwidth]{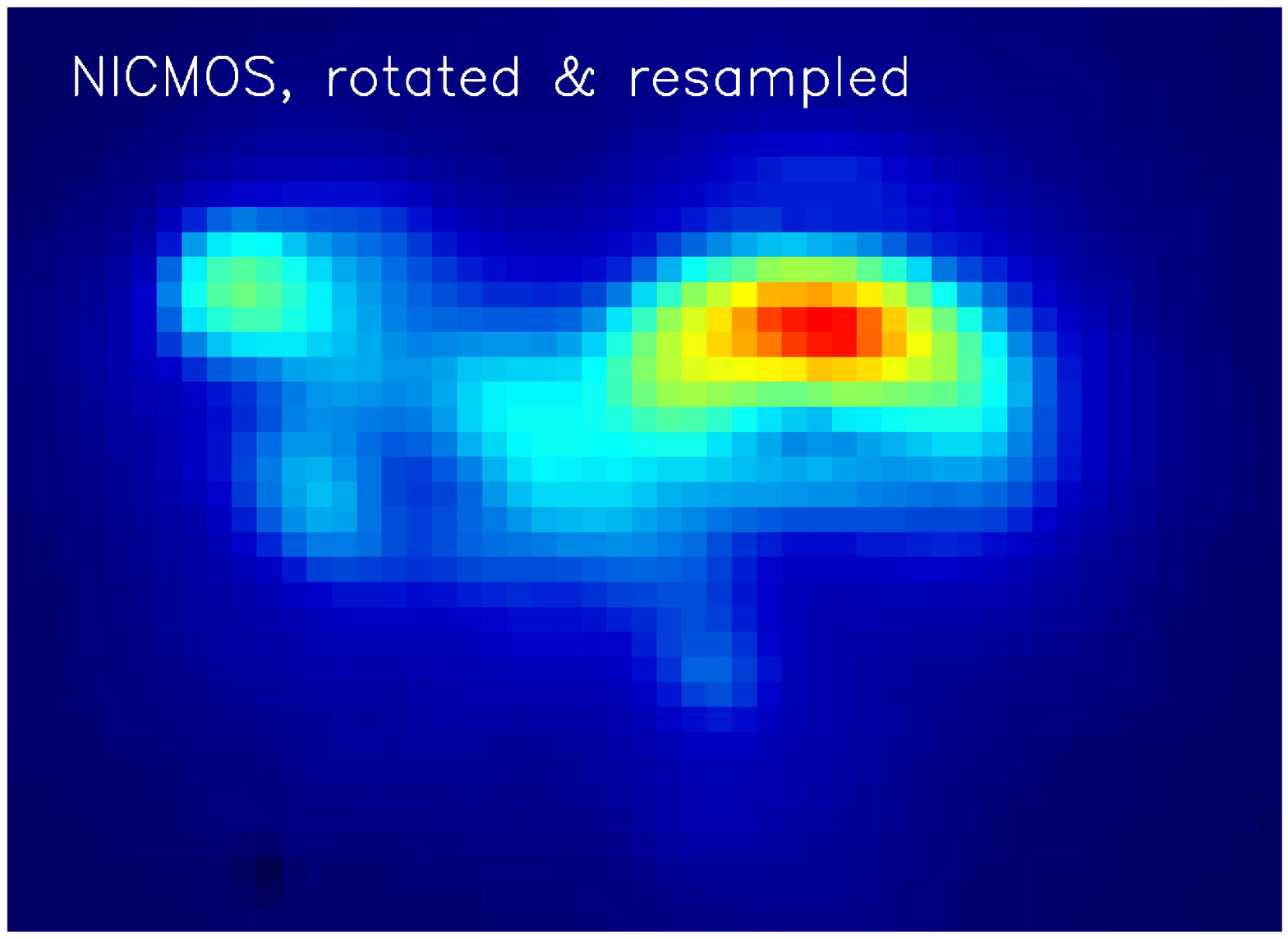} 
\includegraphics[width=0.31\textwidth]{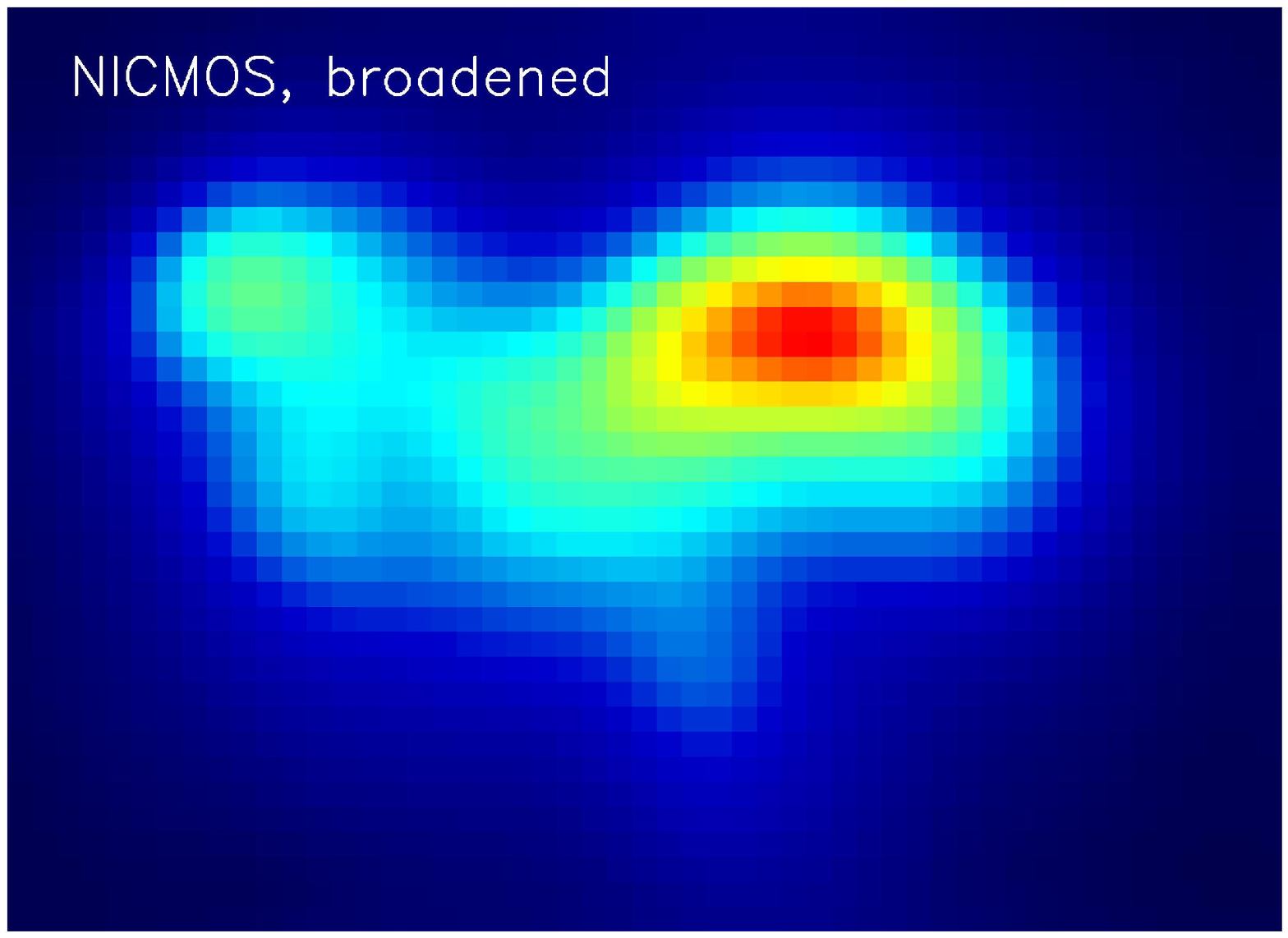}
\caption{We estimate the PSF of our AO data by finding the broadening function which, applied to a higher-resolution image with known PSF, best matches the AO image. The AO PSF is then estimated by convolving this broadening function with the PSF of the high-resolution image. Left: Our SINFONI K-band image. Middle: Archival NICMOS K-band image \citep{scoville98}, rotated and resampled to the SINFONI pixel scale. Right: Best-matching broadened NICMOS image. We find the PSF to be well represented by a slightly asymmetric Gaussian with major axis oriented $4^\circ$ east of north and FWHM 0.30\arcsec$\times$\,0.31\arcsec.\label{fig:psf}}
\end{center}
\end{figure}

In our case, we use \textit{HST} NICMOS 2.16\,$\mu$m observations retrieved from the \textit{HST} archives. Fig.~\ref{fig:psf} shows our SINFONI data, the \textit{HST} data (rotated and resampled to the SINFONI pixel scale), and the broadened \textit{HST} image. We find the PSF to be well represented by an almost perfectly symmetric Gaussian with major axis oriented $4^\circ$ east of north and FWHM 0.30\arcsec$\times$0.31\arcsec -- a good performance for laser guide star assisted adaptive optics without a tip-tilt star \citep{dav08b}.

\subsubsection{Spatial binning}
\label{sec:obs:binning}

In order to achieve a mininum signal-to-noise, we spatially binned each spectral plane of our reduced data cube using an optimal Voronoi tessellation \citep{cap03}, which bins pixels together into groups by accreting new pixels
to each group until a pre-set signal-to-noise-cutoff is reached.

The signal-to-noise cutoff was chosen such as not to compromise spatial resolution in the central regions (i.e.~such that no binning was performed there), but at the same time extending the region in which meaningful analyses can be performed. As a result of this, the outer regions in e.g.~the velocity maps appear in blocks, rather than individual pixels.

\subsubsection{Extracting Stellar Kinematics}
\label{sec:obs:stellar}

The K-band covers the stellar CO absorption bandheads
longward of 2.29\,$\mu$m, whose sharp blue edges are very sensitive to stellar
motions. Here we utilise CO\,2-0 and CO\,3-1 to derive 2D~maps of the stellar velocity and
dispersion. After normalising the spectra with respect to a linear
fit to the line-free continuum, we subtract unity and thus set the continuum level to zero.
A stellar template spectrum, of the M3\,III star HD\,176617, was prepared analogously.
As we discuss in \cite{engel10a}, with this method -- setting the continuum level to zero -- the choice of template star is not critical, with even significantly mis-matched templates resulting in errors of at most only a few percent. Here, the template is matching the bandheads quite well. We furthermore note that finding a single template (or a composite template) that provides an excellent match everywhere is almost impossible to find, due to the large spatial variations in EW$_{CO}$ seen in this system and also e.g.~in NGC\,6240 \citep{engel10a}.

In order to obtain the 2D velocity and dispersion maps, we fit the template to the spectrum of each spatial pixel (spaxel hereafter) by convolving it with a Gaussian, varying the Gaussian parameters until a best match is achieved. The Gaussian position and width then yield the stellar velocity and dispersion at that spaxel. During the fitting, the spectral regions covering the steep edges of the bandhead at all
expected stellar velocities are given five-fold weight, in order to focus the fit on the
kinematics rather than details in the spectrum.
The spectral ranges used to compute the CO$_{2-0}$ bandhead equivalent width (EW$_{CO}$ hereafter) are those of \cite{for00}; these ranges are
also used by the stellar synthesis code STARS employed later. We note that we do not attempt to extract the  Gauss-Hermite terms $h_3$ and $h_4$, since the signal-to-noise across most of our spectra is not sufficient to include these in the fit.

\begin{figure}
\begin{center}
\includegraphics[width=0.35\textwidth]{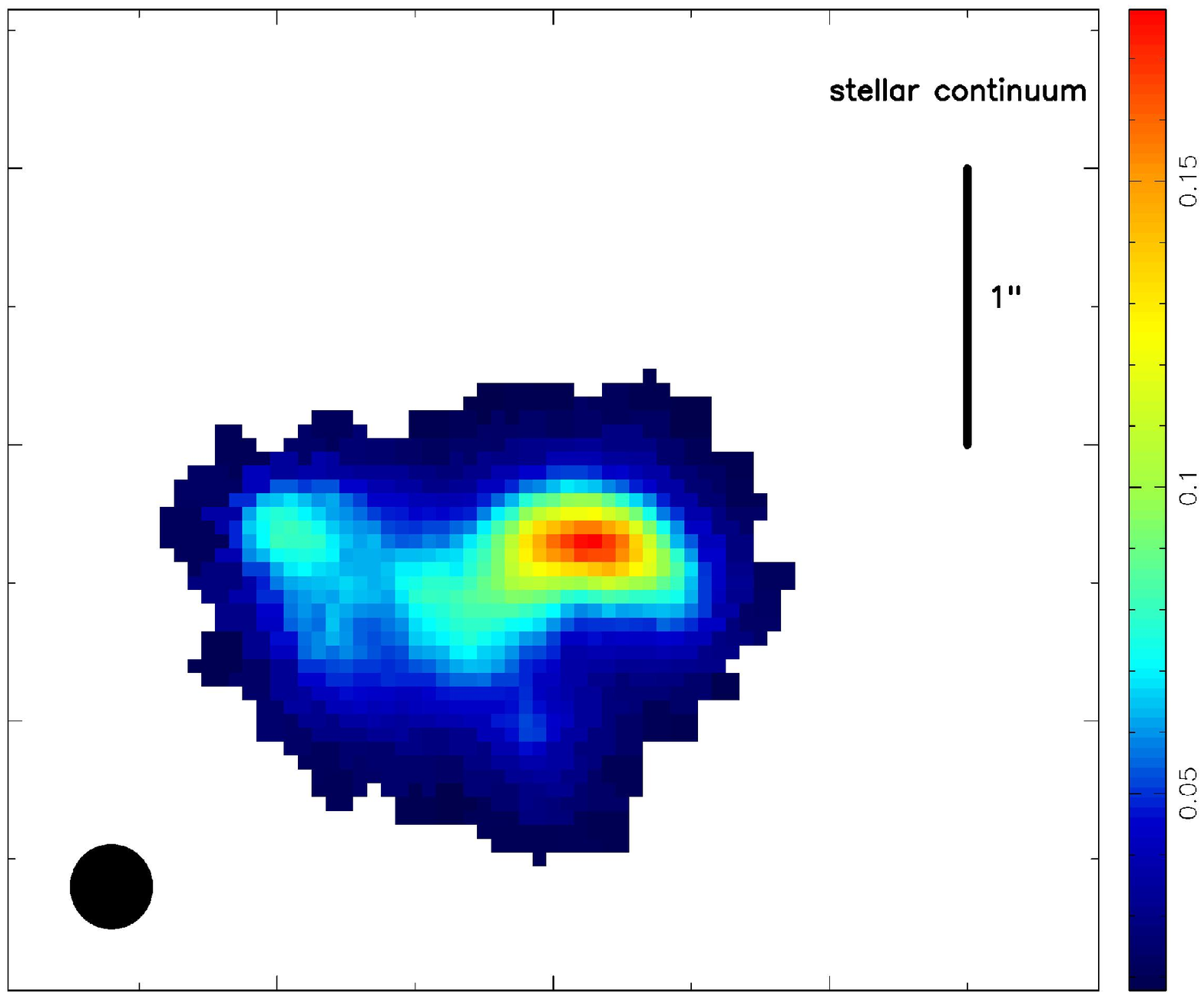} 
\includegraphics[width=0.35\textwidth]{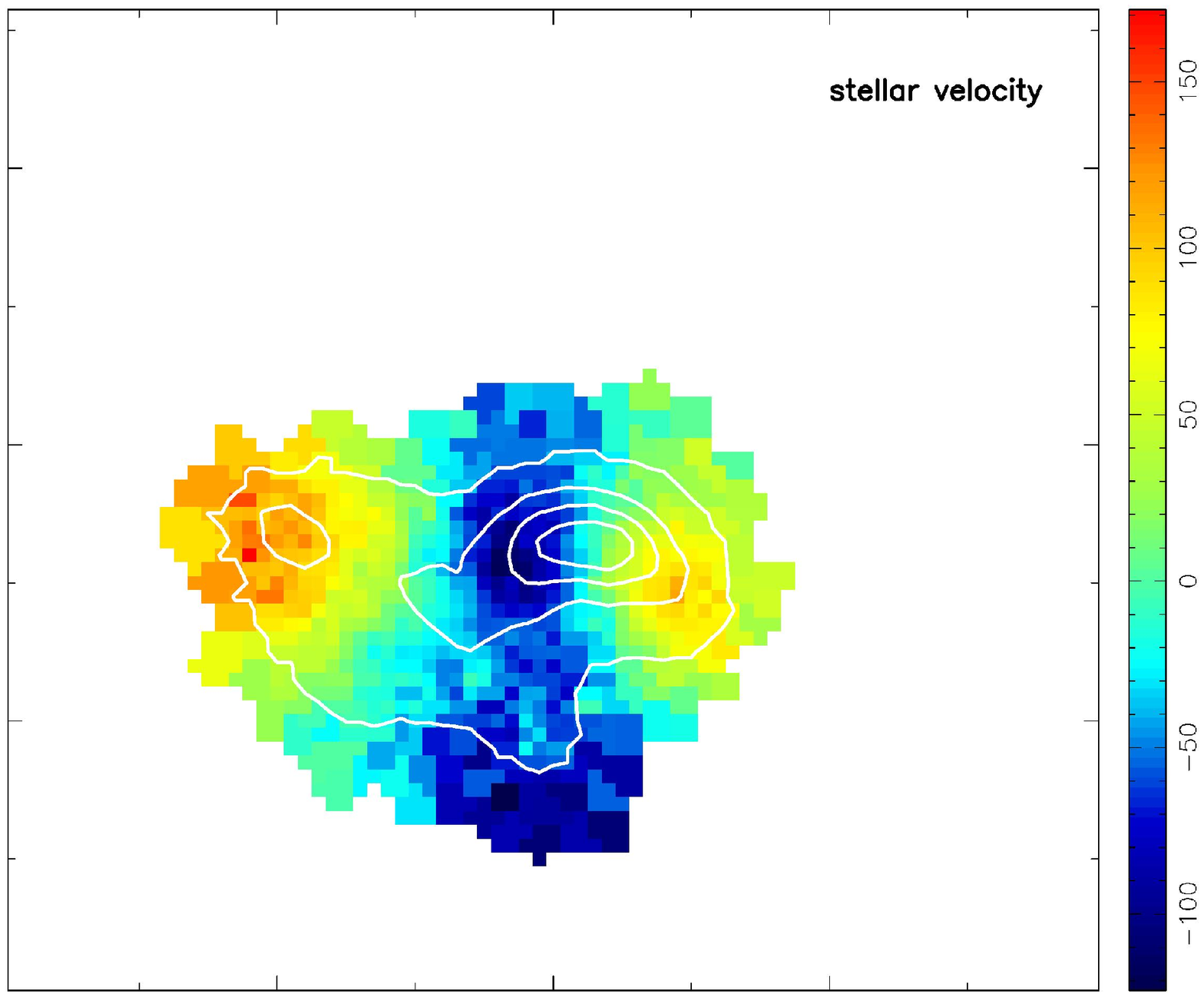} 
\includegraphics[width=0.35\textwidth]{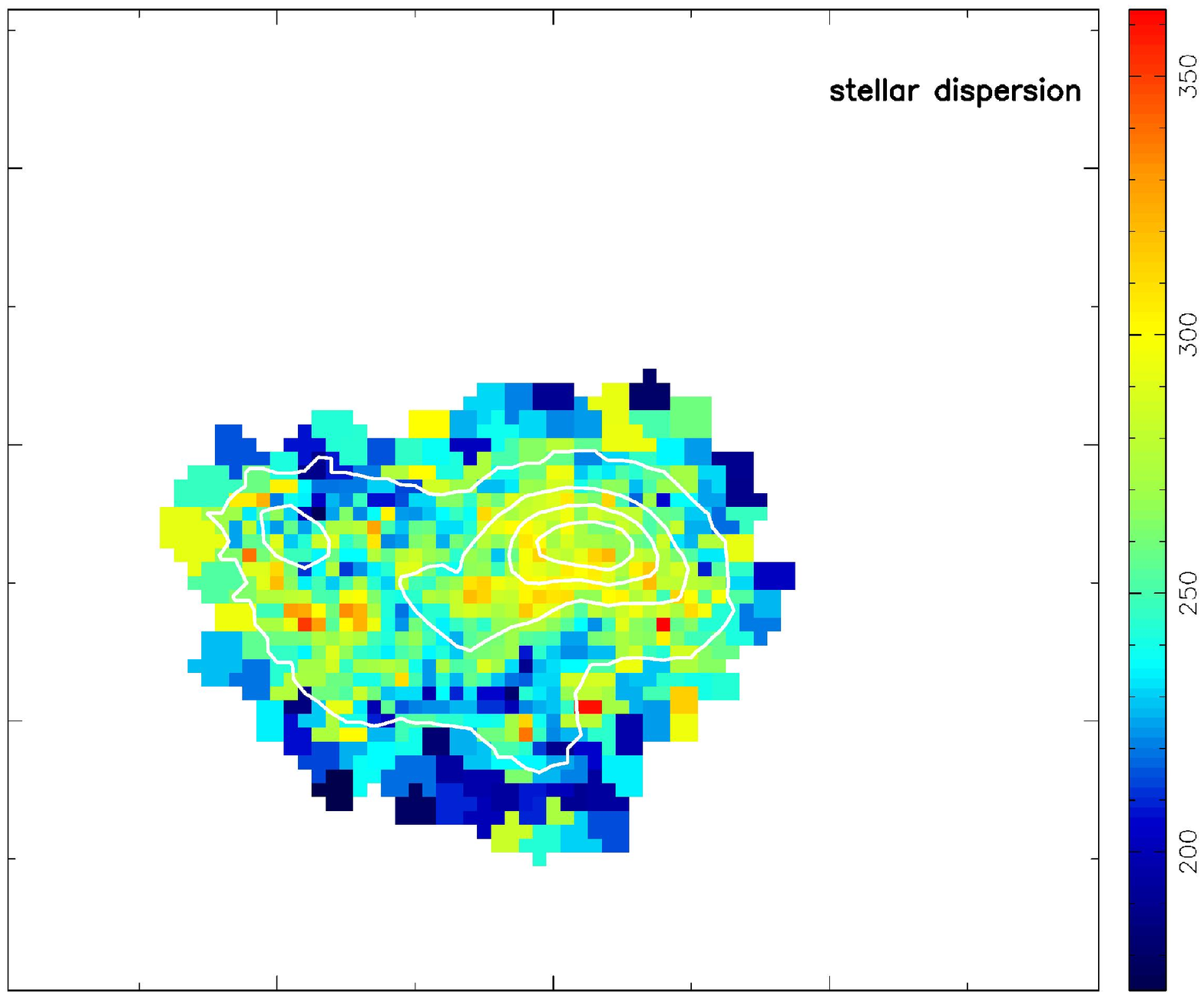} 
\includegraphics[width=0.35\textwidth]{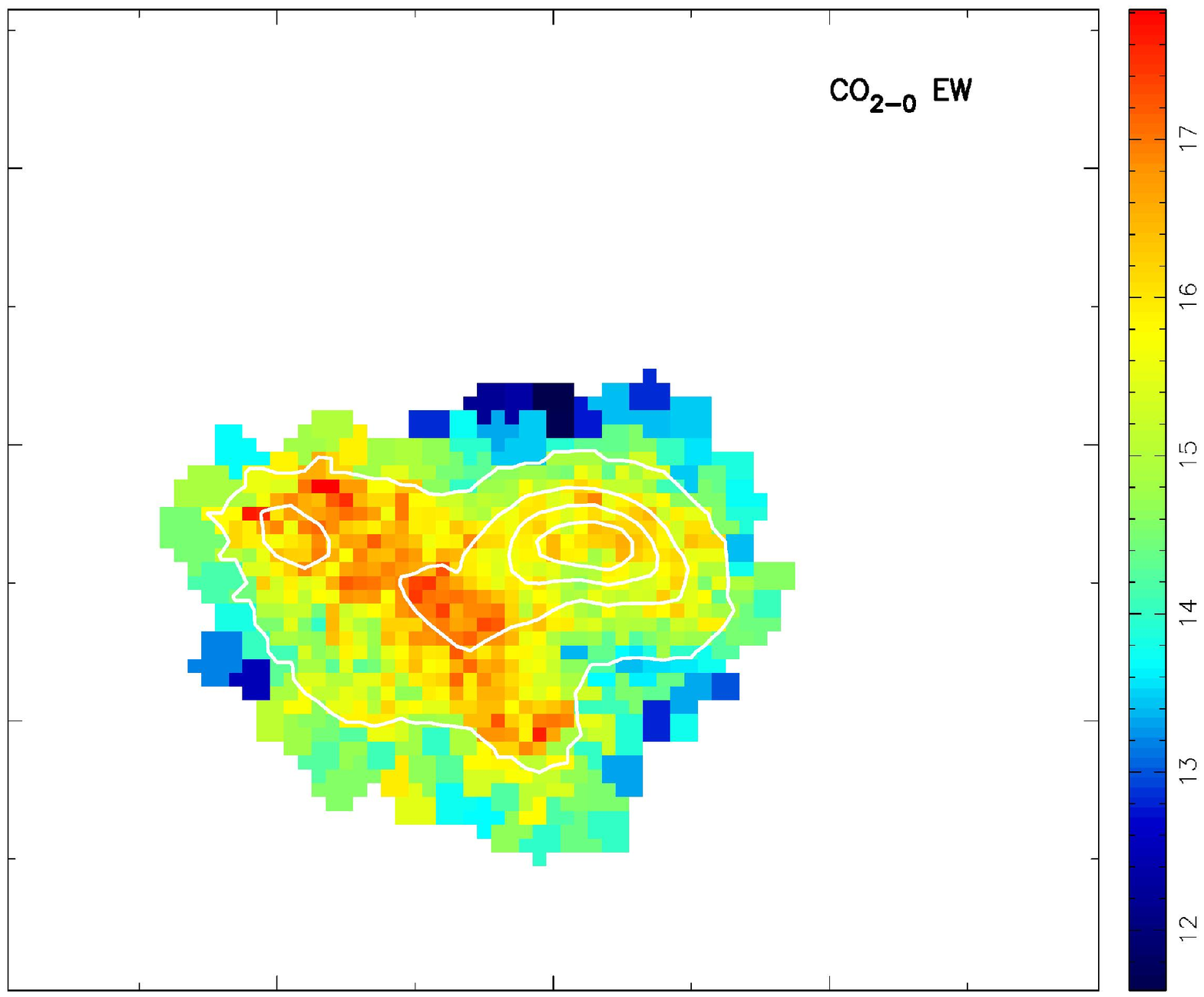} 
\caption{Stellar continuum, velocity, velocity dispersion, and EW$_{CO}$ maps.
White contours trace the continuum.
The bar indicates 1\arcsec, the PSF is displayed in the lower left of panel 1.
North is up and east is to the left.\label{fig:stellar}}

\end{center}
\end{figure}

The resulting final maps of stellar velocity, dispersion, and
EW$_{CO}$ are displayed in Fig.~\ref{fig:stellar}.

\subsubsection{H- \& K-band Gas Tracers}
\label{sec:obs:gas}

\begin{figure}
\begin{center}
\includegraphics[width=0.21\textwidth]{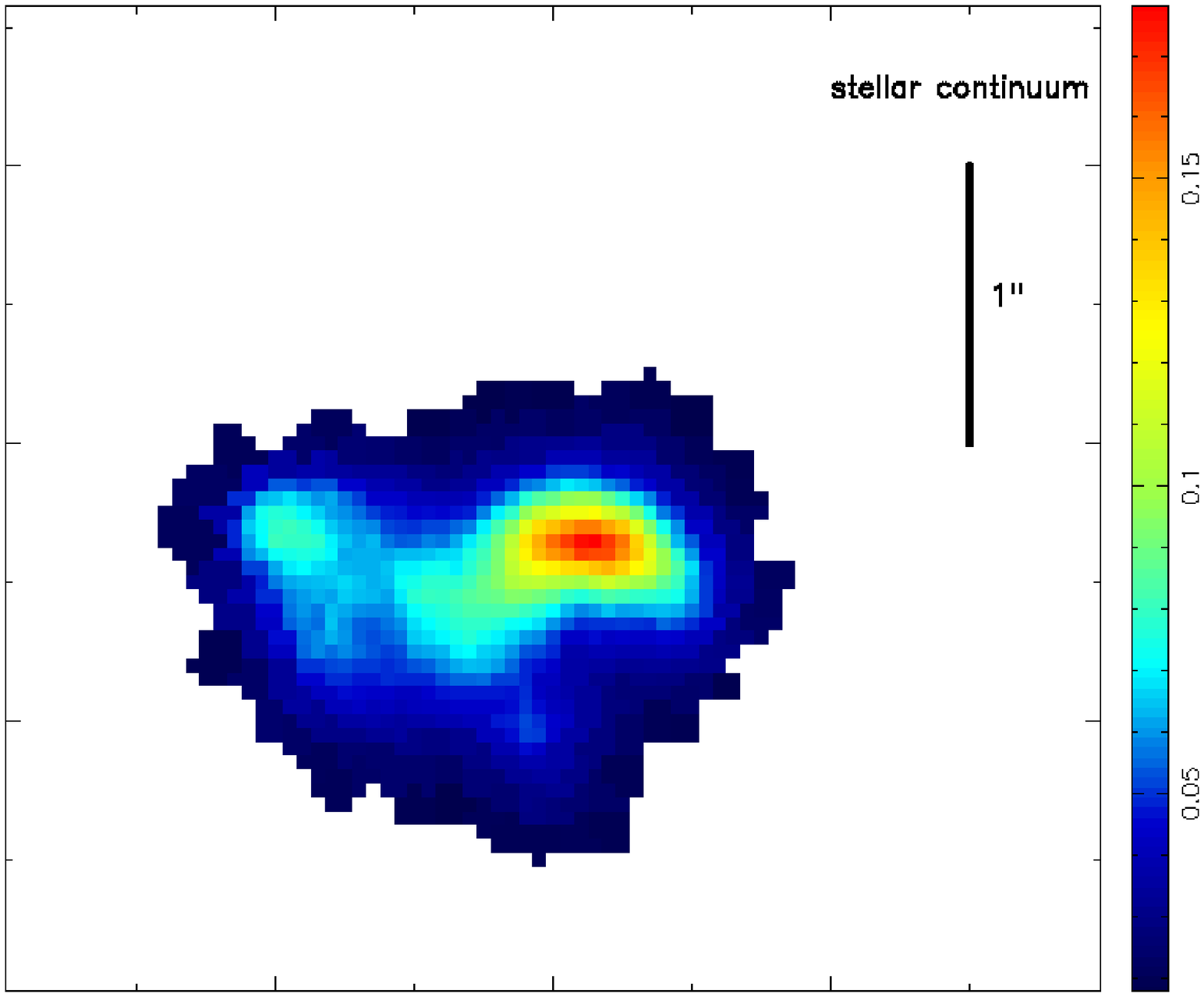}
\includegraphics[width=0.21\textwidth]{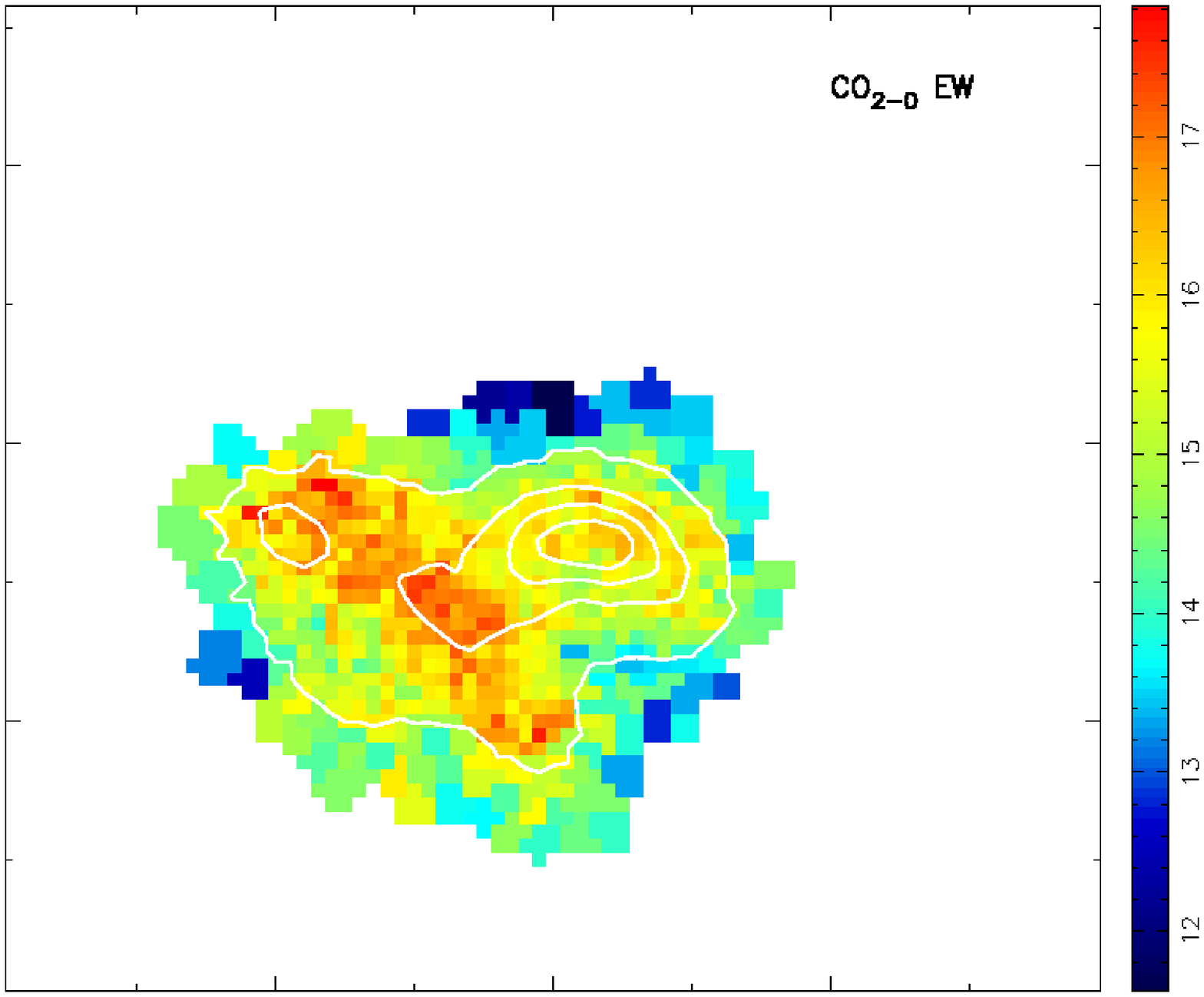}
\includegraphics[width=0.21\textwidth]{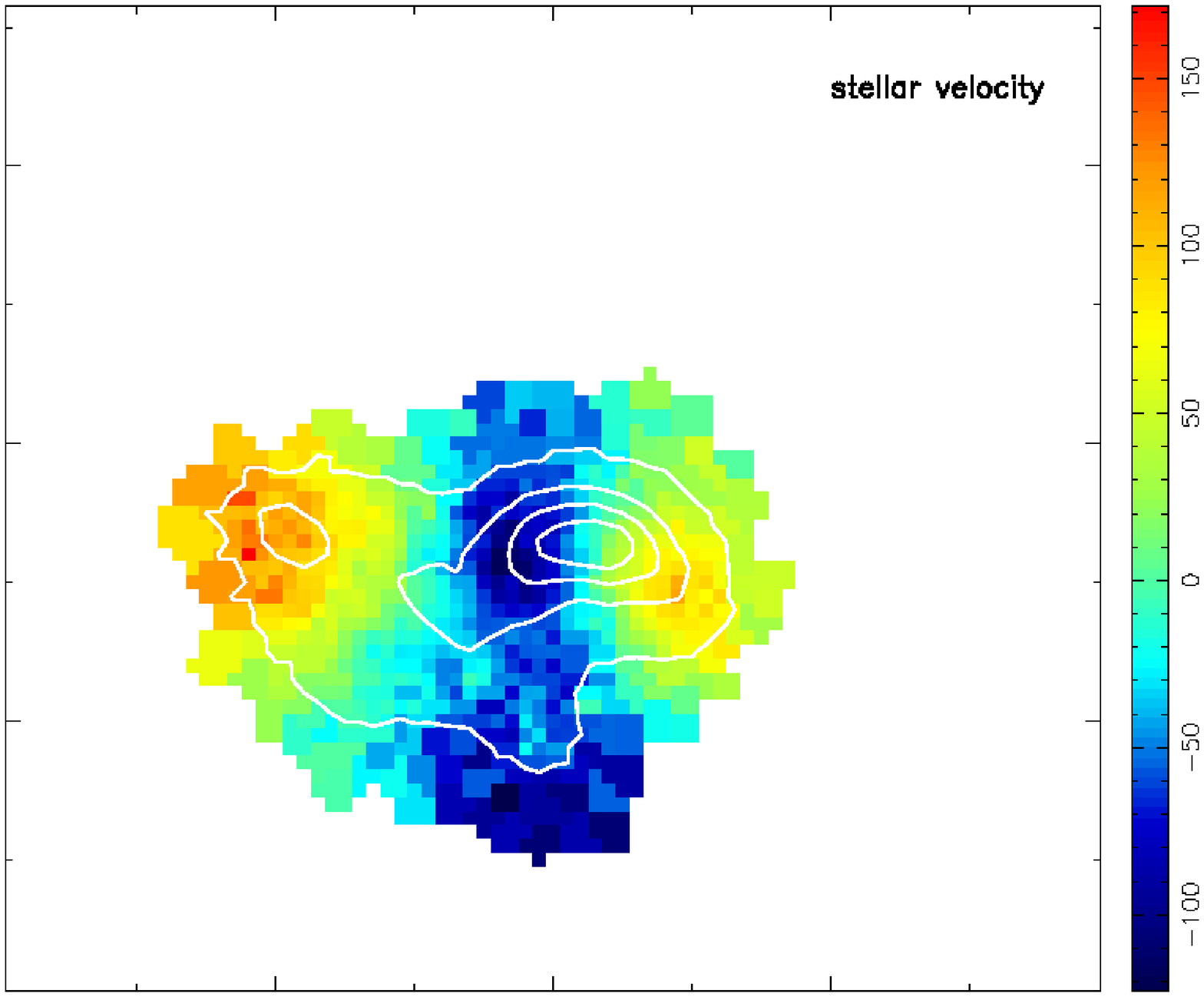}
\includegraphics[width=0.21\textwidth]{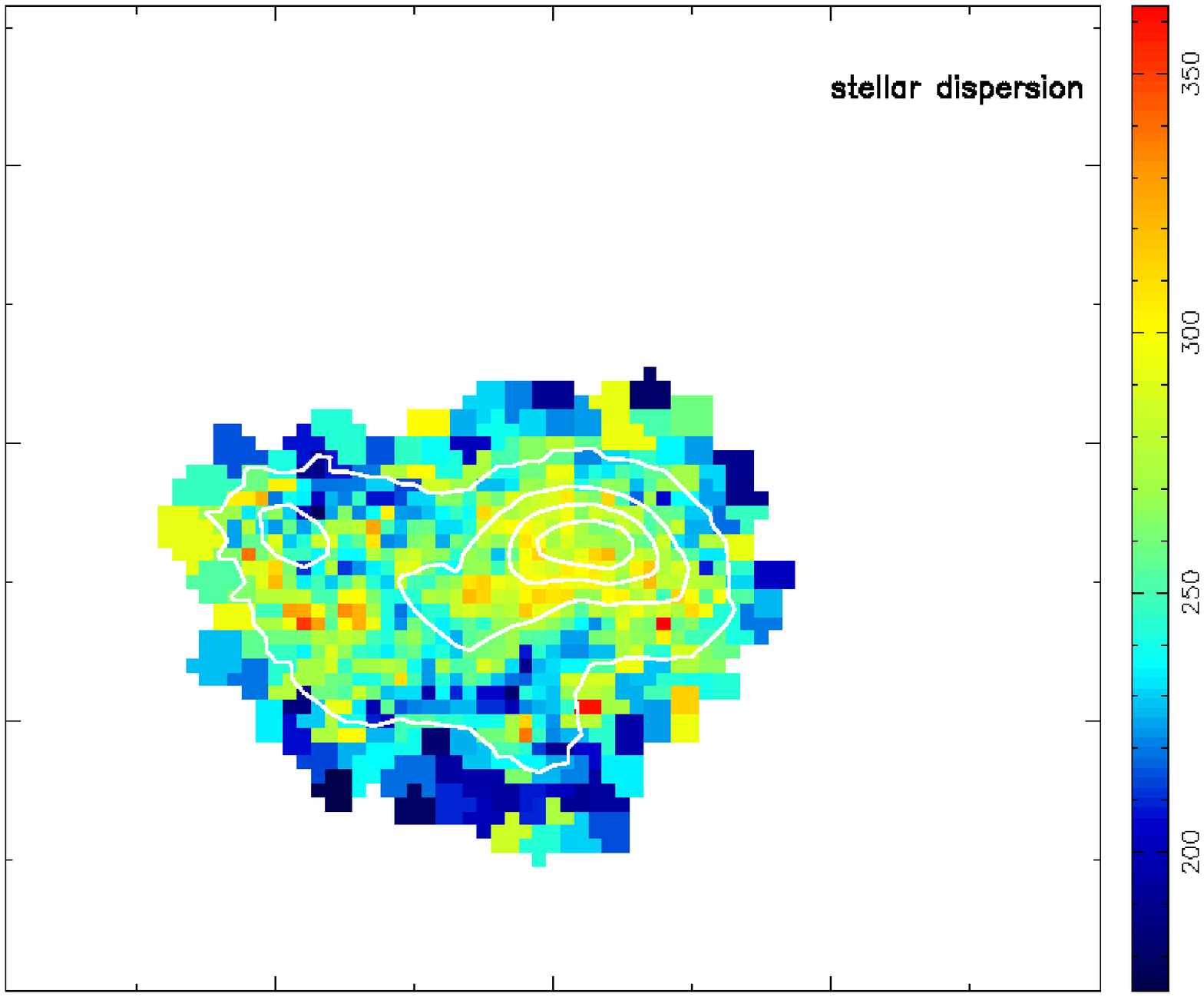}
\includegraphics[width=0.21\textwidth]{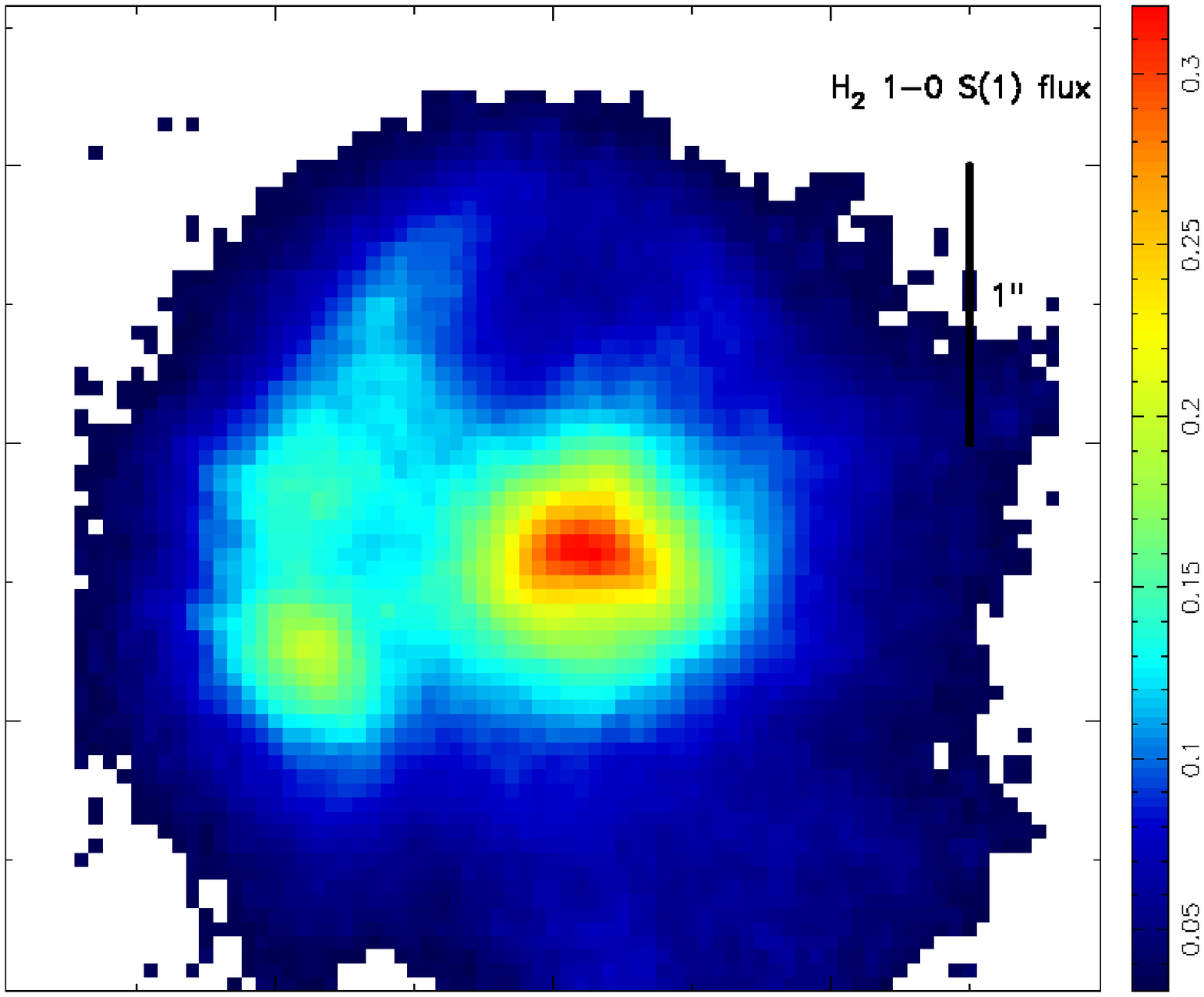} 
\includegraphics[width=0.21\textwidth]{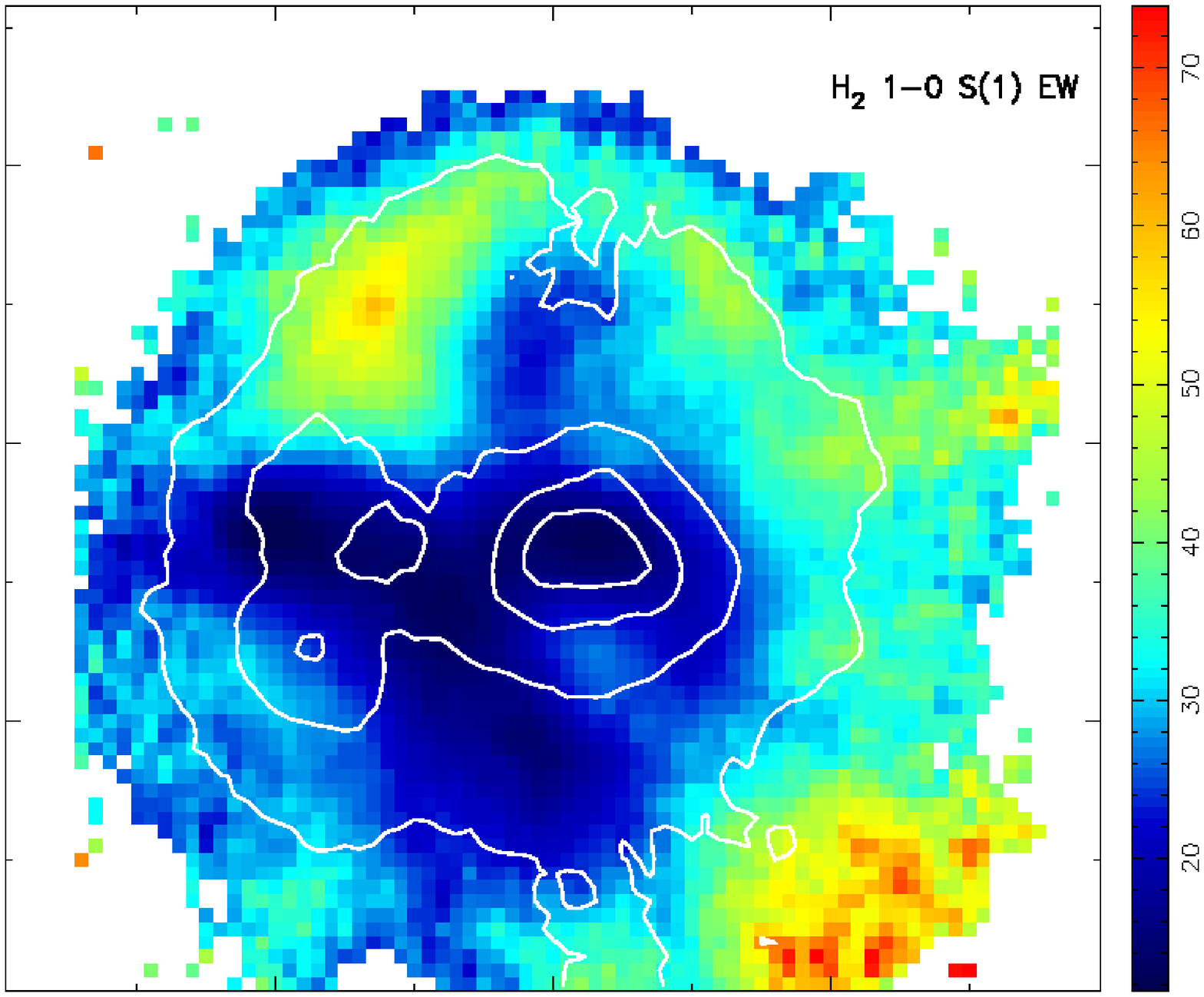} 
\includegraphics[width=0.21\textwidth]{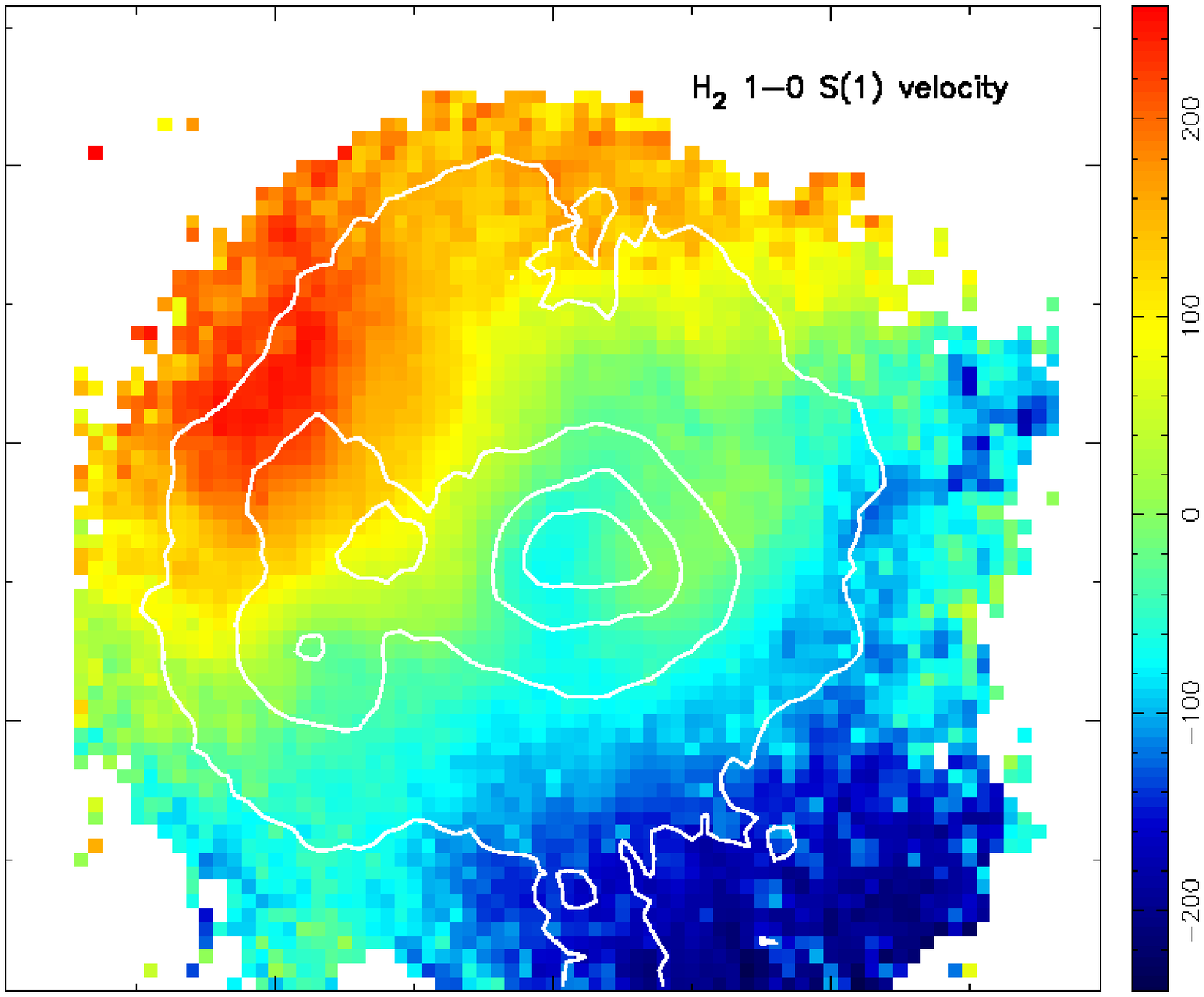}
\includegraphics[width=0.21\textwidth]{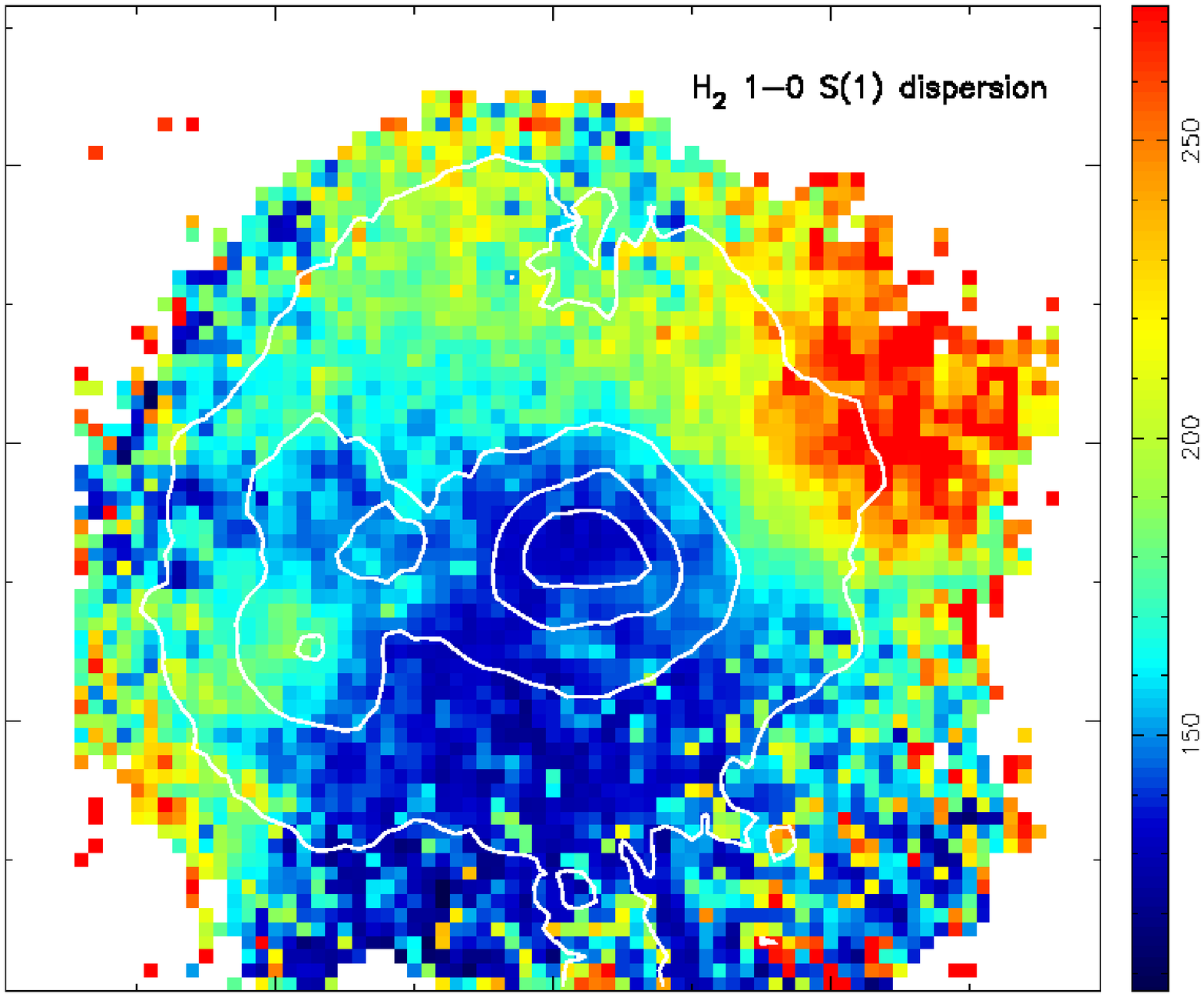}
\includegraphics[width=0.21\textwidth]{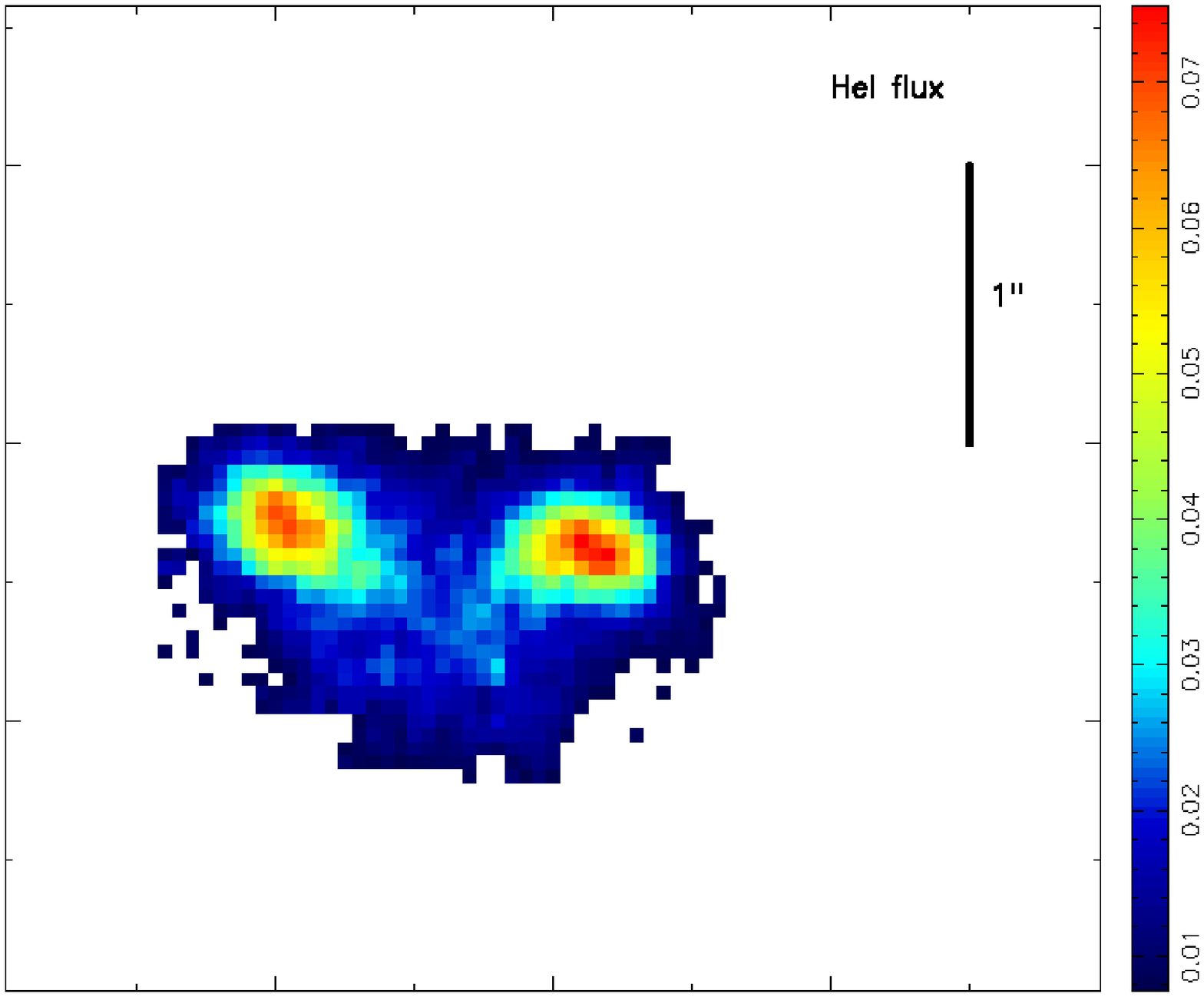} 
\includegraphics[width=0.21\textwidth]{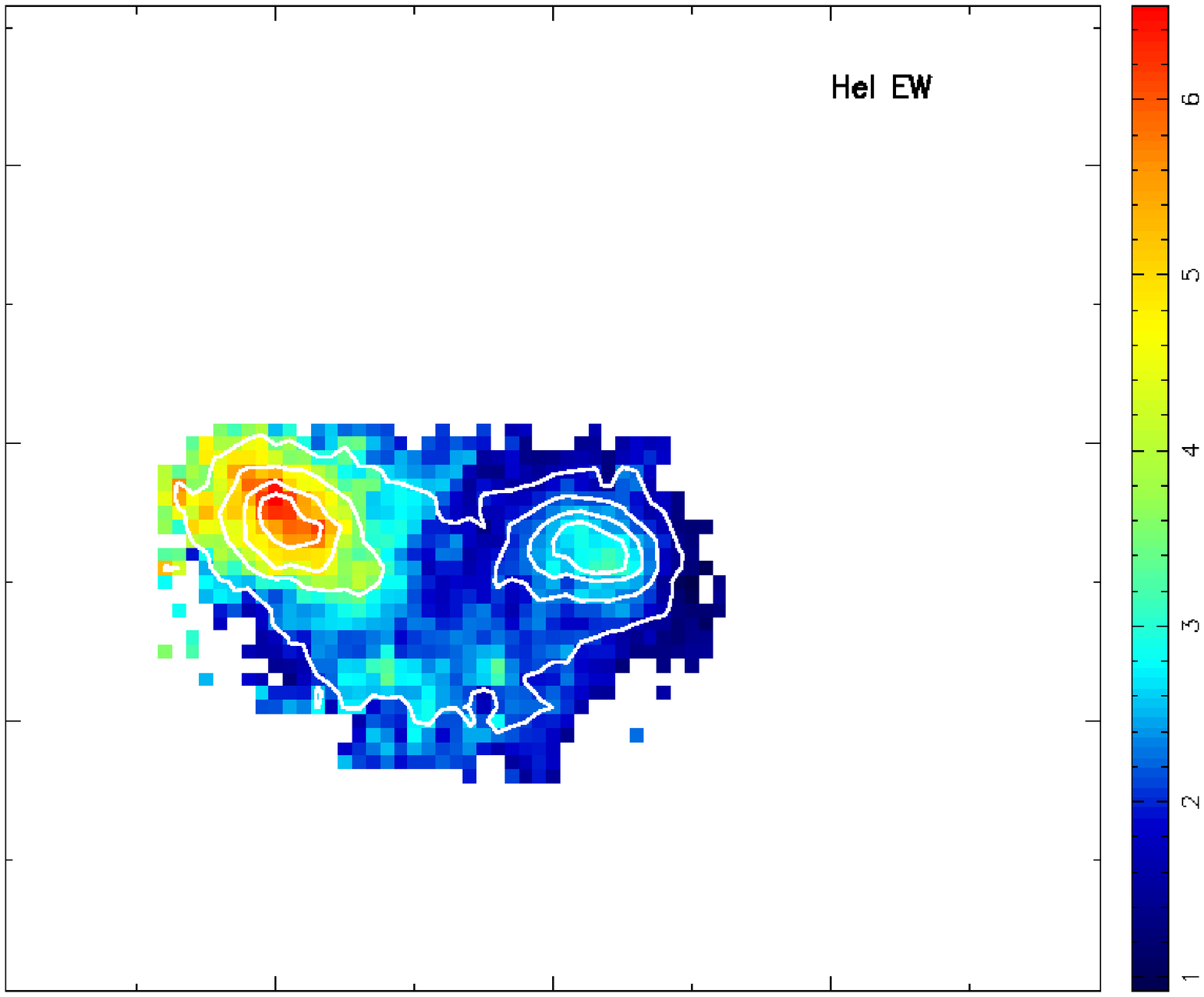} 
\includegraphics[width=0.21\textwidth]{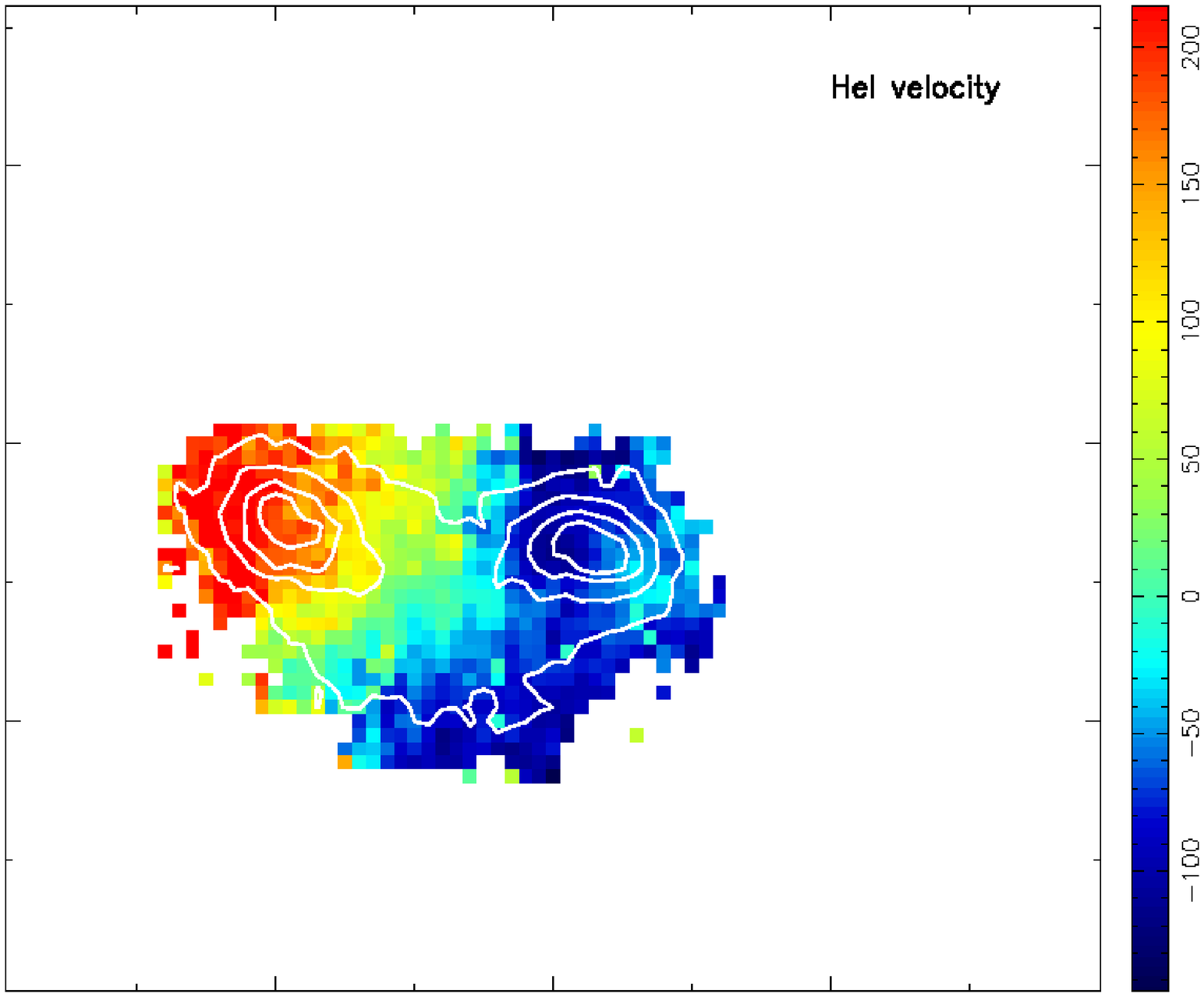}
\includegraphics[width=0.21\textwidth]{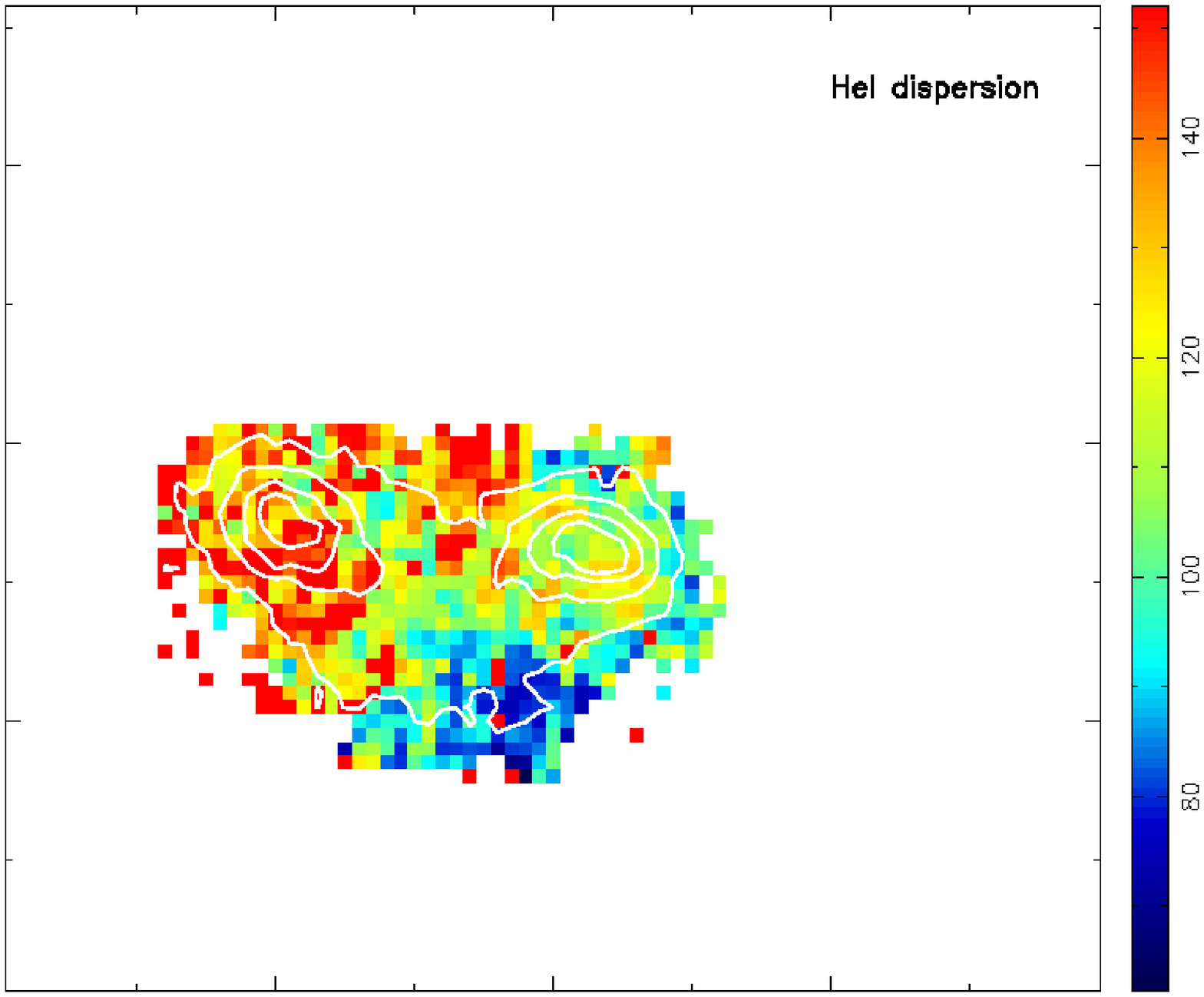} 
\includegraphics[width=0.21\textwidth]{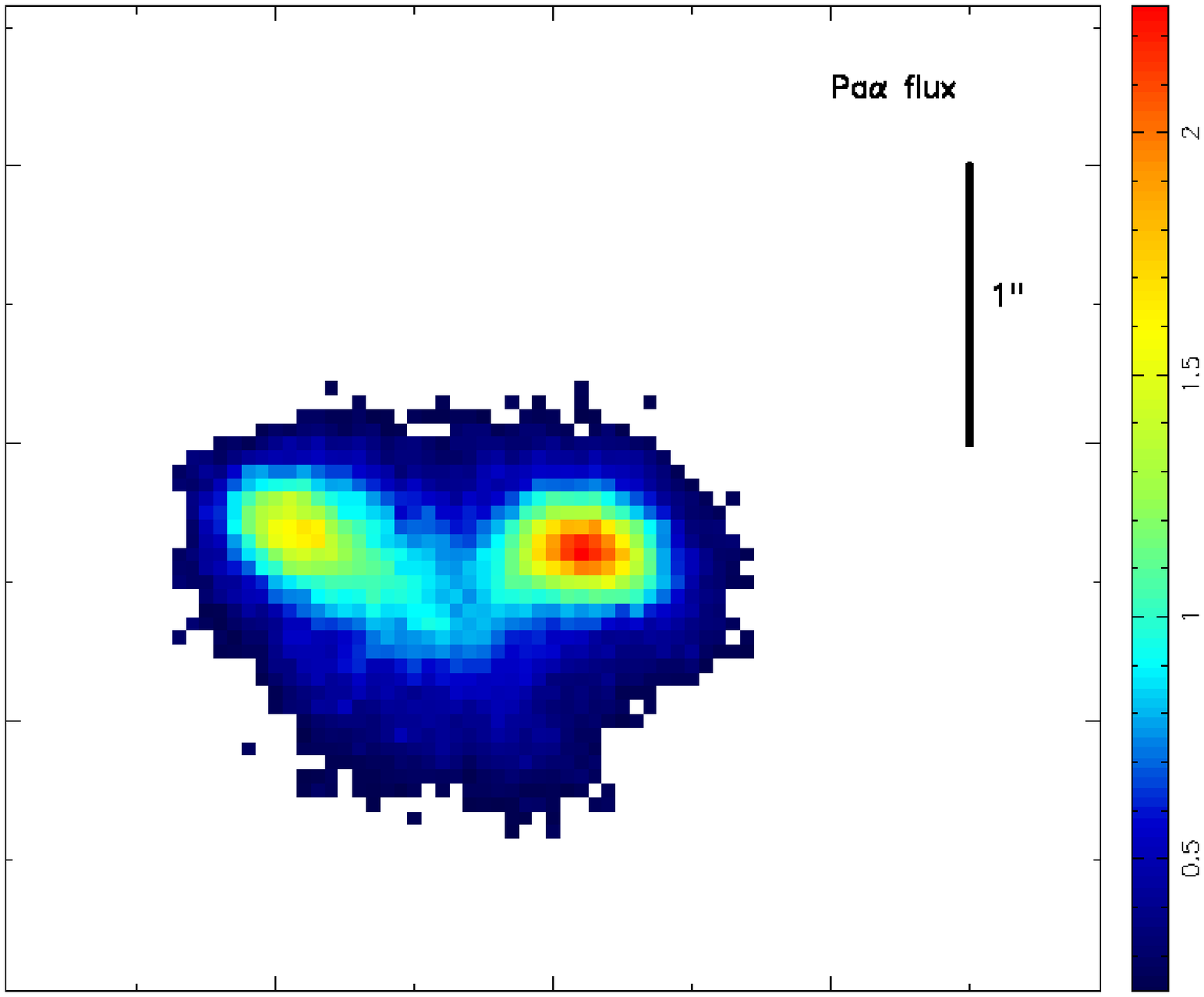}
\includegraphics[width=0.21\textwidth]{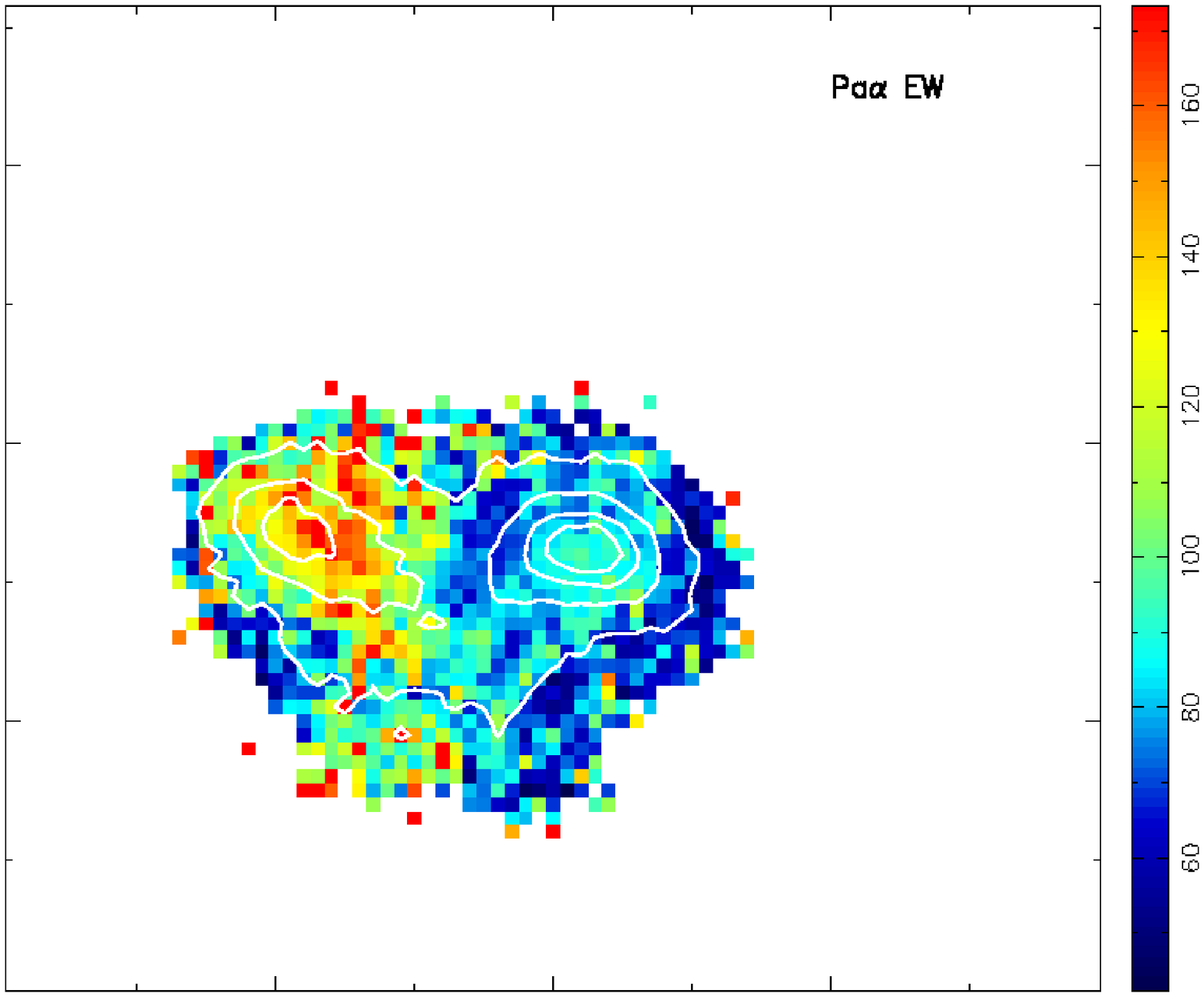}
\includegraphics[width=0.21\textwidth]{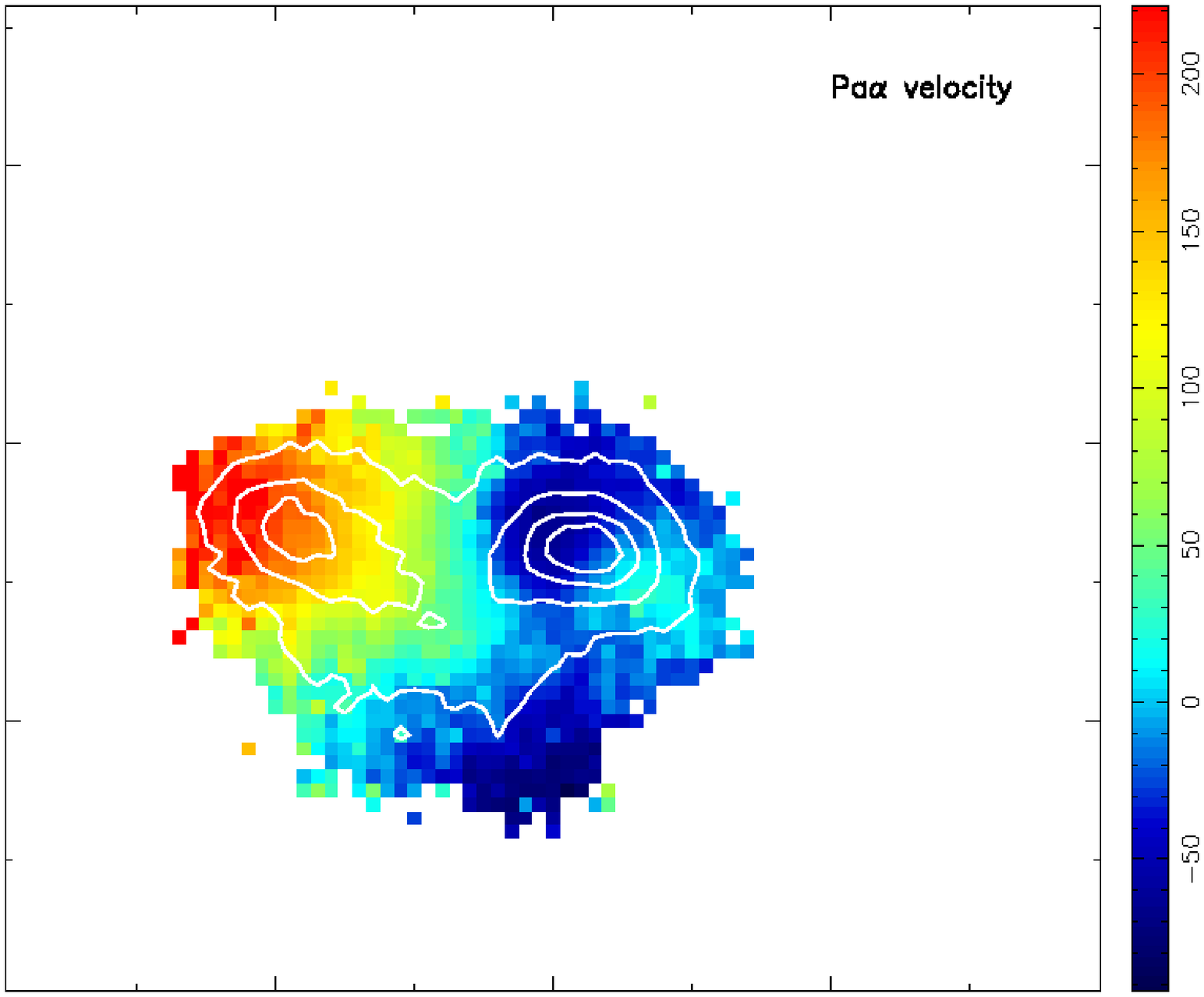}
\includegraphics[width=0.21\textwidth]{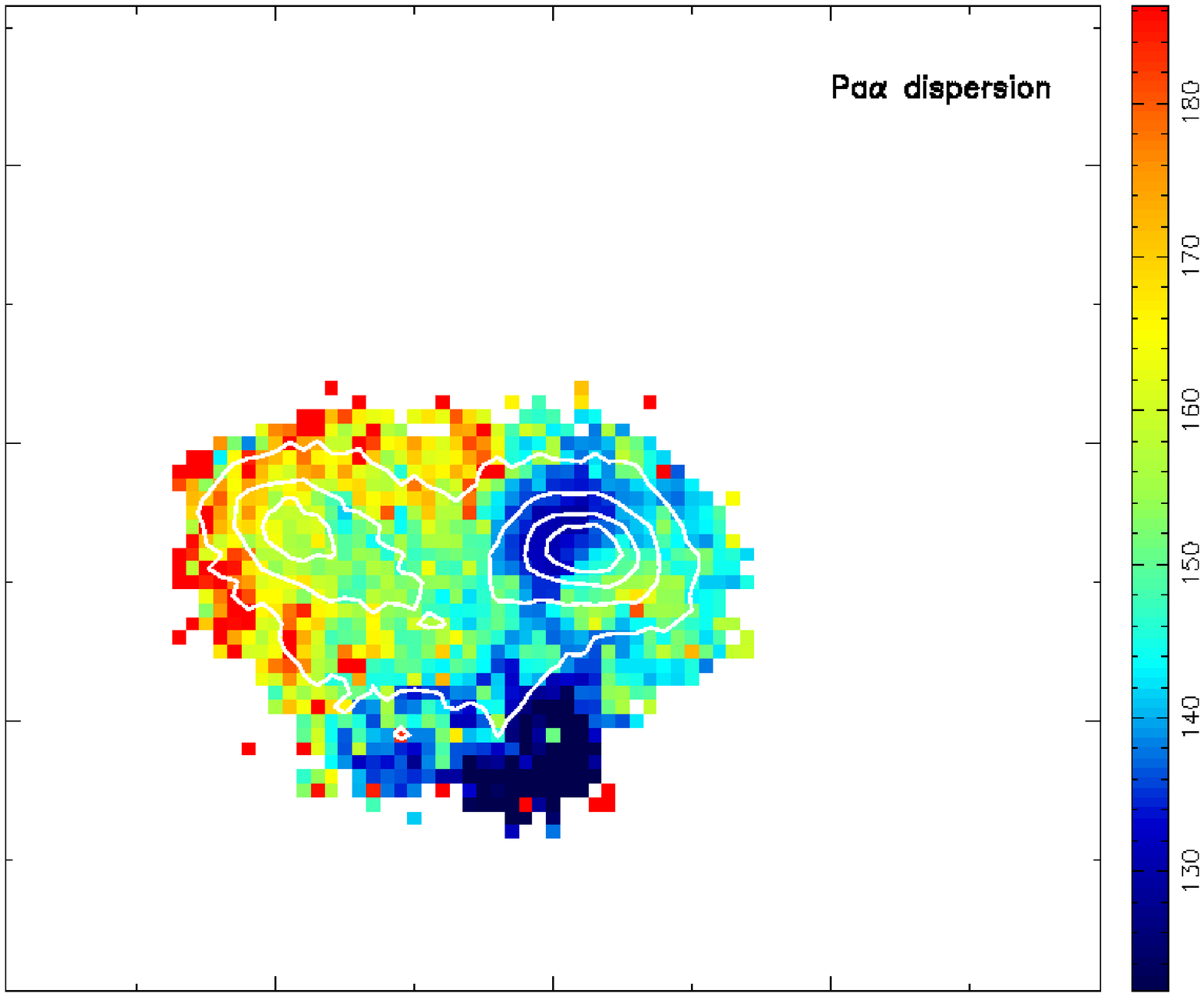}
\includegraphics[width=0.21\textwidth]{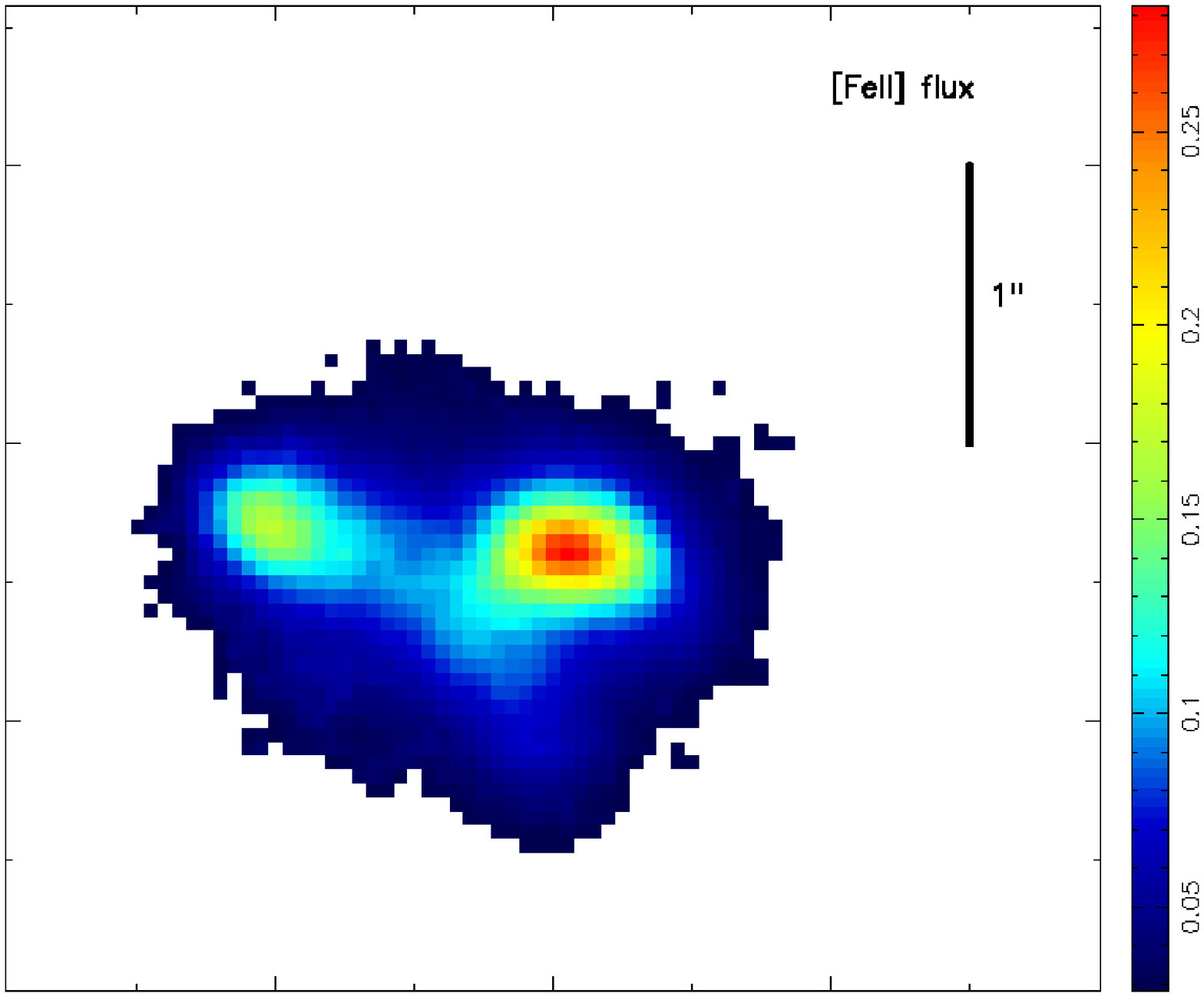}
\includegraphics[width=0.21\textwidth]{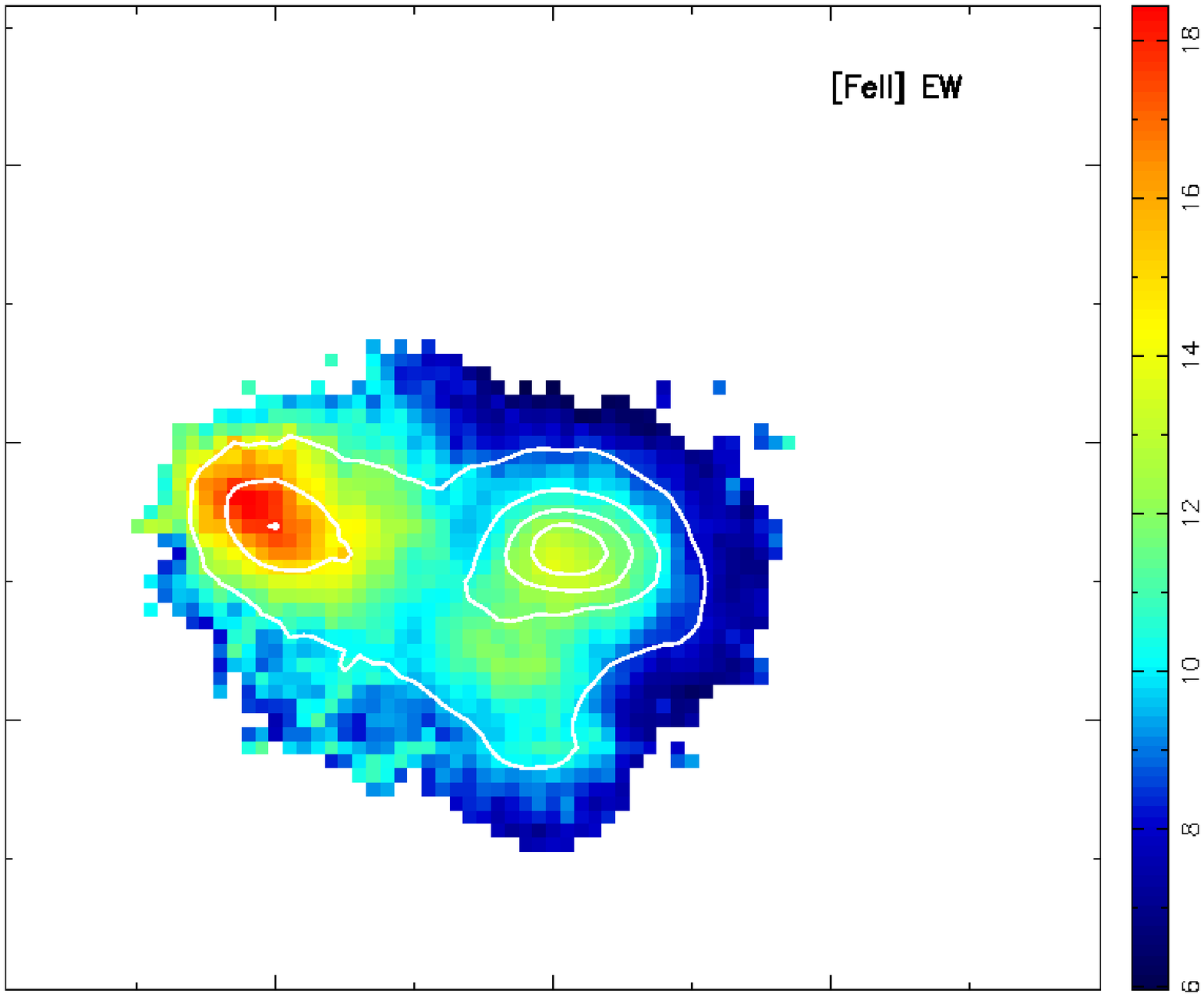}
\includegraphics[width=0.21\textwidth]{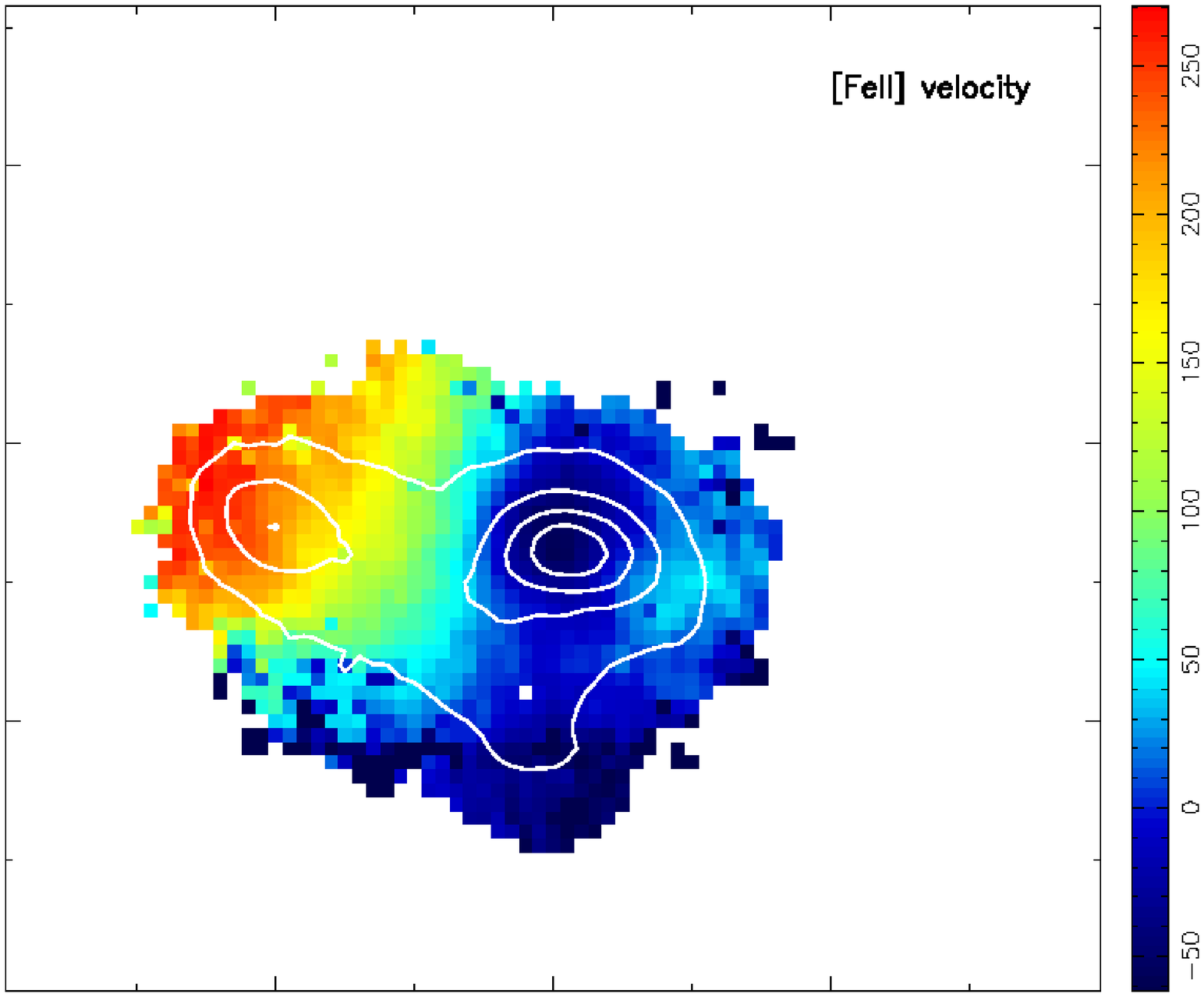}
\includegraphics[width=0.21\textwidth]{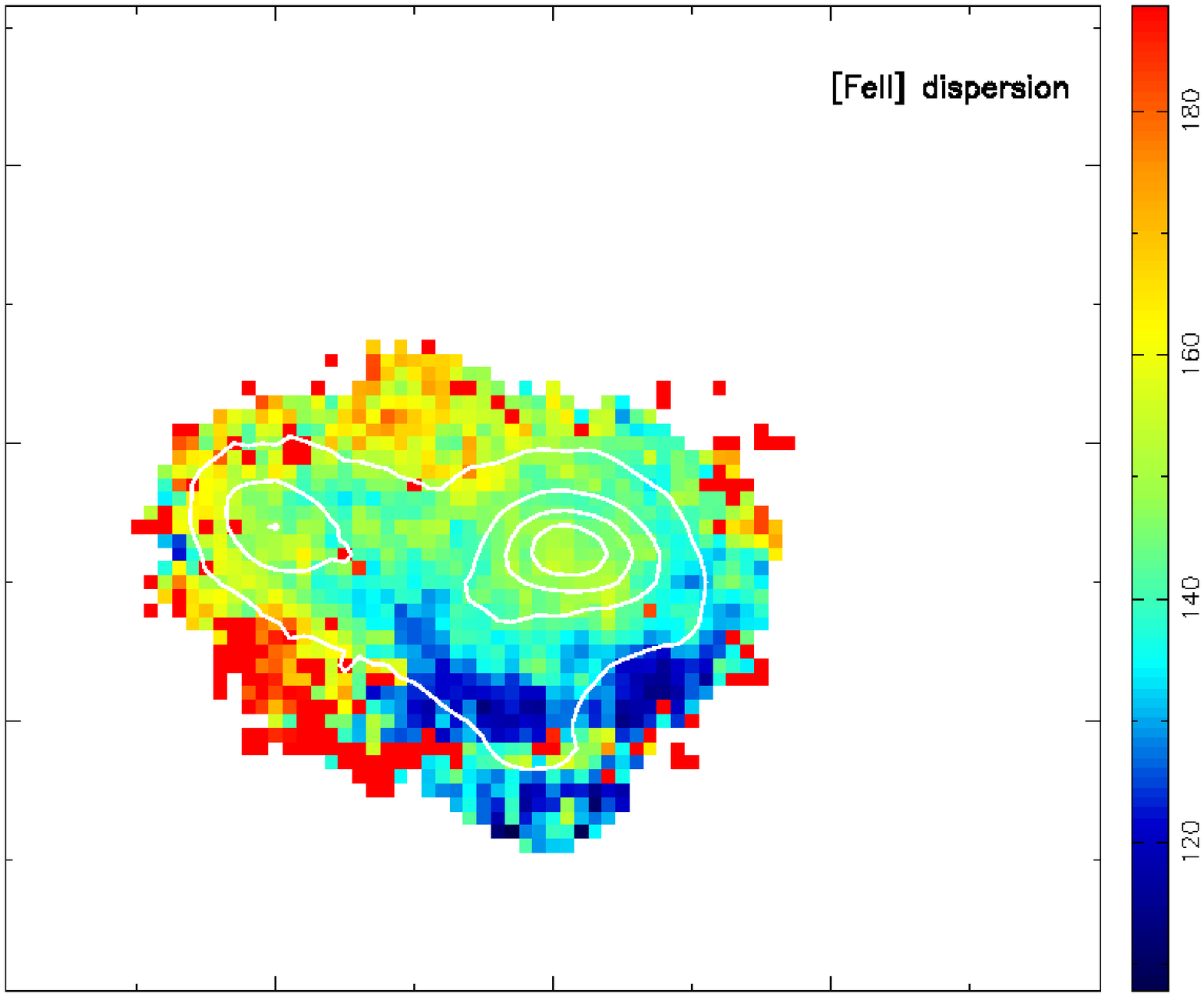}
\includegraphics[width=0.21\textwidth]{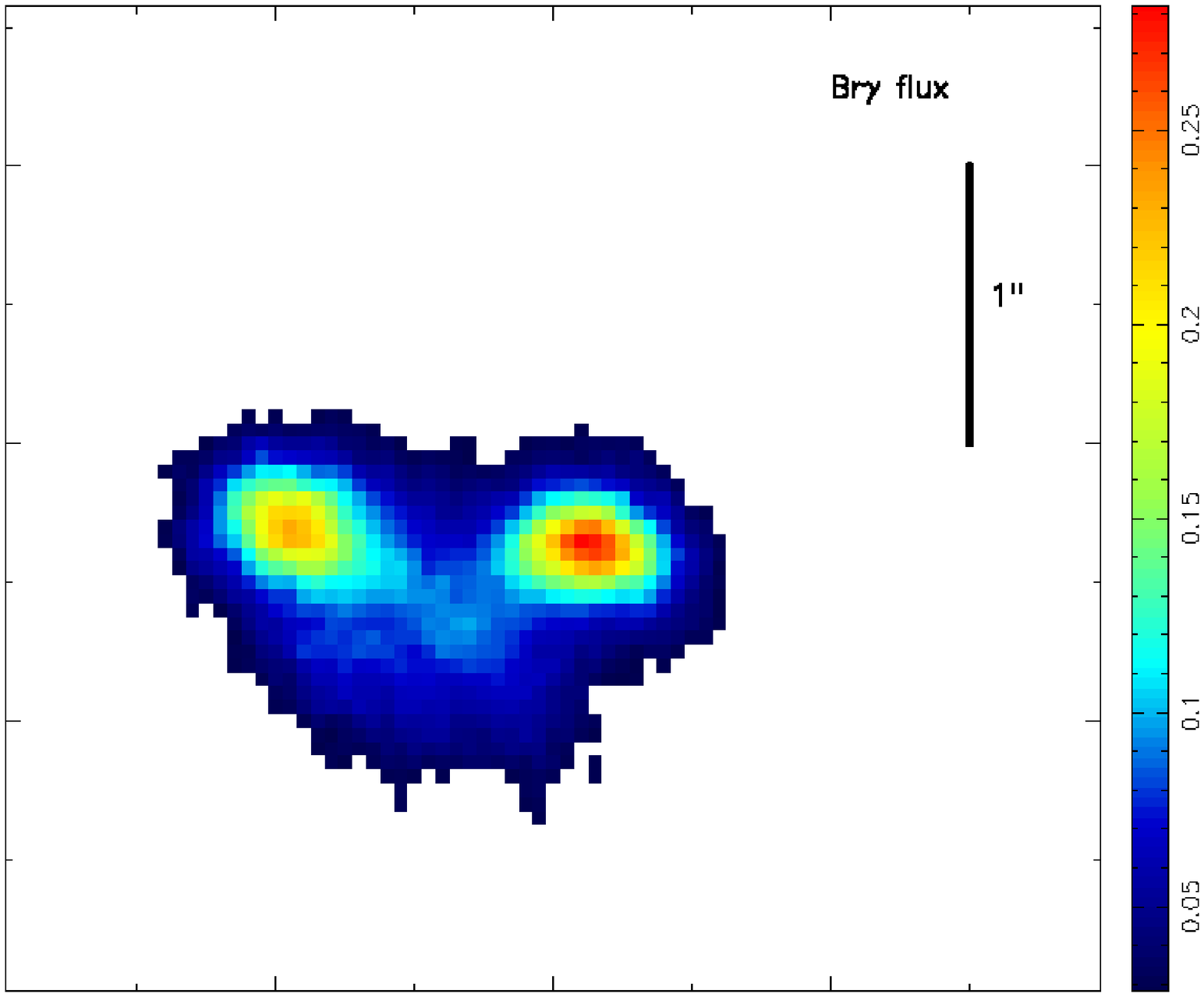}
\includegraphics[width=0.21\textwidth]{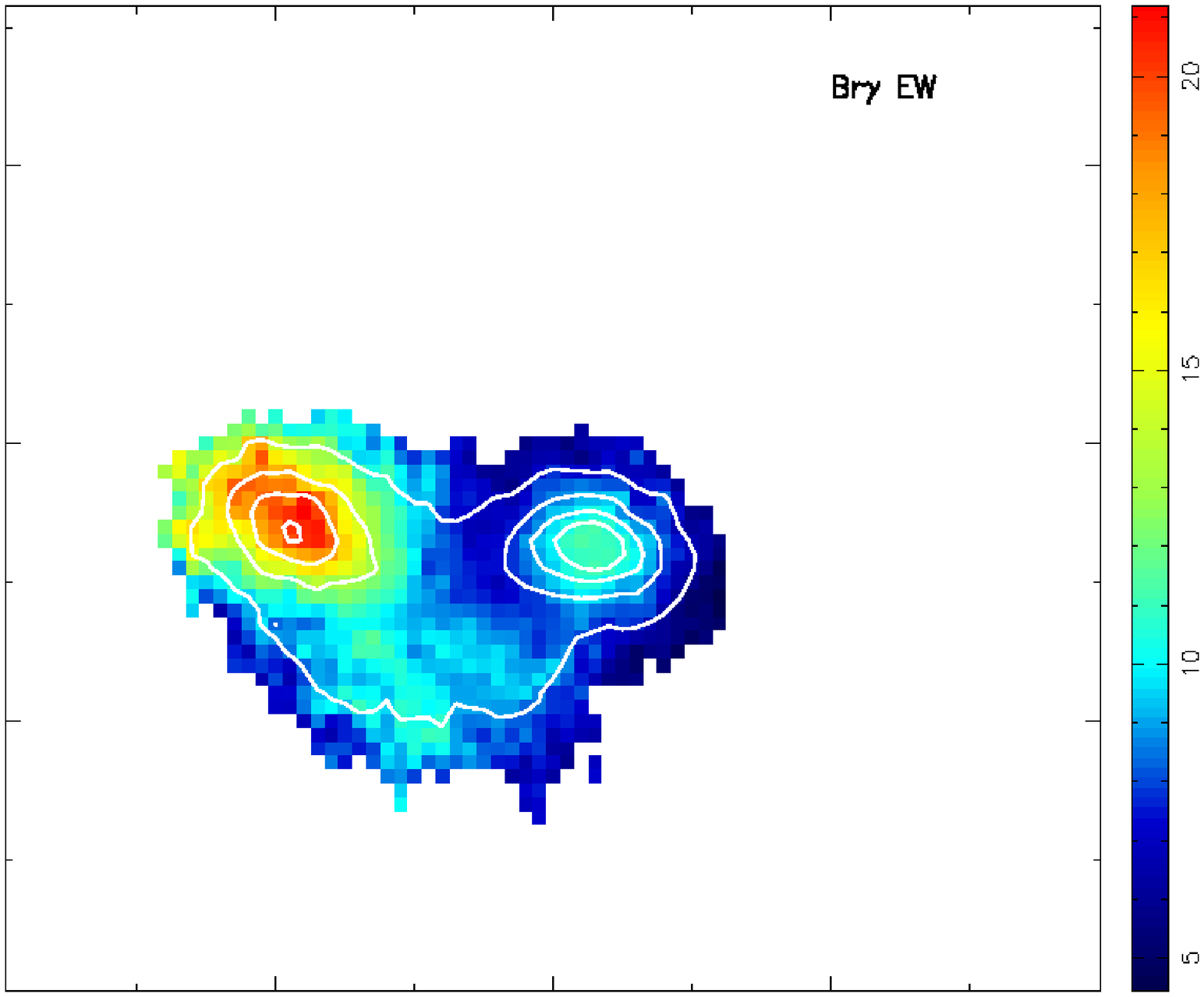}
\includegraphics[width=0.21\textwidth]{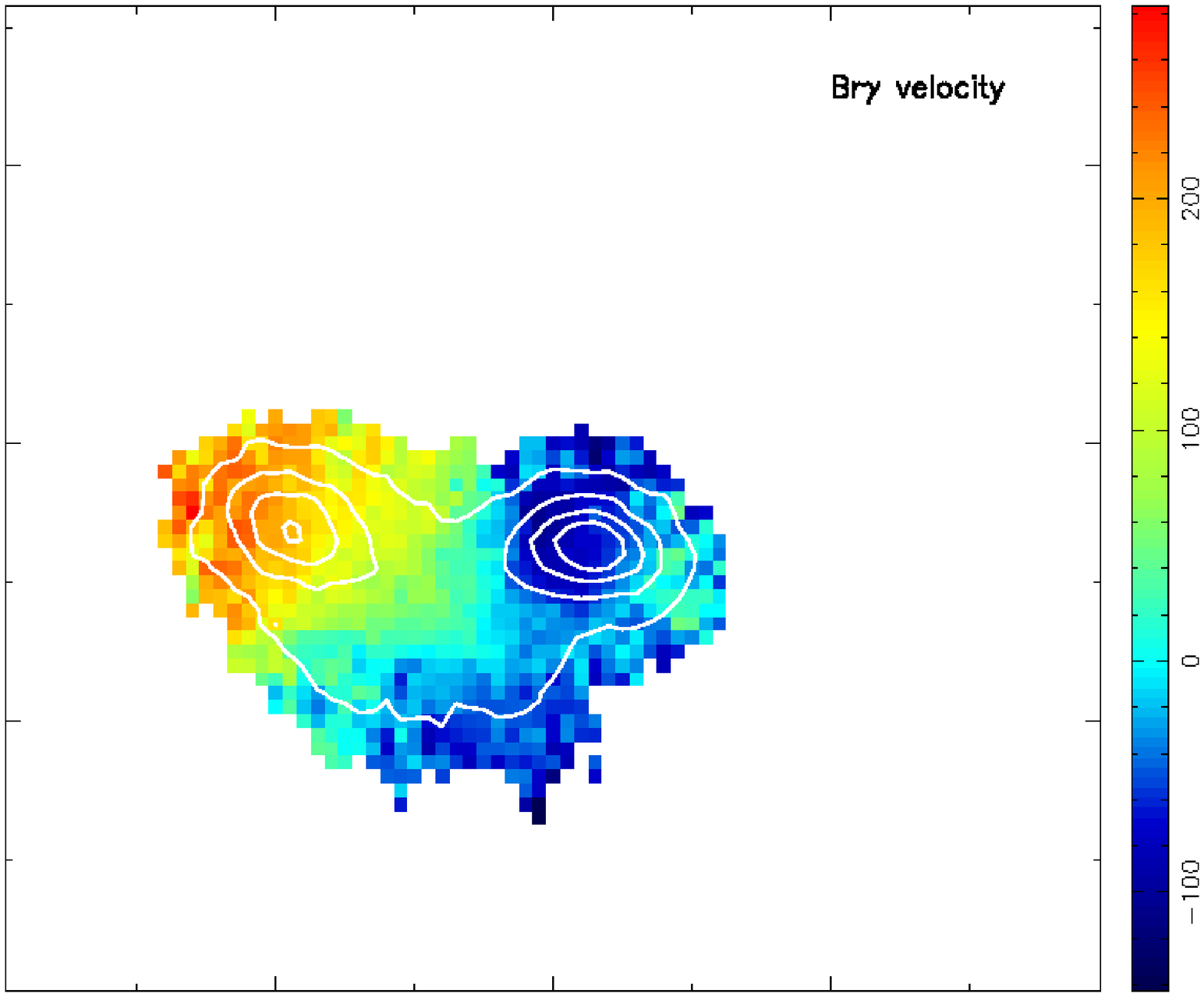}
\includegraphics[width=0.21\textwidth]{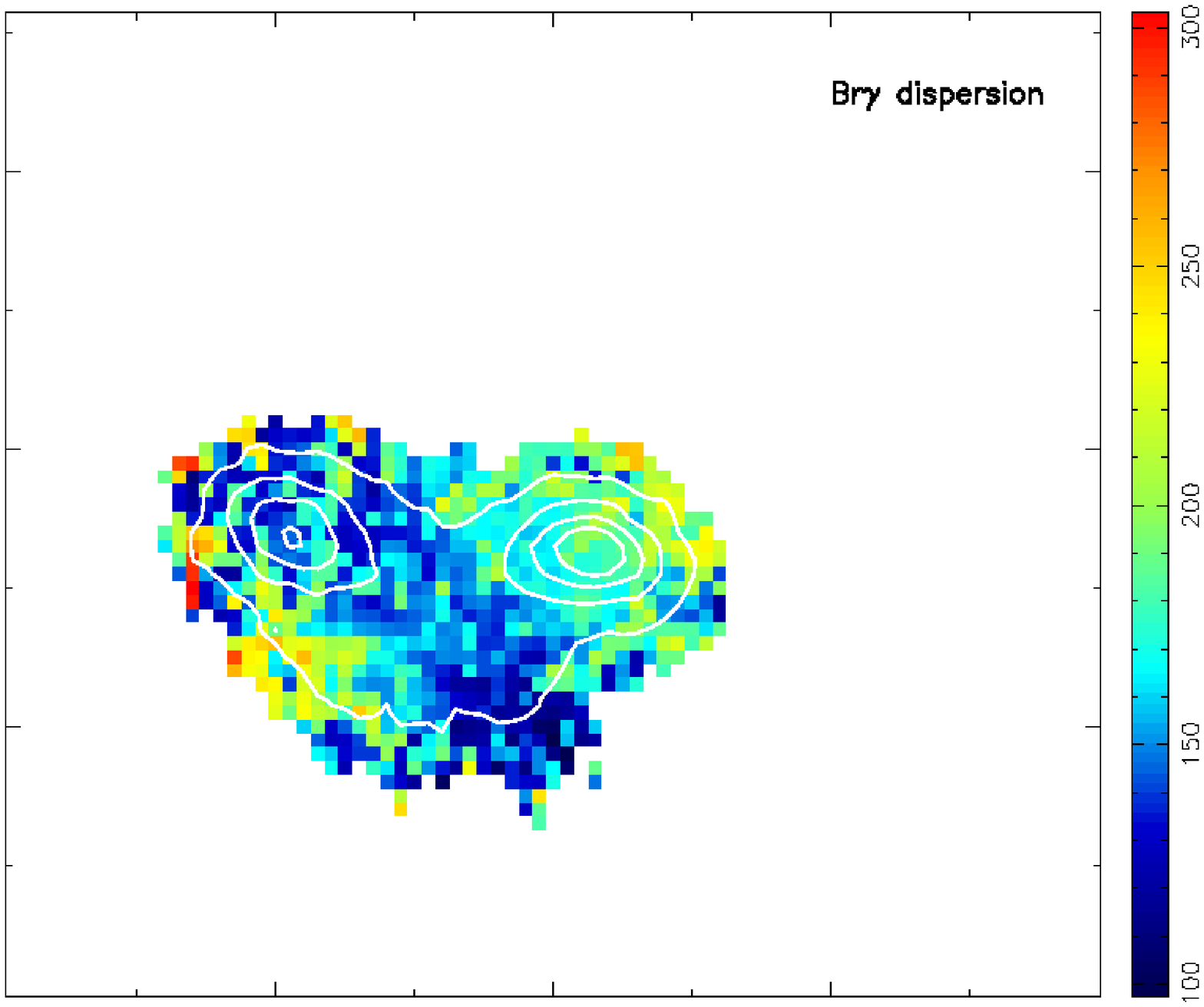}
\includegraphics[width=0.21\textwidth]{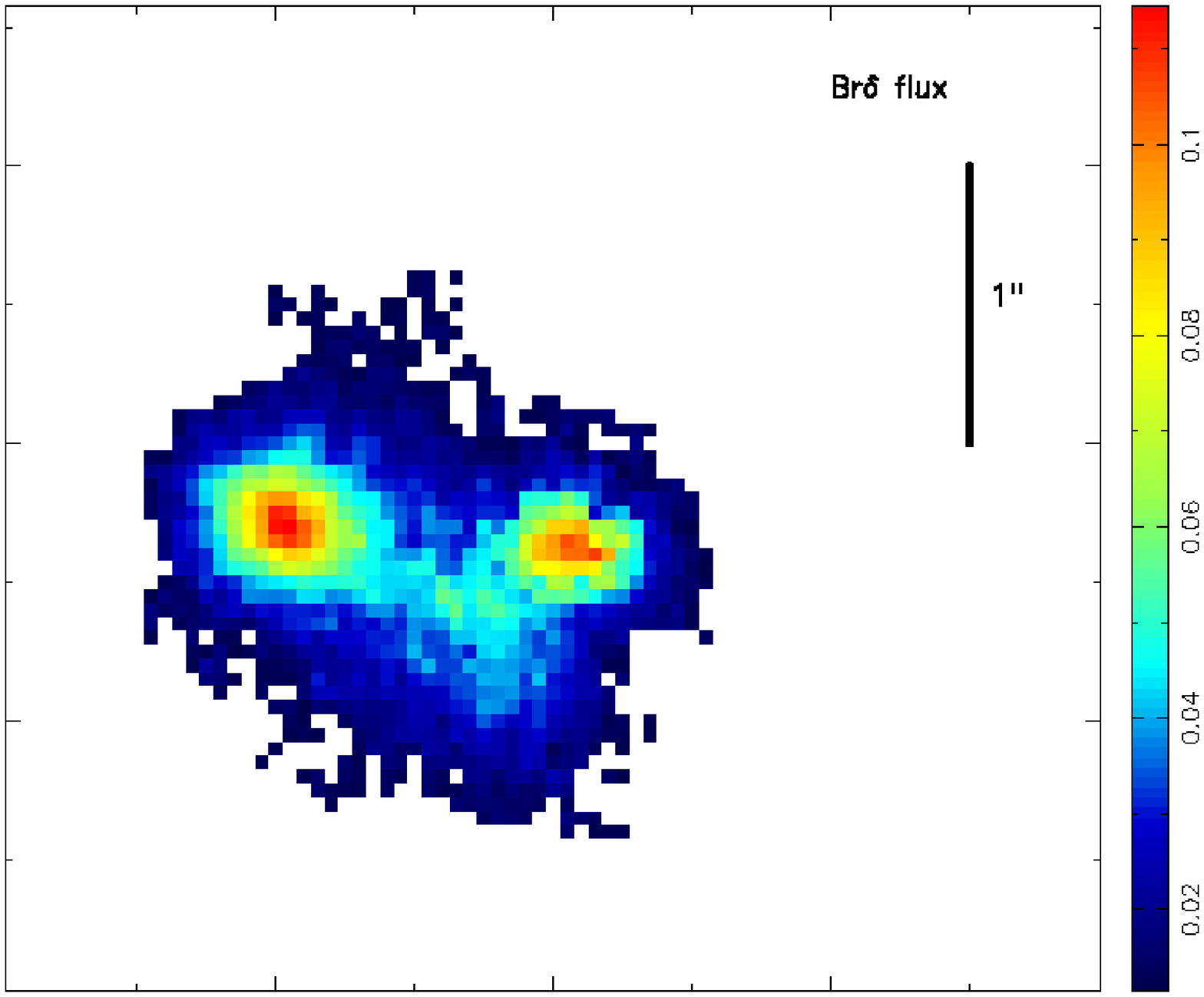} 
\includegraphics[width=0.21\textwidth]{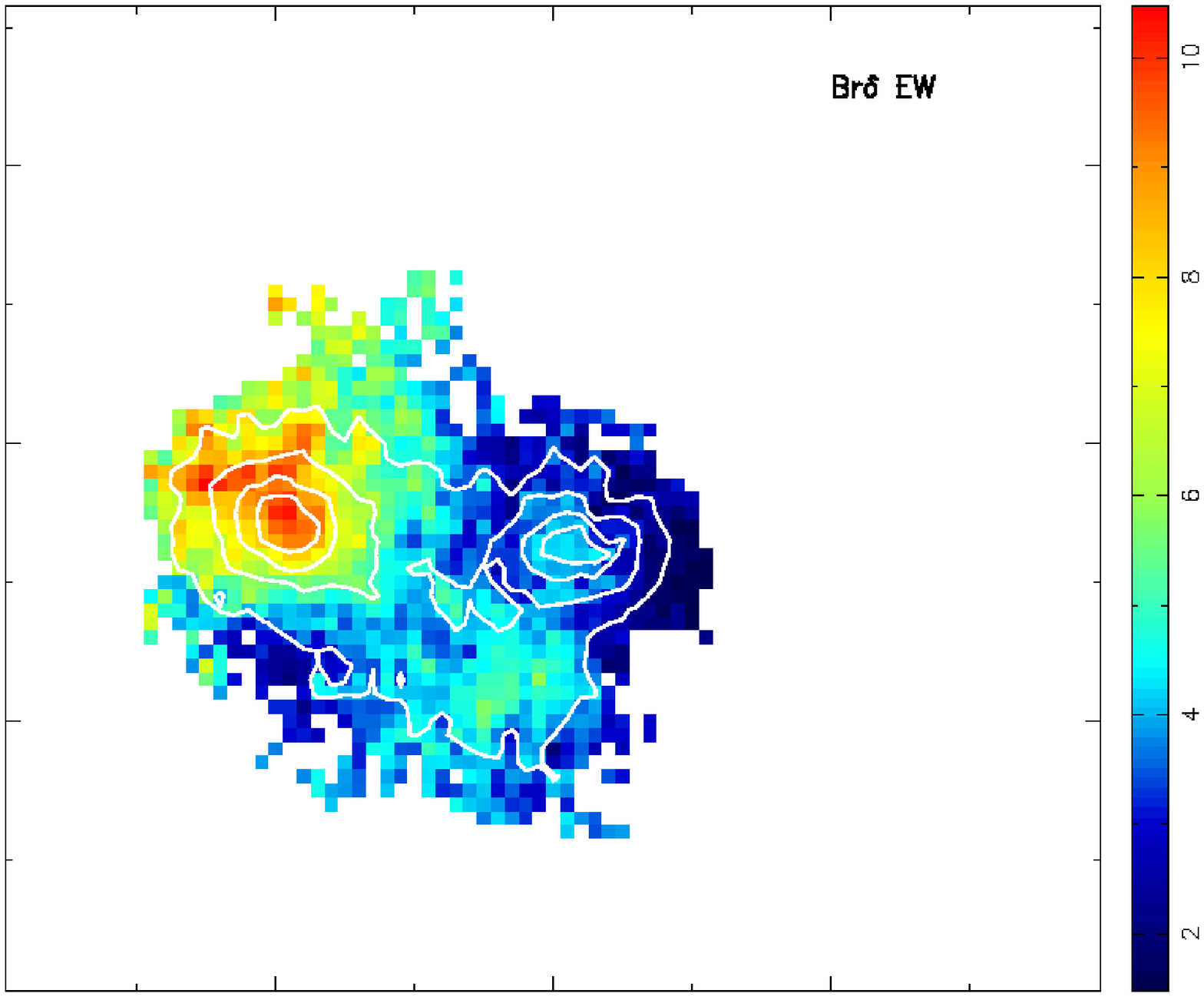} 
\includegraphics[width=0.21\textwidth]{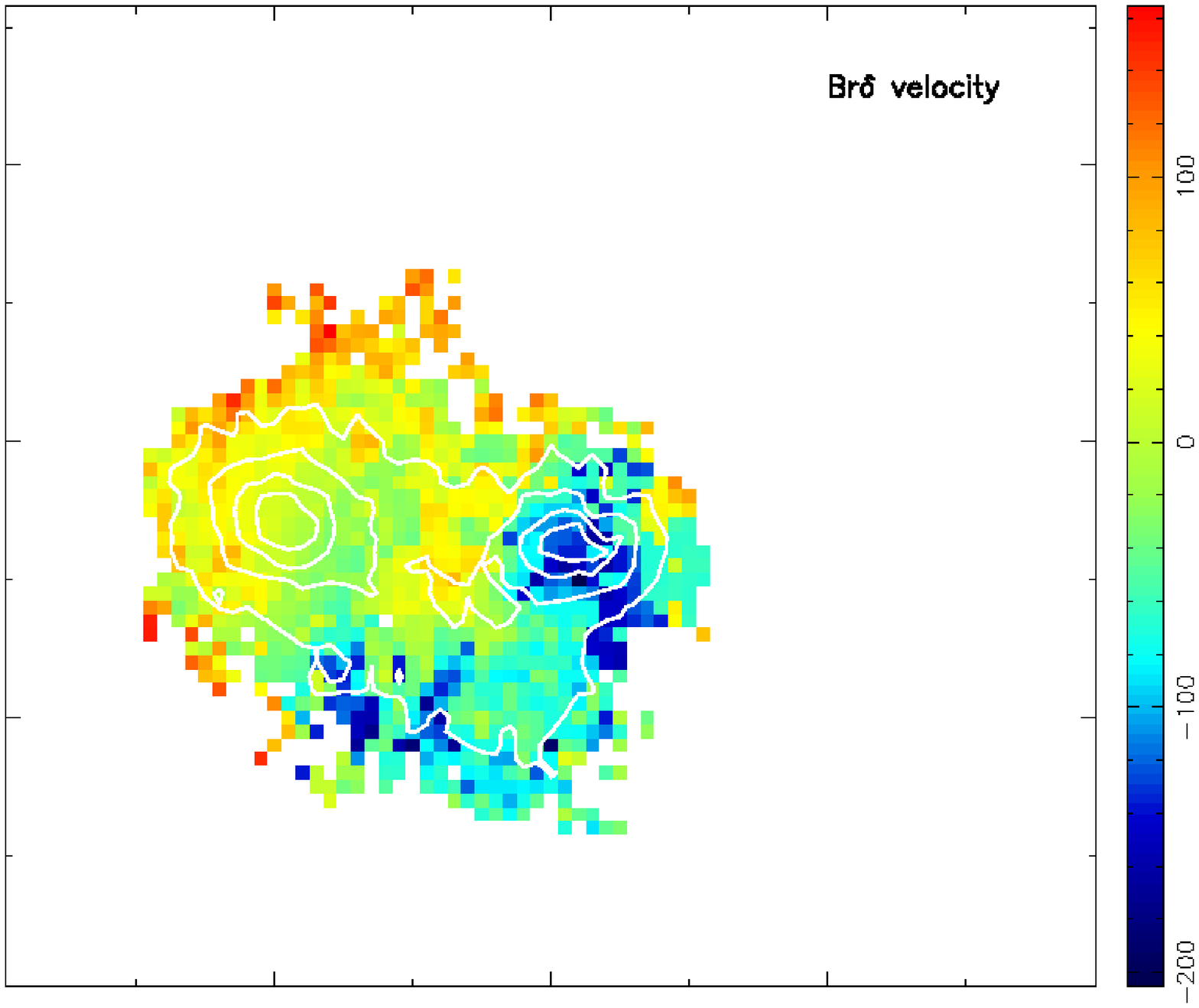} 
\includegraphics[width=0.21\textwidth]{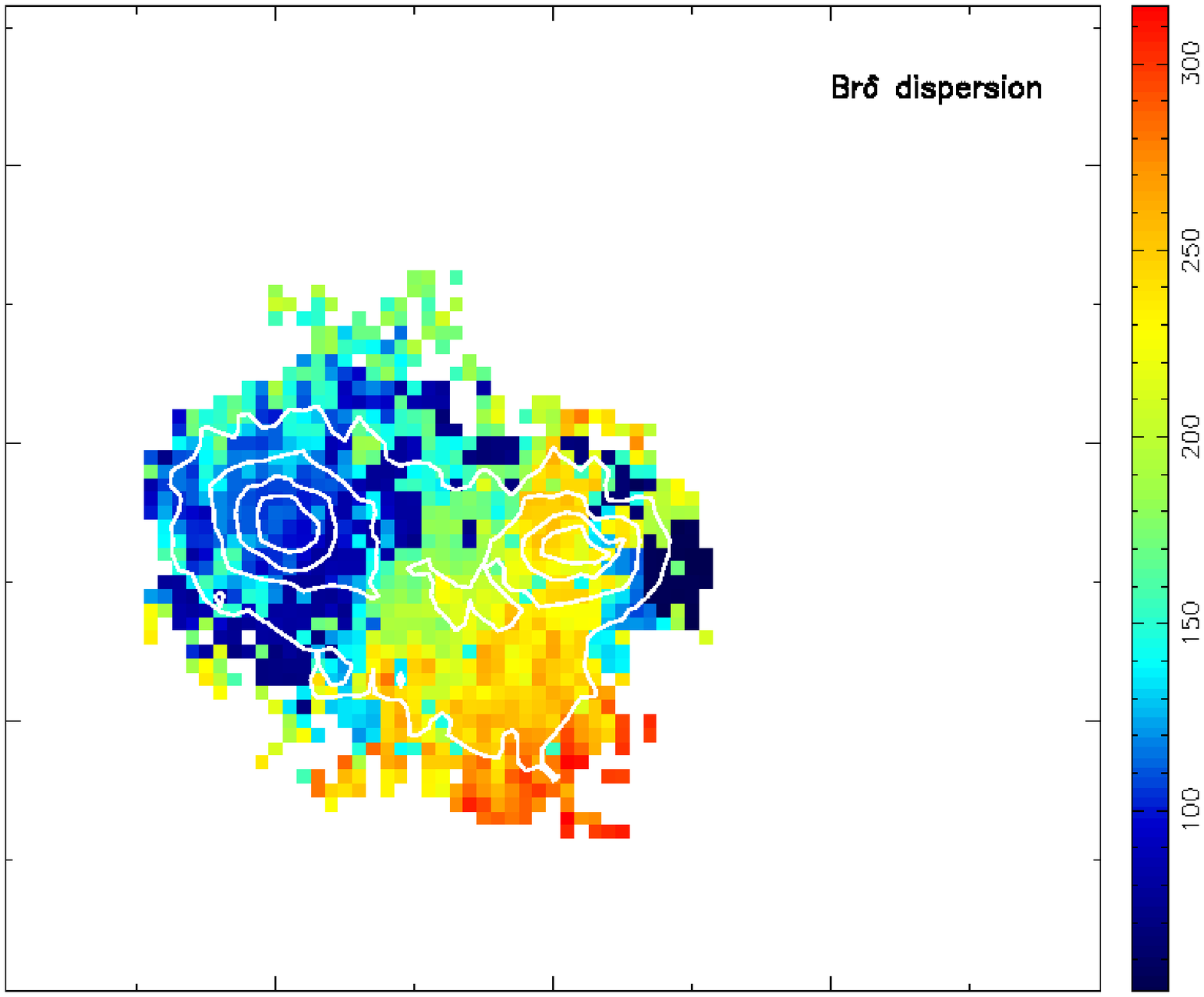}
\caption{Flux, equivalent width, velocity, and velocity dispersion maps for a number of H- and K-band tracers (cf.~\S\ref{sec:obs:gas}).}
\label{fig:all1}
\end{center}
\end{figure}

We derive emission, equivalent width, velocity, and velocity dispersion maps of H$_2$\,1-0S(1) ($\lambda$\,=\,2.1218\,$\mu$m), HeI ($\lambda$\,=\,2.0581\,$\mu$m), Pa$\alpha$ ($\lambda$\,=\,1.8751\,$\mu$m), [FeII] ($\lambda$\,=\,1.6440\,$\mu$m), Br$\gamma$ ($\lambda$\,=\,2.1662\,$\mu$m), and Br$\delta$ ($\lambda$\,=\,1.9445\,$\mu$m), by fitting Gaussians to the lineshapes. The results are shown in Fig.~\ref{fig:all1}.

\subsection{Plateau de Bure Interferometer Data}
\label{sec:bure}

\begin{figure}
\begin{center}
\includegraphics[width=0.35\textwidth]{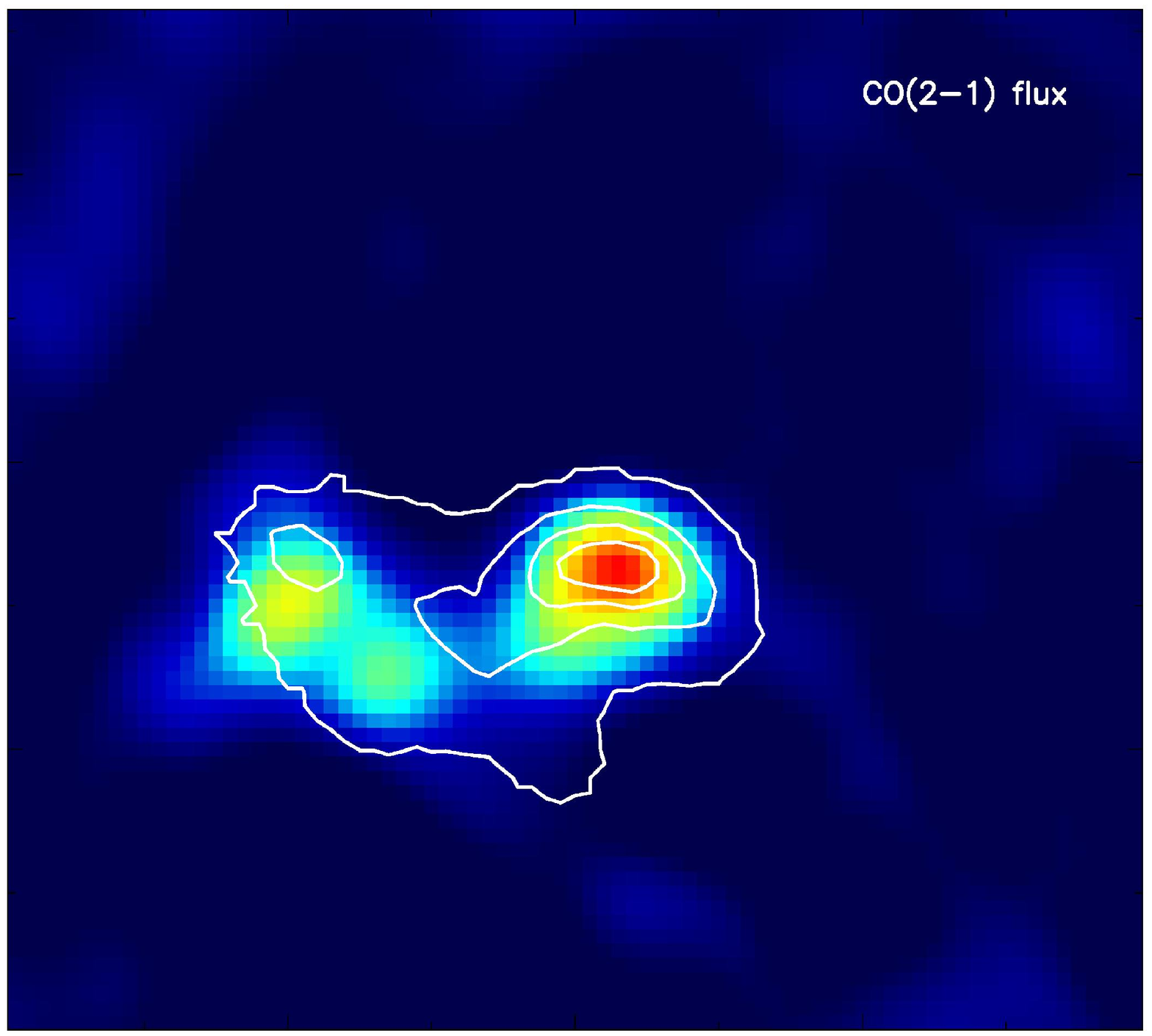} 
\includegraphics[width=0.38\textwidth]{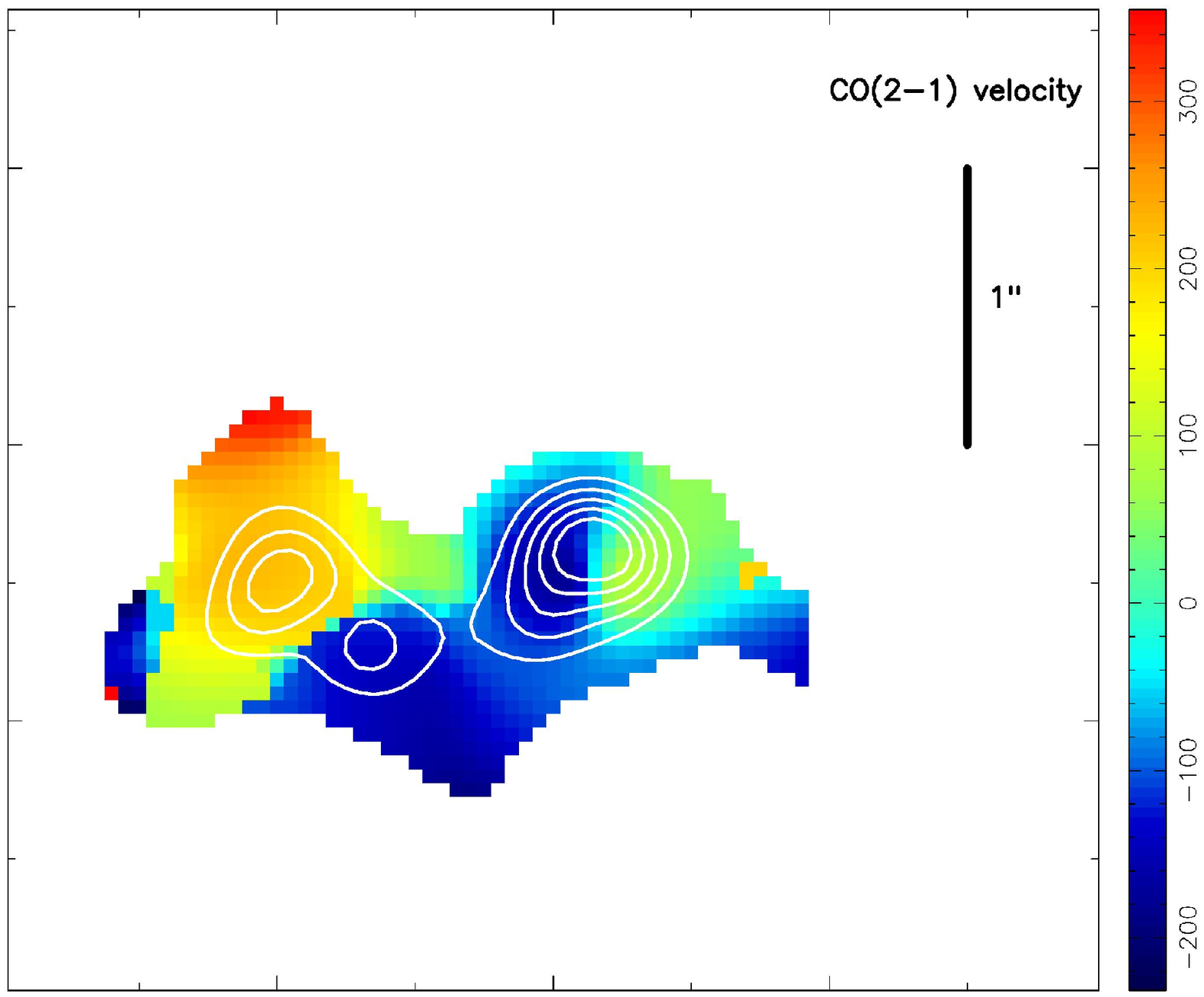} 
\caption{CO(2-1) flux (left, with contours showing the K-band continuum), and velocity map (right, contours tracing the CO(2-1) flux). The bar indicates 1\arcsec. North is up and east is to the left.\label{fig:co}}
\end{center}
\end{figure}

We also analyse $^{12}$CO(J\,=\,2-1) line emission and 1.3\,mm dust continumm emission observations from the IRAM millimetre interferometer, located on the Plateau de Bure, France, at an altitude of 2550\,m \citep{guilloteau92}.  These observations were first analysed and published by \cite{downes07}, and we refer the reader to this publication for the technical details. These authors find evidence for a hot dust source in the western nucleus, which they interpret as being heated by an AGN accretion disk; the best evidence for a black hole in the central region of Arp\,220 to date. Since the absolute astrometry between the PdBI and the SINFONI datasets is not precise, we anchor the SINFONI and PdBI data with respect to each other by assuming that the flux peaks of the western nucleus are coincident. This assumption is justified by the fact that in the western nucleus, all the K-band tracers coincide at the same location (unlike in the eastern nucleus), and hence it is plausible to expect that the cold gas emission also peaks at the same position. We derive a velocity map by fitting Gaussians to the line shapes, the result is displayed in Fig.~\ref{fig:co}.

\section{Star Formation History \& Stellar Populations}
\label{sec:sfh}

\begin{figure}
\begin{center}
\includegraphics[width=0.35\textwidth]{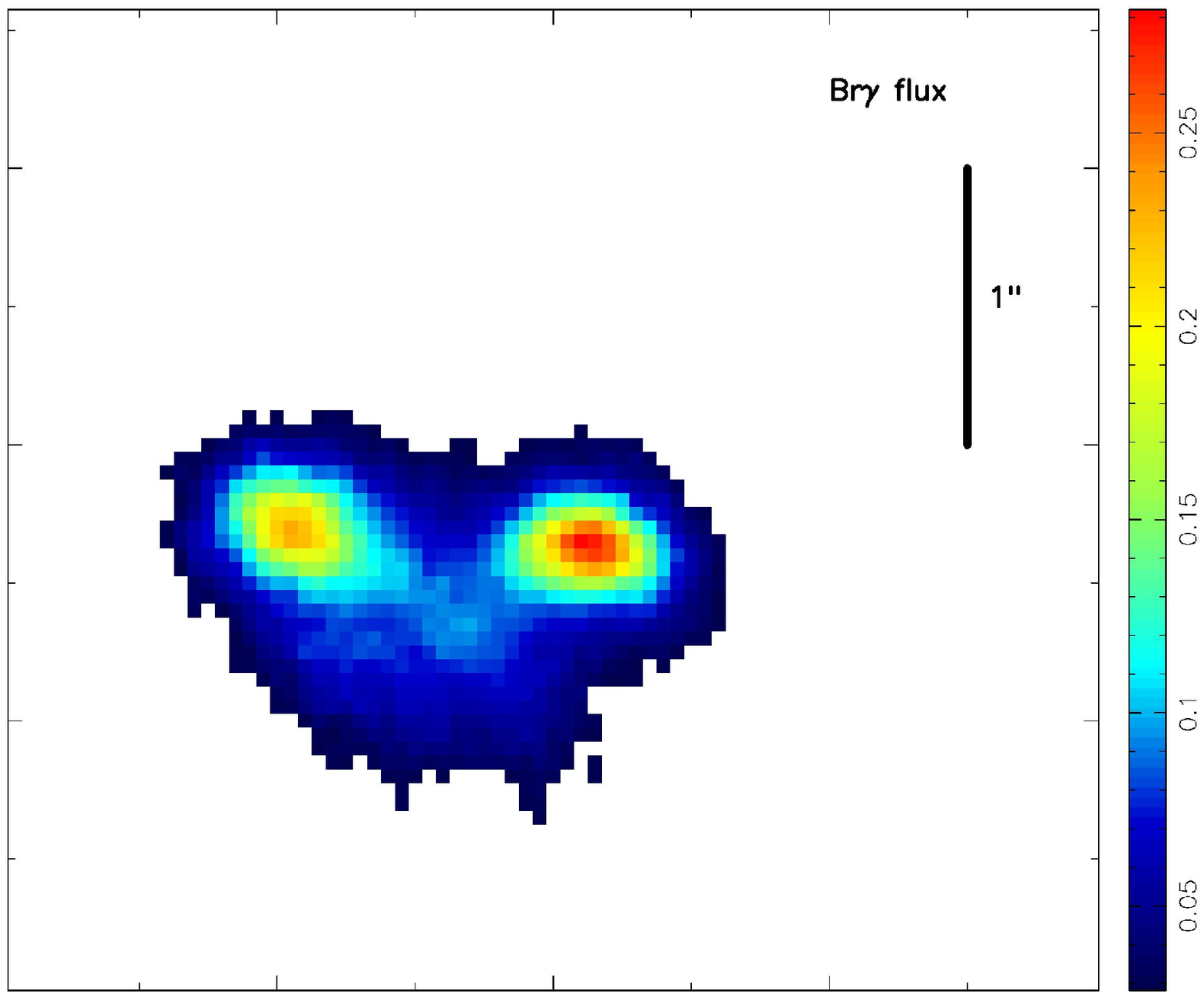} 
\includegraphics[width=0.35\textwidth]{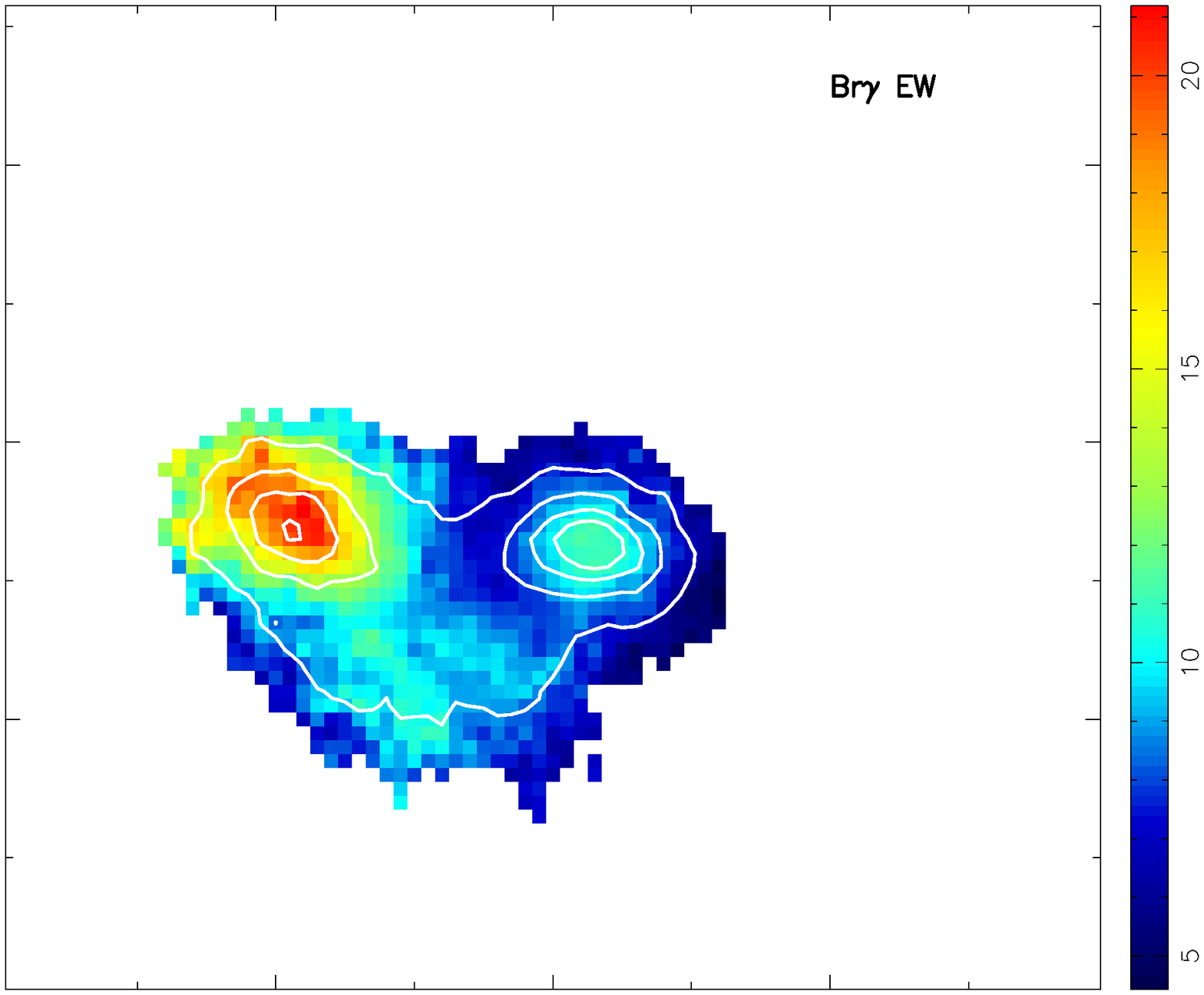} 
\caption{Br$\gamma$ flux (left) and EW (right). Contours trace Br$\gamma$ flux.
The bar indicates 1\arcsec. North is up and east is to the left.\label{fig:brg}}
\end{center}
\end{figure}

\begin{figure}
\begin{center}
\includegraphics[width=0.45\textwidth]{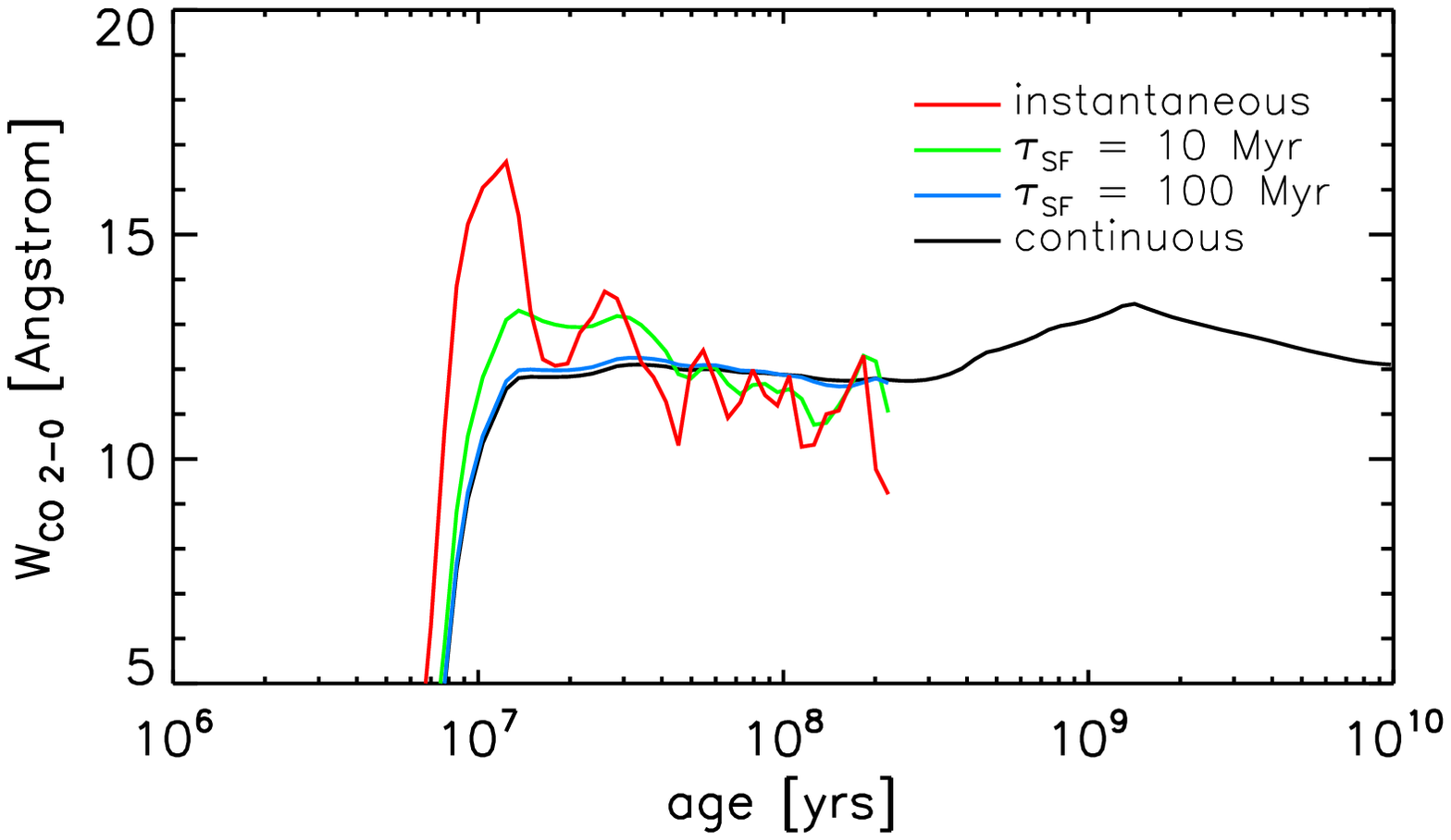} 
\includegraphics[width=0.45\textwidth]{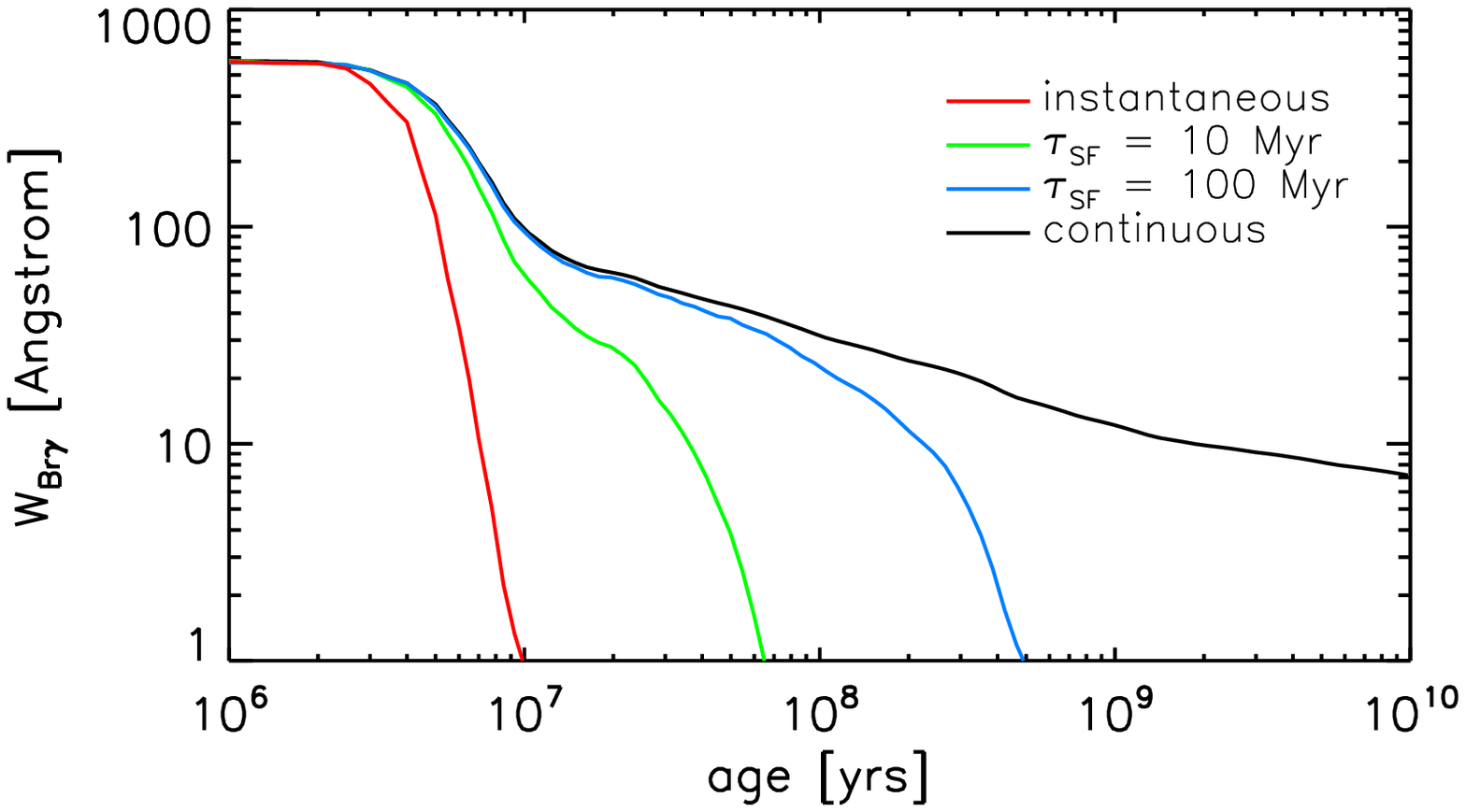} 
\includegraphics[width=0.45\textwidth]{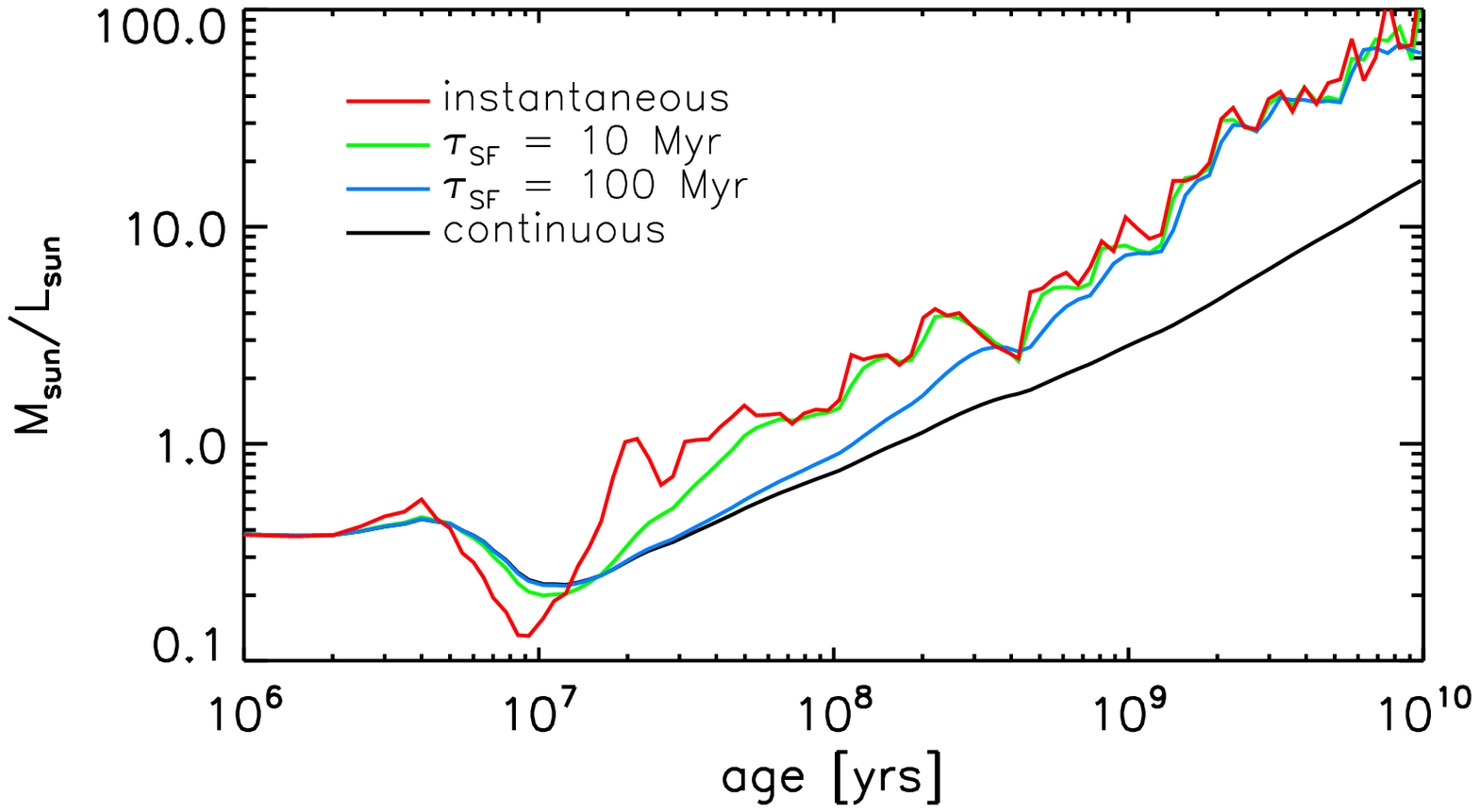}
\includegraphics[width=0.45\textwidth]{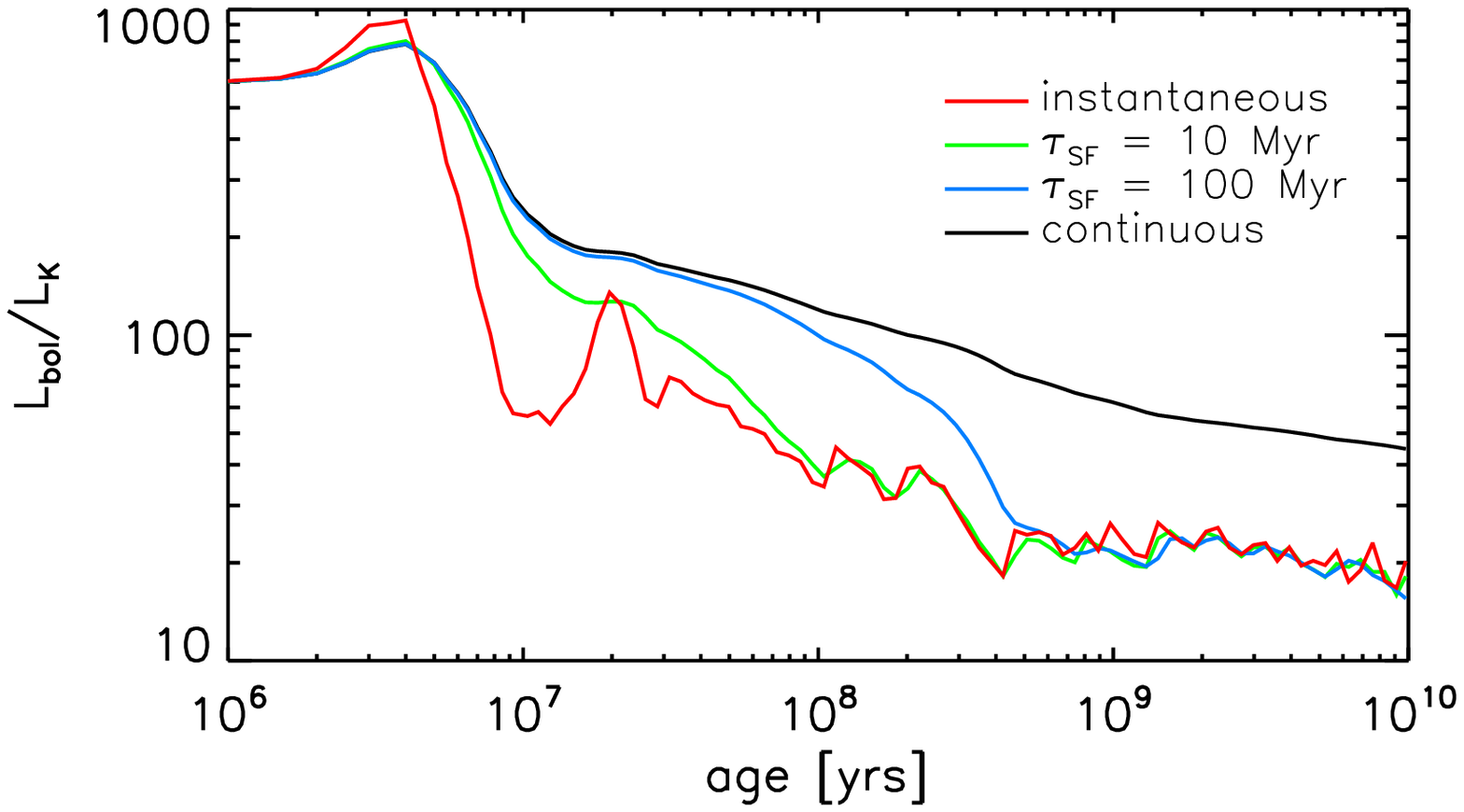}
\caption{Evolution of EW$_{CO}$, EW$_{Br\gamma}$, $L_{bol}/L_K$, and $M/L_K$, computed with STARS, for an instantaneously decaying starburst (red), starbursts with decay timescale 10\,Myr (green) and 100\,Myr (blue), and continuous star formation (black). As can be seen, only a $\sim$\,10\,Myr old `delta'-starburst can produce EW$_{CO}\gtrsim$\,14\,\AA.\label{fig:stars}}
\end{center}
\end{figure}

In this Section, we investigate the star formation history, and the composition of the stellar population, using observables and the stellar synthesis code STARS \citep{sternberg98,ste03,for03,dav03,dav05,dav06,dav07}. 
In the following, we use spatially averaged measurements to investigate the composition of the stellar population. Since we are summing over a large spatial area, we can be sure to cover a large number of individual star clusters (which may have started forming stars at slightly different times), and hence to be probing the average star formation properties. Our primary diagnostics are the EW$_{CO}$ and the Br$\gamma$ flux and EW$_{Br\gamma}$. The EW$_{CO}$ map is shown in the last panel of Fig.~\ref{fig:stellar}, the Br$\gamma$ maps are displayed in Fig.~\ref{fig:brg}. In Fig.~\ref{fig:stars}, we show the evolution of EW$_{CO}$, EW$_{Br\gamma}$, and $L_{bol}/L_K$ for a number of starbursts with different decay timescales. We furthermore use dynamical mass estimates of the nuclei (details can be found in \S\ref{sec:WN:dynmass} and \S\ref{sec:EN:dynmass}) and predictions of the mass-to-light ratios of different stellar populations (Fig.~\ref{fig:stars}). In what we follows, we show that a single stellar population cannot account for the observed diagnostics. Instead we find that three distinct populations are required; a 10\,Myr old starburst accounting for over half the observed luminosity, a significant contribution from an old ($\gtrsim$\,1\,Gyr) stellar population, and a rather smaller population of very young stars, responsible for the Br$\gamma$ emission.

\subsection{A 10\,Myr old Starburst...}

We observe EW$_{CO}$ to be $\gtrsim$\,14\,\AA\ everywhere, and $\gtrsim$\,16\,\AA\ on the nuclei; as can be seen from Fig.~\ref{fig:stars}, only an instantaneously decaying (`delta'-) starburst 10$\pm$2\,Myr old can produce such large values of EW$_{CO}$. Do these stars account for the entire stellar luminosity and mass, or are there significant contributions from other stellar populations?
We can estimate the luminosity contribution of any stars not belonging to this 10\,Myr old starburst population by noting that the theoretically possible upper limit to EW$_{CO}$ is 17\,\AA. This allows us to calculate the luminosity contribution from a stellar population with smaller EW$_{CO}$ that may be diluting it. We note that this is strictly speaking only an upper limit, since we do not know the intrinsic EW$_{CO}$ of the starburst population. However, this intrinsic value likely is quite close to 17\,\AA, since this is what is measured in between the two nuclei, where the dilution from any putative older stellar population associated with the nuclei would be minimal. Hence our `upper limit' to the non-starburst luminosity is in fact more likely to be a reasonable estimate.

\subsection{... \& a younger stellar population?}

Both the presence of radio supernovae \citep{parra07} and our Br$\gamma$ maps clearly indicate the presence of young stars -- only very young ($\lesssim$\,8\,Myr) stars are hot enough to excite hydrogen sufficiently to emit Br$\gamma$ flux. An important question is whether this population contributes to the K-band continuum. We thus first investigate the possibility that a population of stars younger than 10\,Myr may be diluting the CO bandheads of the 10\,Myr starburst population; these stars would have  EW$_{CO}$=\,0\,\AA\ (Fig.~\ref{fig:stars}). In this case, the younger stars would contribute $\sim$\,10\% of the K-band luminosity.
Below (\S\ref{sec:extinction}), we calculate that if the assumption of younger ($<$\,10\,Myr) stars diluting the CO bandheads is correct, then we are only seeing $\approx$\,20\% of the actually emitted (no reddening-correction) K-band luminosity. For a 10\,Myr old instantaneously decaying starburst, STARS predicts a K-band mass-to-light ratio of $\approx$\,0.2\,M$_\odot$/L$_\odot$ (Fig.~\ref{fig:stars}). Since we know the luminosity of the 10\,Myr old stellar population under this scenario to be $\sim$\,90\%, we can therefore calculate its expected mass (correcting the measured K-band luminosity for the western nucleus for the missing light). Using our dynamical mass estimate (\S\ref{sec:WN:dynmass}, and assuming a 10\% gas fraction), we can then calculate the expected mass-to-light ratio of the non-starburst stellar population of the western nucleus; we find that it must be M/L$_K$\,$\sim$\,75\,M$_\odot$/L$_\odot$. The same calculation for the eastern nucleus yields an even larger M/L$_K$\,$\sim$\,99\,M$_\odot$/L$_\odot$. The expected M/L$_K$ for stars $\lesssim$\,10\,Myr old is $\lesssim$\,0.1\,M$_\odot$/L$_\odot$ -- a divergence of three orders of magnitude. We therefore reject the assumption of a younger stellar population diluting the bandheads of the 10\,Myr old starburst population.

\subsection{... \& an older stellar population!}

Stars older than $\sim$\,20\,Myr have an EW$_{CO}$\,$\approx$\,12\,\AA; a fractional luminosity contribution of  26-44\% would be required to achieve the bandhead dilution (Table~\ref{tab:stellar} lists the results for both the western nucleus and the entire nuclear region). In this case, we are only seeing $\approx$\,7\% of the actually emitted K-band luminosity (\S\ref{sec:extinction}; this is different to the fraction calculated above under the assumption of a younger stellar population, since the theoretically expected $L_{bol}/L_K$ is different in this scenario). Calculating the expected mass-to-light ratio for the older stellar population in the same way as above, we find that it must have M/L$_K$\,$\approx$\,7.4\,M$_\odot$/L$_\odot$ -- which is what is expected for a several Gyr old stellar population. The same is found for the eastern nucleus, here an M/L$_K$\,$\approx$\,9.8\,M$_\odot$/L$_\odot$ is required. We therefore conclude that the assumption of a significantly ($\gtrsim$\,1\,Gyr) older stellar population alongside the 10\,Myr old starburst in the central kpcs of Arp\,220 is correct. We note that we still require a third population of stars younger than $\sim$\,7\,Myr (which would have EW$_{Br\gamma}$\,$\sim$\,600\,\AA, Fig.~\ref{fig:stars}), to account for the Br$\gamma$ emission. However, their luminosity contribution would be minimal; their intrinsic EW$_{Br\gamma}$\,$\sim$\,600\,\AA\ has to be diluted by a factor $\gtrsim$\,75 to reach the measured EW$_{Br\gamma}$\,$\lesssim$\,8\,\AA.

This is in good agreement with what \cite{rod08} find based on optical spectroscopy of the central region of Arp\,220; namely near-equal contributions from a $\leq$\,0.1\,Gyr and a 0.5--0.9\,Gyr old stellar population. \cite{wilson06} derive photometric age estimates of clusters from \textit{HST UBVI} data and find two age classes; $\lesssim$\,10\,Myr and 70--500\,Myr.

\section{Reddening \& Extinction}
\label{sec:extinction}

\subsection{Reddening}
\label{sec:ext:red}

Like most ULIRGs, Arp\,220 contains significant amounts of dust. In the K-band, the effect of this is mostly a wavelength-dependent extinction due to scattering and absorption, with longer-wavelength radiation being less severely affected. The effect of this is `reddening', a steepening of the continuum level towards longer wavelengths. Another effect pertaining to the presence of dust is the re-emission of dust-absorbed stellar light. However in the H- and K-band this would only be observable if very hot dust were present. Since any continuum emission due to hot dust would lead to dilution of the CO absorption bandheads, and since we are seeing absorption depths very near the theoretically possible maximum of $\sim$\,17\,\AA, we can exclude the possibility of any significant hot dust emission. 

\begin{figure}
\begin{center}
\includegraphics[width=0.40\textwidth]{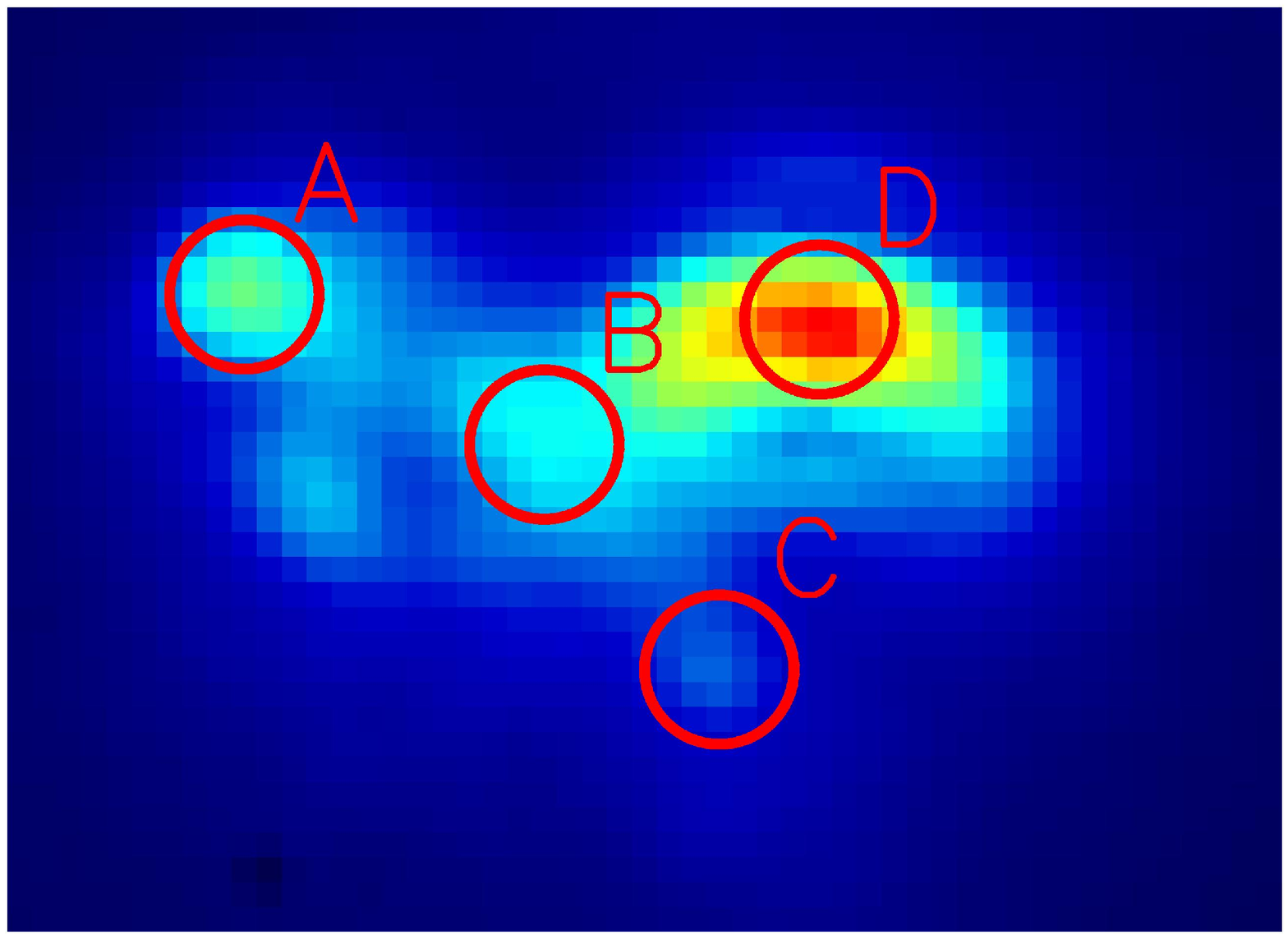} 
\includegraphics[width=0.58\textwidth]{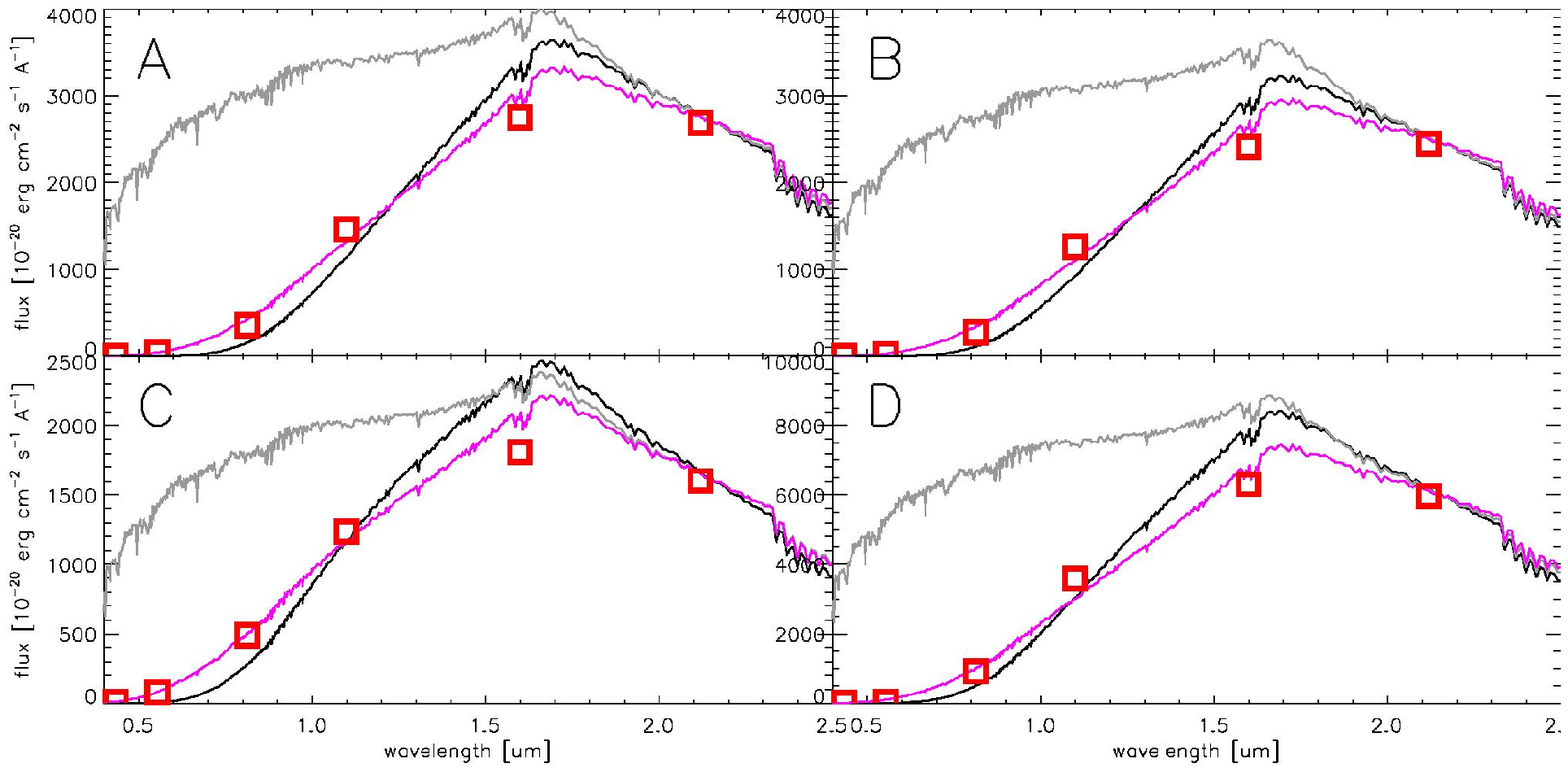} 
\caption{Comparison of different extinction models: We fitted a synthesised stellar population spectrum to \textit{HST} photometric data points extracted in 0.15\arcsec\ radius apertures in a range of locations across our field of view (4 examples shown), by reddening the synthetic SED until it matched the data. Grey: mixed extinction, black: screen extinction, magenta: \cite{calzetti00} reddening law. As can be seen, the \cite{calzetti00} reddening law produces the best fit to the data.\label{fig:hst}}
\end{center}
\end{figure}

When accounting for the obscuring effects of dust, the dust is most commonly assumed to either be spatially uniformly mixed with the stars (`mixed model'), or to form a screen between the observed stars and the observer. Or one can use the empirically derived reddening law of \cite{calzetti00}. In cases of strong obscuration, the mixed model may be insufficient to account for the reddening, since it saturates beyond a certain level of extinction. In order to determine which method is appropriate for Arp\,220, we obtained archival \textit{HST} images spanning a wavelength range of 0.45\,-2.12\,$\mu$m, and extracted photometric data points at a number of locations using 0.15\arcsec\ radius apertures. We then reddened a synthetic stellar population spectrum (corresponding to the star formation history determined in \S\ref{sec:sfh}), using each reddening prescription outlined above, until it fit those data points.
The results can be seen in Fig.~\ref{fig:hst} -- as expected, the mixed model was insufficient to account for the strong extinction of Arp\,220. The screen model results in a decent match, but the \cite{calzetti00} reddening law clearly produces the best results. We note that \cite{engel10a} arrive at the same conclusion for NGC\,6240. Here, the screen extinction law also yields a somewhat lower $F_{obs}/F_{em}$ than the \cite{calzetti00} reddening law.

\begin{figure}
\begin{center}
\includegraphics[width=0.35\textwidth]{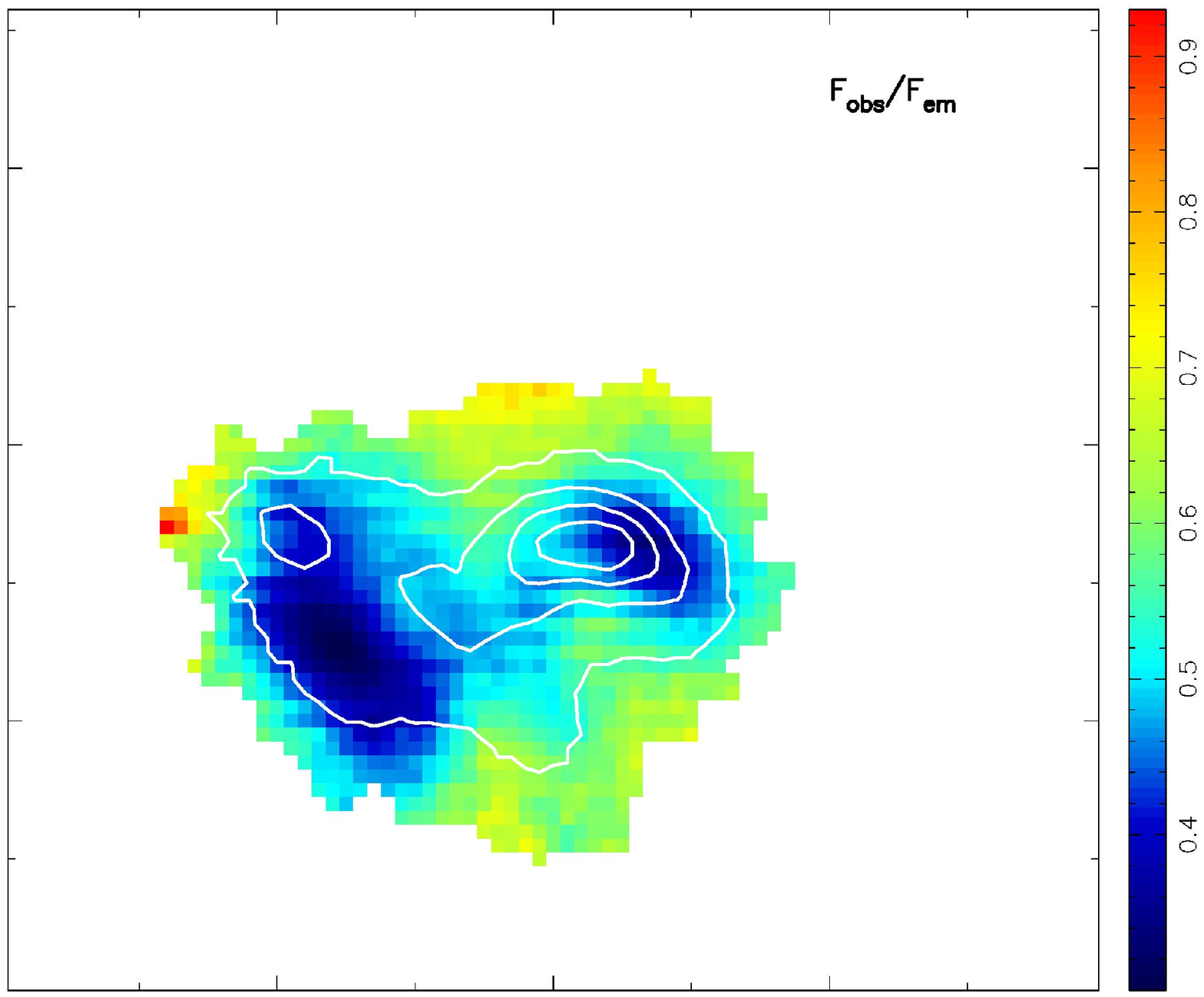} 
\includegraphics[width=0.35\textwidth]{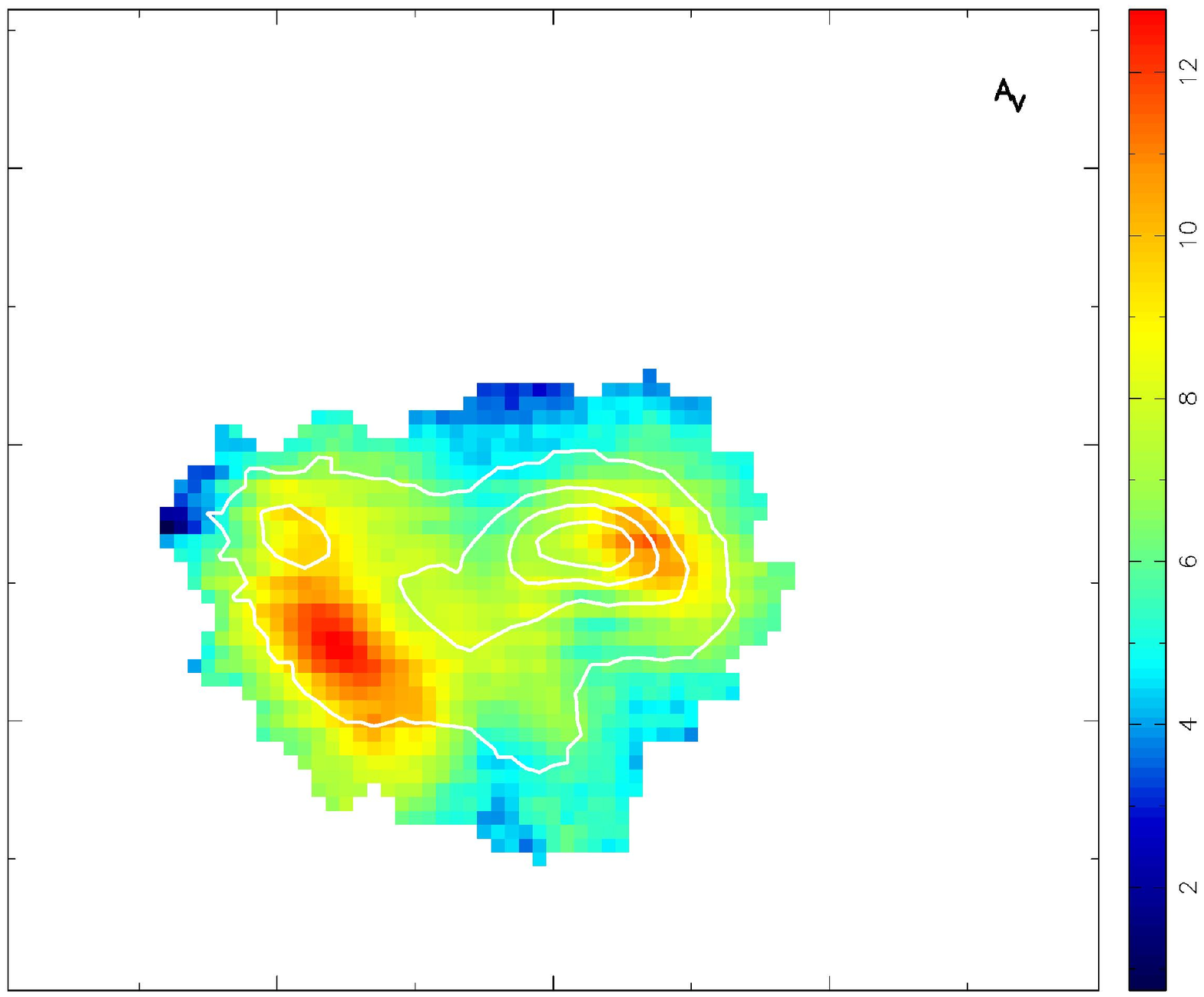} 
\caption{Left: $F_{obs}/F_{em}$ in the K-band, derived by fitting a stellar template to the line-free continuum using the \cite{calzetti00} reddening law. Right: corresponding optical extinction $A_V$. Contours trace the continuum.\label{fig:ext}}
\end{center}
\end{figure}

In order to quantify the reddening in Arp\,220, we fitted a stellar template to the line-free continuum, using the \cite{calzetti00} reddening law. In Fig.~\ref{fig:ext}, we show the resulting map of $F_{obs}/F_{em}$ at 2.12\,$\mu$m, and of the corresponding optical extinction $A_V$; obtained by calculating $F_{obs}/F_{em}$ at 0.55\,$\mu$m and converting to extinction via $A_V$\,=\,$-2.5$\,$\log$($F_{obs}/F_{em}$).

\begin{figure}
\begin{center}
\includegraphics[width=0.35\textwidth]{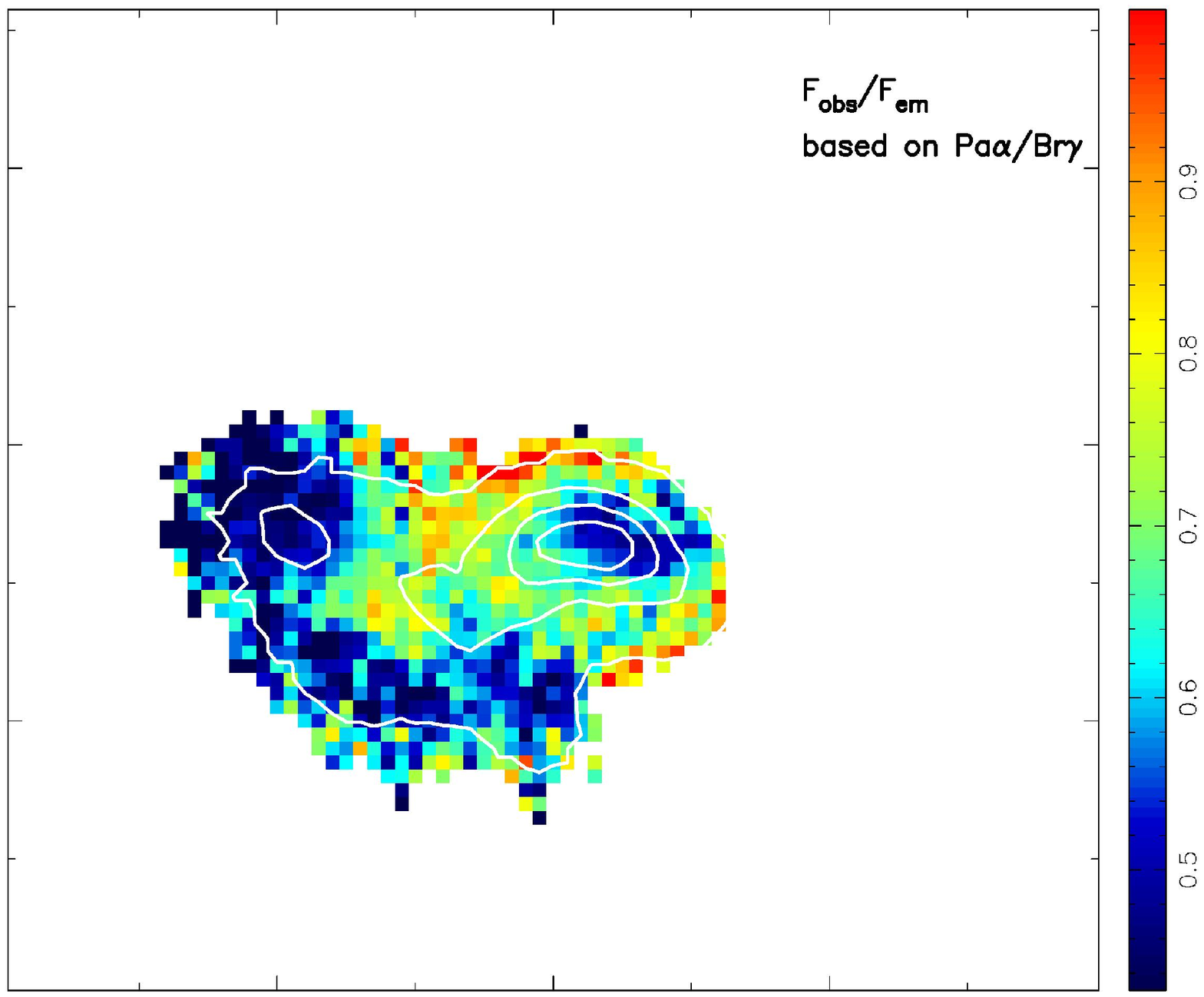} 
\includegraphics[width=0.35\textwidth]{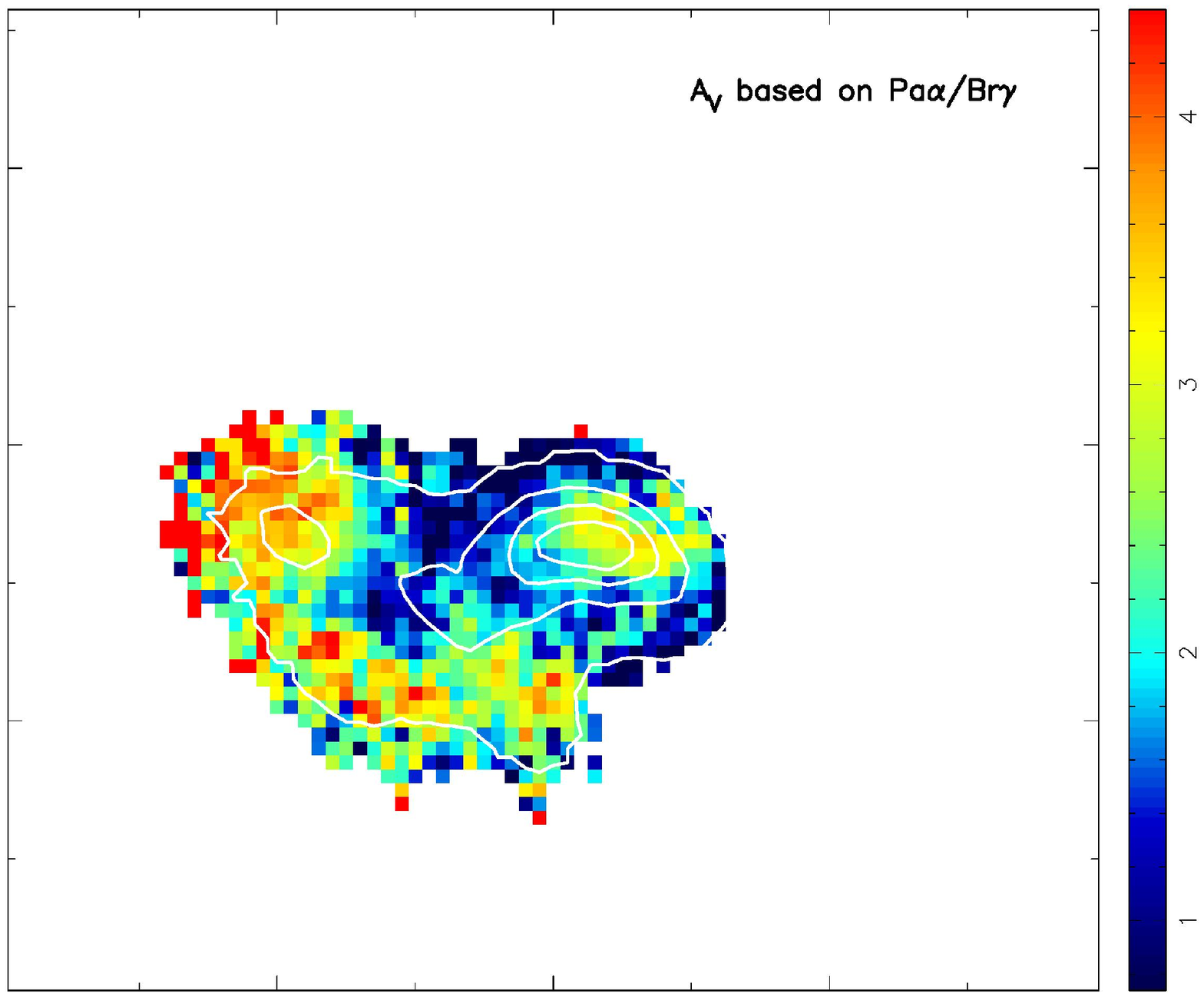} 
\caption{Left: $F_{obs}/F_{em}$ in the K-band, derived by comparing the observed flux ratio of Pa$\alpha$ and Br$\gamma$ to the theoretically expected value, again using the \cite{calzetti00} reddening law. Right: corresponding optical extinction $A_V$. Contours trace the continuum.\label{fig:ext_lr}}
\end{center}
\end{figure}

We also derived the reddening using molecular hydrogen emission lines. For case B recombination at $T$\,=\,10,000\,K and $n_e$=\,10$^4$\,cm$^{-3}$, the line flux ratio of Pa$\alpha$ and Br$\gamma$ is 12.2; by comparing this to the actually observed line ratio, the relative reddening can be deduced. Again using the \cite{calzetti00} reddening law, we thus measured $F_{obs}/F_{em}$ and derived $A_V$, both shown in Fig.~\ref{fig:ext_lr}. As can be seen, the two extinction measurements agree well both in regard to strength and spatial variation of the extinction. The 1.3\,mm continuum map of \cite{downes07} could give us valuable information about the spatial distribution of the obscuring dust, however unfortunately only two barely resolved point sources are detected at their sensitivity level. \cite{downes07} estimate the optical depth for the western nucleus to be $\tau_{WN}$(1.3\,mm)\,$\geq$\,0.7; adopting their assumptions also for the eastern nucleus, we estimate $\tau_{EN}$(1.3\,mm)\,$\geq$\,0.15; this is by extension also an upper limit of the off-nuclear $\tau$(1.3\,mm).
For $\tau$\,$\propto$\,$\lambda^{-2}$, $\tau$\,$\sim$\,0.15 implies that all emission shortwards of 500\,$\mu$m is optically thick. There appear to be two different regimes: the extremely opaque central $\sim$50\,pc which are fully obscured at near infrared wavelengths ($\tau$\,=\,1 at 500\,$\mu$m), and the outer parts where we can detect reddened near-IR emission (and where we find $\tau$\,=\,1 is at $\sim$\,2.5\,$\mu$m). This plausibly explains why even after applying the reddening correction, we do not recover all the near-IR luminosity: we are simply not seeing all that is emitted in the central region. We note that this is qualitatively similar to what we found in NGC\,6240 \citep{engel10a}: over most of the disk the near-infrared obscuration is moderate, but increases dramatically in the central tens of parsecs of each nucleus.

\subsection{Extinction}
\label{sec:ext:ext}

The significant reddening quantified in \S\ref{sec:ext:red} is indicative of a substantial amount of stellar light emitted in the near-infrared in the central region of Arp\,220 which is completely absorbed, and re-emitted at longer wavelengths. Here, we attempt to quantify the amount of `missing light' in the K-band, by comparing the infrared luminosity, calculated from \textit{IRAS} flux measurements, to the stellar bolometric luminosity calculated from our measured L$_K$ and the L$_{bol}$/L$_{K}$ expected from our stellar population analysis (\S\ref{sec:sfh} \& Table~\ref{tab:stellar}). We use the \cite{sanders03} \textit{IRAS} fluxes from the Revised Bright Galaxy Sample, and calculate the infrared flux via F$_{IR, 8-1000\micron}$\,=\,1.8\,$\times$\,10$^{-14}$\,$\times$(13.48$\times$f$_{12}$ + 5.16$\times$f$_{25}$ + 2.58$\times$f$_{60}$ + f$_{100}$)\,W\,m$^{-2}$, with f$_{12}$ the IRAS flux density in Jy at 12\,$\mu$m etc \citep{sandersmirabel96}. We then convert this to infrared luminosity, L$_{IR}$\,=\,3.127\,$\times$\,10$^7$\,D$^2$\,F$_{IR}$\,L$_\odot$  (D in pc). This yields L$_{IR,8-1000\mu m}$\,=\,1.37\,$\times$\,10$^{12}L_\odot$. 
\cite{nardini10} perform a 5--8\,$\mu$m spectral analysis of Arp\,220 using the Infrared Spectrograph onboard \textit{Spitzer}; they find a 17\% AGN contribution to the bolometric luminosity of Arp\,220, which we need to correct for in order to find the stellar bolometric luminosity. However, this is largely offset by the translation of infrared to bolometric luminosity, which for ULIRGs is typically $L_{bol}$=\,1.15\,$\times$\,$L_{IR}$ \citep{kimsanders98}. This yields a total stellar bolometric luminosity of 1.31\,$\times$\,10$^{12}L_\odot$. 
We then measure the K-band luminosity of Arp\,220 using the archival \textit{HST} NICMOS images already utilised in \S\ref{sec:obs:psf}, finding L$_{K}$\,=\,(1.75$\pm$0.1)$\times$10$^9$L$_\odot$ (for comparison, our SINFONI data yield L$_{K}$\,=\,(1.34$\pm$0.07)$\times$10$^9$L$_\odot$ for the SINFONI FoV, implying that $\sim$\,80\% of the total L$_K$ are emitted in the nuclear region -- cf.~\citealt{wynn93}). 
This results in a $L_{bol}/L_K$ of $\sim$\,750, whereas our earlier stellar population analysis (\S\ref{sec:sfh}) requires $\sim$\,150 (theoretically expected for 10\,Myr old starburst plus younger stellar population with the derived relative luminosity contributions; cf.~Table \ref{tab:stellar}) or $\sim$\,50 (10\,Myr old starburst plus older stellar population) -- implying that in the K-band we are missing a factor of $\sim$\,5 or $\sim$\,15 of light, respectively. As we show in \S\ref{sec:sfh}, the assumption of a population younger than 10\,Myr is inconsistent with our dynamical mass estimate, and we can therefore exclude this alternative. We thus conclude that in the K-band, we are only observing $\sim$\,7\% of the actually emitted (no reddening-correction) stellar light.
This is consistent with \cite{rod08}, who find that the bolometric luminosity derived from optical wavelength observations is an order of magnitude smaller than the mid- to far-infrared luminosity, and are led to conclude that `most of the on-going star formation in the nuclear region is hidden by dust'.
Further support for our result is lend by the supernova rate estimates of \cite{rov05}; these authors derive SNR\,$\sim$\,0.7\,yr$^{-1}$ for the western nucleus. STARS predicts a supernova rate of 10$^{10}L_{K}^{-1}$\,yr$^{-1}$ for a 10\,Myr old instantaneous starburst, which for our corrected K-band luminosity, of which $\sim$\,75\% are attributed to the 10\,Myr old starburst, implies a supernova rate $\approx$\,1.0\,yr$^{-1}$.





\section{Western Nucleus}
\label{sec:WN:analysis}

\subsection{Stellar \& CO(2-1) Kinematics}
\label{sec:WN:kin}
\begin{figure}
\begin{center}
\includegraphics[width=0.85\textwidth]{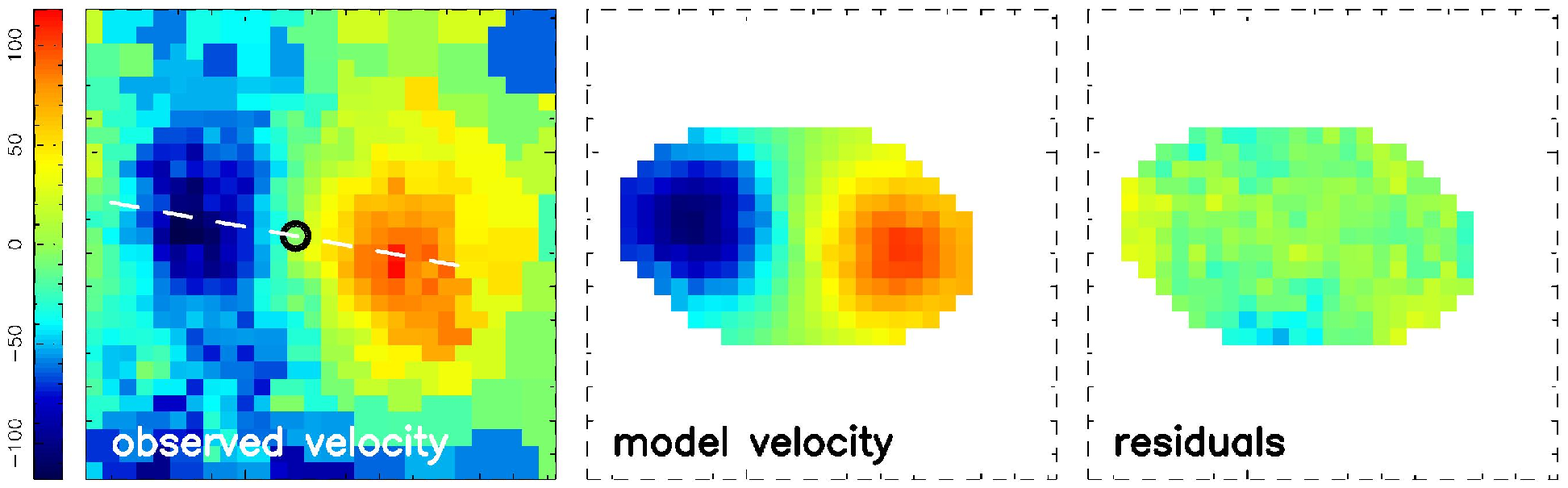} 
\caption{Western Nucleus: Stellar velocity field, best-fitting disk model, and residuals. The kinematic centre is marked by a black circle, and the major axis is indicated with a white dashed line. 1\,pixel corresponds to 0.05\arcsec.\label{fig:WNmodeldisk}}
\end{center}
\end{figure}

As Fig.~\ref{fig:stellar} shows, the stellar kinematics exhibit a clear rotational signatures around the western nucleus. In order to extract a rotation curve and find the kinematic centre of the stellar rotation, we fit an inclined disk model to the velocity map (Fig.~\ref{fig:WNmodeldisk}). The best-fitting model has an inclination q\,=\,0.62$\pm$0.06, and a position axis (PA hereafter) of 9.8$\pm$0.4deg south of west, with average residuals of 12.2\,km\,s$^{-1}$. We also derive a dispersion profile by azimuthally averaging the 2D dispersion map. Note that we do not suggest that the stars in the western nucleus are moving in a thin disk -- we simply parameterise the rotation field in terms of circular orbits with a fixed centre, PA, and inclination, in order to derive a rotation curve and kinematic centre.

\begin{figure}
\begin{center}
\includegraphics[width=0.85\textwidth]{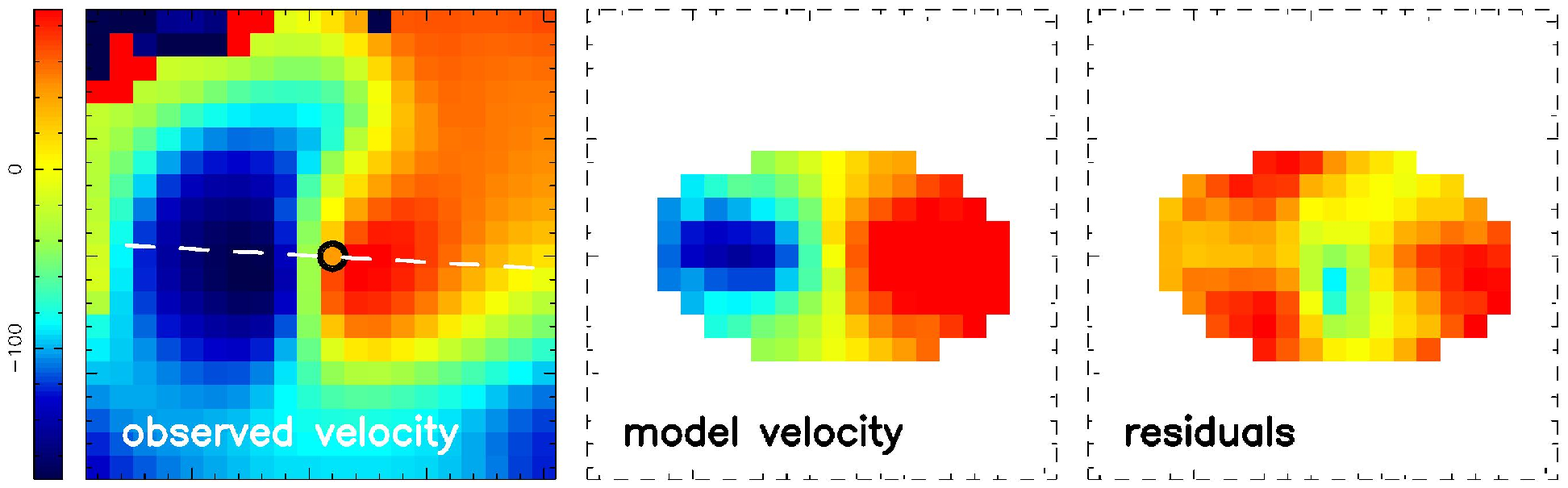} 
\caption{Western Nucleus: CO(2-1) velocity field, best-fitting disk model, and residuals. The kinematic centre is marked by a black circle, and the major axis is indicated with a white dashed line. 1\,pixel corresponds to 0.05\arcsec.\label{fig:codisk}}
\end{center}
\end{figure}

The CO(2-1) also displays a regular rotation pattern around the western nucleus, which we model analogous to the stellar velocity analysis. The best-fitting model (Fig.~\ref{fig:codisk}) indicates an inclination q\,$\approx$\,0.56 and PA\,$\approx$\,3deg south of west, with the kinematic centre coincident with that of the stellar velocity field within the uncertainties. The stellar and CO(2-1) kinematics in the western nucleus thus seem to agree well.




\subsection{Kinematic Centre \& Hot Dust Emission}
\label{sec:WN:BH}

\cite{downes07} detect emission from a hot, compact dust ring just south of the CO(2-1) emission peak of the western nucleus, which they interpret as being heated by an AGN accretion disk. We locate the position of the continuum emission using the relative position of the dust emission peak and the CO(2-1) emission, and assuming that the peak of the stellar continuum in the western nucleus is coincident with that of the CO(2-1) emission (as outlined in \S\ref{sec:bure}). 
Intriguingly, the putative AGN position thus derived is coincident with the kinematic centre of the stellar and CO(2-1) rotation found in \S\ref{sec:WN:kin}. This strongly suggests that the stars in the western nucleus are moving in a gravitational potential with either a supermassive black hole \citep{downes07} or an extremely dense, young starburst \citep{sakamoto08} at its centre.

\subsection{Dynamical Mass Estimate}
\label{sec:WN:dynmass}

In principle, our data are of high enough quality to derive the dynamical mass via Jeans modelling. However, as we found in \S\ref{sec:extinction}, Arp\,220 is severely affected by extinction in the K-band. Since we do not know the relative three-dimensional distribution of the obscuring material, it is impossible to derive the intrinsic kinematics from the observed two-dimensional line of sight projections. We therefore have to content ourselves with a simpler estimate of the dynamical mass. We adopt the approach taken by \cite{bender92}, which is based on the tensor virial theorem \citep{BT87}, and refer the reader to appendix B of \cite{bender92} for a detailed discussion. According to this, the appropriate dynamical mass formula for the relative stellar velocities and dispersions of the western nucleus is M$_{dyn}$\,=\,1.12$\times$3$\times \sigma_0^2 \times R / G$. Using the stella rotation and dispersion profiles derived in \S\ref{sec:WN:kin}, this yields an enclosed dynamical mass out to $\sim$\,100\,pc (180\,pc) of 6.3\,$\times$\,10$^9M_\odot$ (9.0\,$\times$\,10$^9M_\odot$).


Since the CO data probe significantly deeper than the K-band stellar kinematics, we use these to obtain another mass estimate. We use the code described in \cite{cresci09} to model the CO emission as arising from a thick rotating disk, taking beam smearing effects into account. We achieve the best match to the observed flux, velocity, and dispersion maps with a mass distribution in which the majority ($\gtrsim$\,80\%) of the mass is concentrated within a radius of $\sim$\,0.15\arcsec\ (50\,pc). An even more concentrated mass distribution (point mass) also gives a good match (although the more extended distribution is preferred) -- the 0.3\arcsec\ beam size prevents a definitive differentiation here. The total mass predicted within a radius of 100\,pc is 1.6\,$\times$\,10$^9L_\odot$, although we note that our model underpredicts the central dispersion by $\sim$\,60\,km\,s, and hence this value probably is somewhat too low. Since our modelling cannot definitively exclude a significant mass distribution from a central point mass (i.e., black hole), and since the presence of such a non-luminous mass would have ramifications for our analysis in \S\ref{sec:sfh}, we note here that the black hole mass expected from the $M_{BH}$-$\sigma$-relation of \cite{tremaine02} is 1.4\,$\times$\,10$^8M_\odot$; i.e.~less than 10\% of the total dynamical mass.

\cite{downes98} estimate a gas mass for the western nucleus of 6$\times10^8M_\odot$, based on 1.3\,mm dust emission, implying a gas fraction of $\sim$10\%, which agrees well with what is expected for local ULIRGs and starburst galaxies (\citealt{hicks09} and references therein).


\section{Eastern Nucleus}
\label{sec:EN:analysis}

\subsection{Stellar \& Gas Kinematics}
\label{sec:EN:kin}

\begin{figure}
\begin{center}
\includegraphics[width=0.5\textwidth]{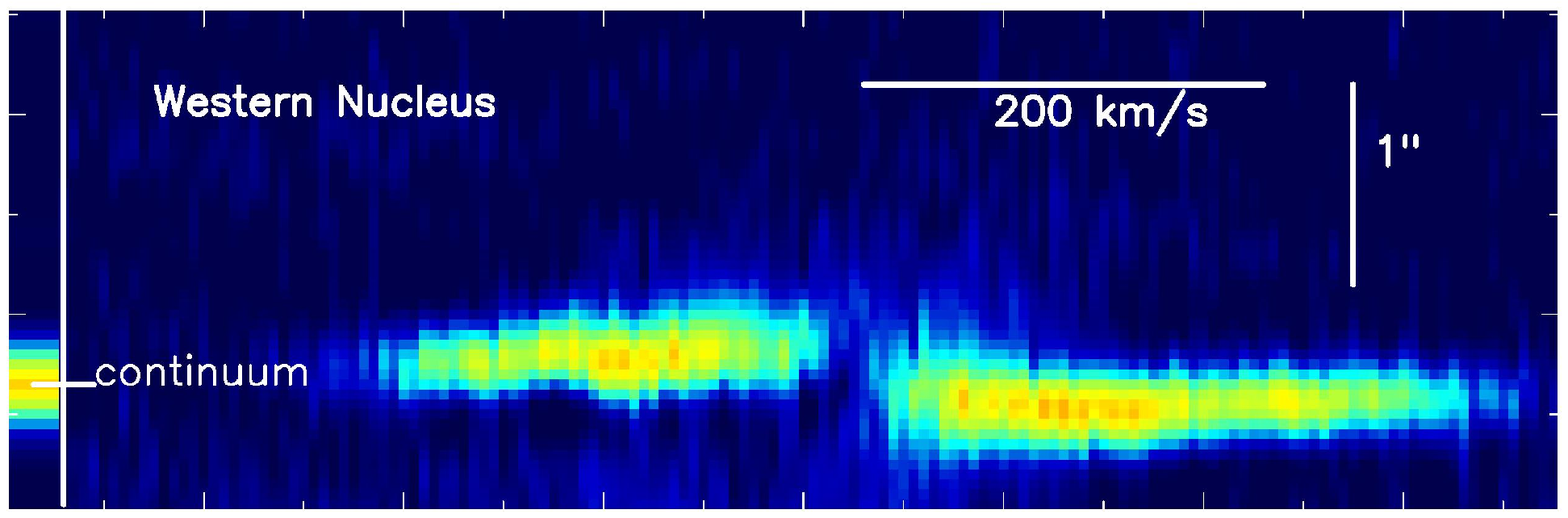} \\
\includegraphics[width=0.5\textwidth]{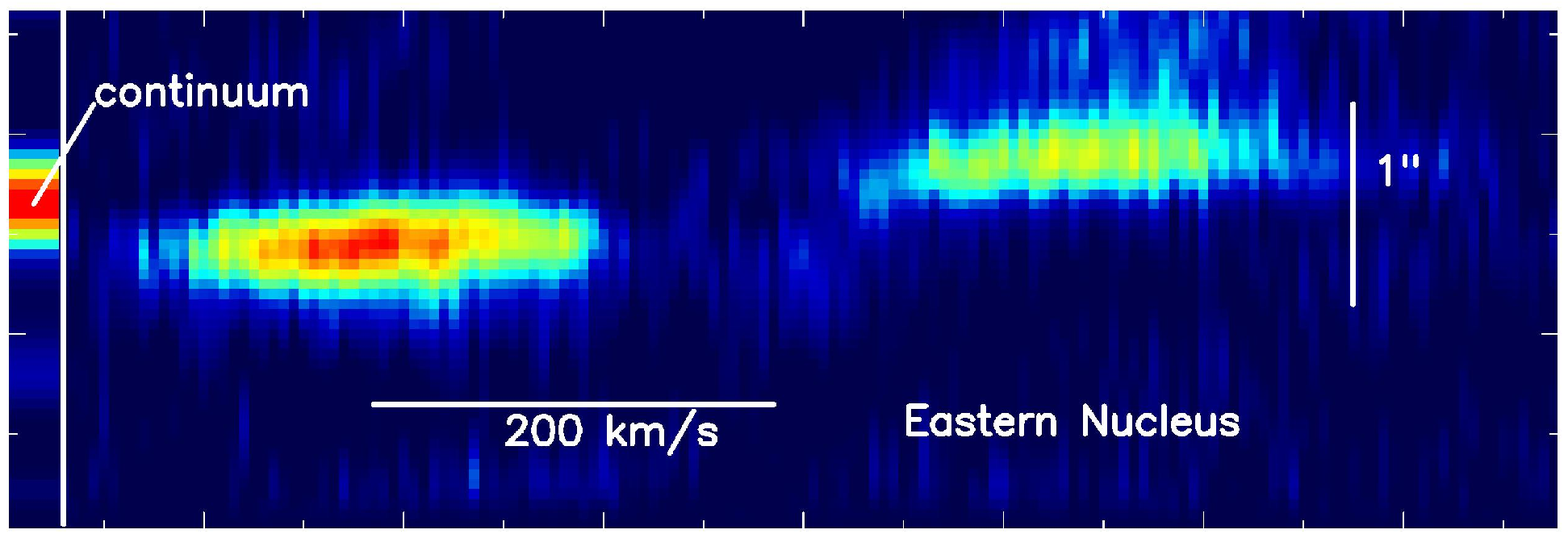} \\
\includegraphics[width=0.5\textwidth]{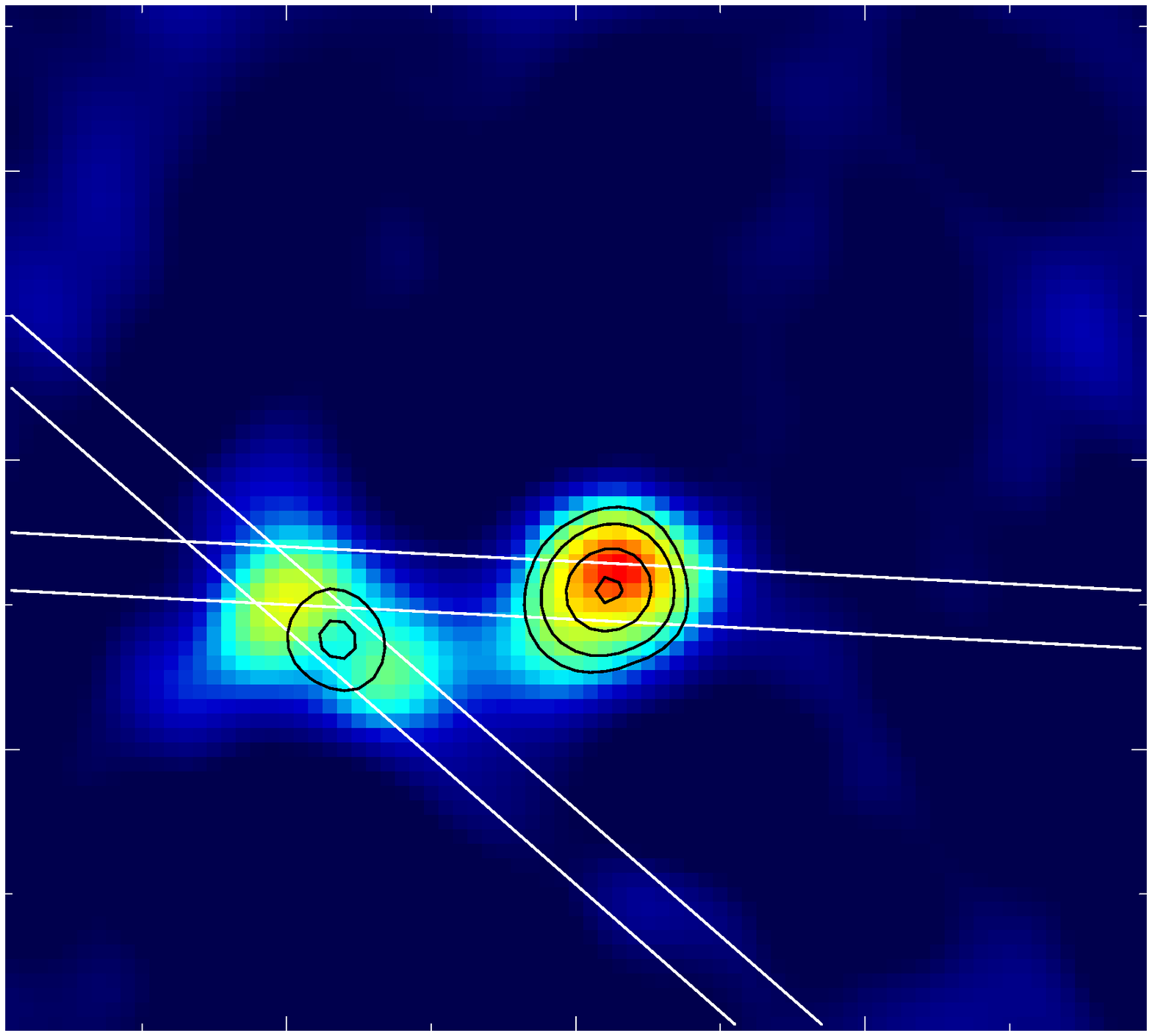} 
\caption{Position-Velocity diagrams across the western and eastern nucleus. On the left is a CO(2-1) flux map indicating the position of the slits, with 1.3\,mm continuum emission overlaid in contours. On the right are the P-V diagrams for western (top) and eastern (bottom) nucleus. As can be seen, there is no line emission in the central region of the eastern nucleus, coincident with the peak in dust emission.\label{fig:coPV}}
\end{center}
\end{figure}

\begin{figure}
\begin{center}
\includegraphics[width=0.55\textwidth]{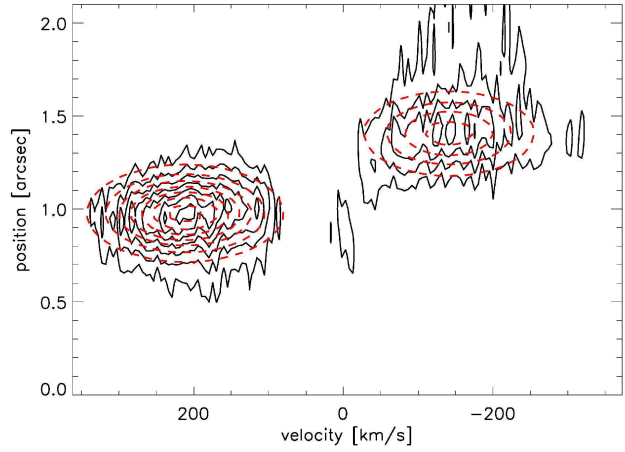} 
\caption{Position-Velocity diagram across the eastern nucleus (solid black contours) and from model (dashed red contours). Both sets of contours are displaying the same flux levels. As can be seen, the model (two unresolved sources orbiting the continuum emission peak) provides an excellent fit to the data.\label{fig:coPVmodel}}
\end{center}
\end{figure}

The stellar kinematics of the eastern nucleus are more difficult to interpret; although a strong velocity gradient is clearly present, pinning down a rotational major axis is much less obvious. The CO(2-1) kinematics are even more intriguing; the velocity map appears as a step-function, with the velocity changing abruptly from -135 to +205\,km\,s$^{-1}$. In order to investigate this, and any differences to the western nucleus, we produced P-V diagrams across each nucleus, also mapping the 1.3\,mm continuum emission. As can be seen (Fig.~\ref{fig:coPV}), the P-V diagram of the western nucleus is fairly typical of a rotating disk or spheroid, with only a slight flux gap in velocity-space near the centre of rotation. The eastern nucleus however appears markedly different, with a gap of almost 200\,km\,s$^{-1}$ between the two emission regions. To understand this, we construct a model of the eastern nucleus, using the simple \textit{ansatz} that we are seeing two unresolved sources at different velocities (we note that at this stage we do not make any assumptions about the actual gas distribution, only the actually observed emission -- as we discuss below, the gas is in fact most likely in a disk, but we can only see emission from two unresolved sources). After convolving a model data cube with the PSF of the CO observations, we extract a flux map and P-V diagram and compare it to the data. As Fig.~\ref{fig:coPVmodel} shows, the match achieved is striking -- supporting our hypothesis that we are indeed seeing two compact, unresolved emission sources. Moreover, as Fig.~\ref{fig:coPV} shows, the peak of the continuum emission lies at the centre of rotation, and hence the gravitational potential. We constrain the mass enclosed, by assuming that the two sources are on circular orbits, and calculating the enclosed mass for a range of inclinations of the rotation plane. Excluding the most extreme inclinations (q\,$<$\,0.25 \& q\,$>$\,0.9), we derive a dynamical mass of (1.8$\pm$0.5)\,$\times10^9M_\odot$ within 0.22\arcsec, or 81\,pc. However, this must be regarded as a lower limit (unless the gas is in a perfectly thin disk), since it neglects dispersion. 

\begin{figure}
\begin{center}
\includegraphics[width=0.85\textwidth]{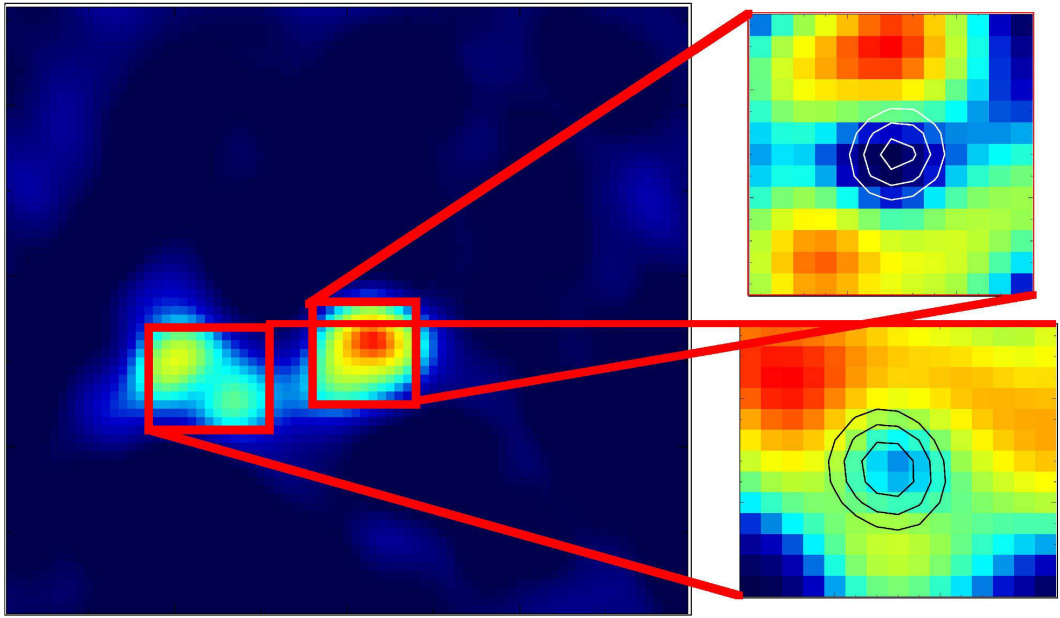} 
\caption{Integrated CO(2-1) flux map, with insets showing the CO line flux from a narrow velocity range near systemic velocity; contours mark mm-continuum emission. As can be seen, in those wavelength intervals, the (continuum-subtracted) line emission drops markedly towards the continuum peaks. This is a clear indication for continuum absorption by the line-emitting gas.\label{fig:abs}}
\end{center}
\end{figure}

This appears strongly suggestive of a gas disk centred on the continuum emission. However, one might wonder why in this case we are not seeing strong CO(2-1) emission from the centre. There clearly are significant amounts of gas present in the line of sight to the continuum peak -- \cite{sakamoto09} find the HCO$^+$ emission from the centre to display P-Cygni profiles. This, as they elaborate, is indicative of outflowing gas absorbing the continuum emission. We confirm the presence of absorbing gas by noting that the (continuum-subtracted) CO line emission at velocities near systemic appears ring-like around the continuum peak (Fig.~\ref{fig:abs}) -- a clear signature of continuum-absorption. It is therefore plausible that colder gas in the line of sight also causes a drop in CO line emission through line self-absorption. This also implies that the geometry of the eastern nucleus is similar to that of the western nucleus, in that there is a very dense, compact obscured central region about 50\,pc across which is completely obscured from the far-IR to shorter wavelengths; and that around this there is a smaller, but still significant, mass distributed over a large region with much lower mass surface density, and hence much lower obscuration.

\subsection{Dynamical Mass Estimate}
\label{sec:EN:dynmass}

The stellar kinematics of the eastern nucleus do not allow a straightforward measurement of a rotation major axis, we therefore instead use the kinematic centre and PA of the CO(2-1) velocity field (Fig.~\ref{fig:co}) to measure the stellar rotational velocity and dispersion at a radius of $\sim$\,230\,pc (100\,pc). Analogous to \S\ref{sec:WN:dynmass}, this yields an enclosed mass of 11.2\,$\times$\,10$^9M_\odot$ (5.8\,$\times$\,10$^9M_\odot$).
As outlined in \S\ref{sec:EN:kin}, the CO kinematics suggest a lower limit to the dynamical mass of (1.8$\pm$0.5)\,$\times10^9M_\odot$ within 81\,pc. We note that this indicates very similar masses for western and eastern nucleus within 100\,pc radius, and, due to the more centrally concentrated distribution, that at smaller radii the enclosed mass of the eastern nucleus is in fact larger than that of the western nucleus. 
\cite{downes98} estimate a gas mass of 1.1\,$\times$\,10$^9M_\odot$ for the eastern nucleus, again implying a gas fraction of $\sim$\,10\%.

\subsection{Nature of Eastern Nucleus}
\label{sec:EN:nature}

\begin{figure}
\begin{center}
\includegraphics[width=0.75\textwidth]{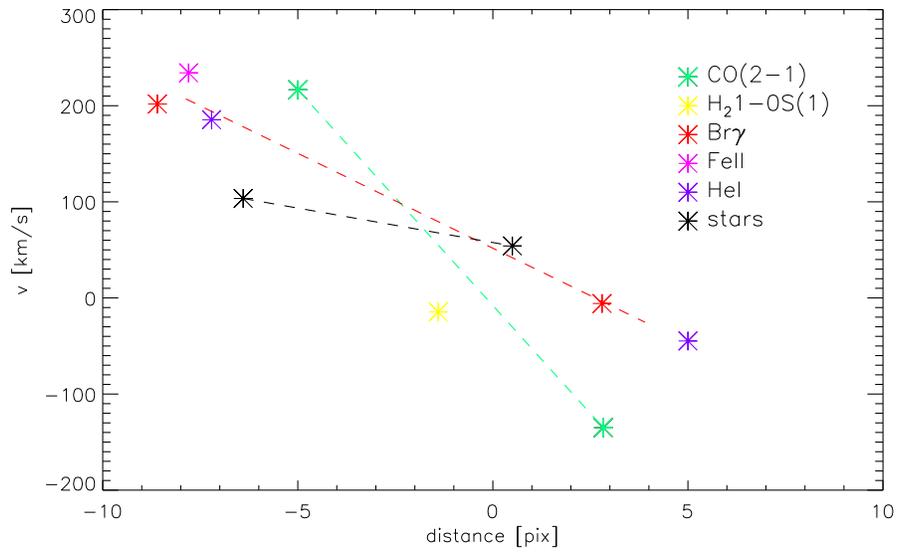} 
\caption{Position and velocities of CO(2-1) and various K-band tracers in the eastern nucleus, along the CO(2-1) rotation major axis. CO(2-1) displays the largest velocity differential, followed by the K-band gas tracers, and finally the stars. This is interpreted as arising due to the different depths probed by the various tracers.\label{fig:ENxv}}
\end{center}
\end{figure}

As our data show, different tracers display different morphologies at the eastern nucleus (cf.~also the high-resolution Keck maps of \citealt{soifer99}); raising the question whether we are seeing a kinematically coherent structure, or something different (a gas streamer, e.g.). 
The 2D velocity fields of the CO(2-1) and various K-band tracers all appear broadly similar, suggesting that they are all governed by the same gravitational potential. One way to look at this is to directly compare the velocities of the different tracers. Fig.~\ref{fig:ENxv} plots the velocity of the emission peak of a number of tracers along the CO(2-1) rotation major axis; as can be seen, the velocity differential is largest for CO(2-1), followed by the K-band gas tracers, and somewhat smaller still for the stars. Since the 2D velocity fields of these all appear qualitatively broadly similar, these differences are most likely due to the different depths probed by the CO(2-1) and the K-band tracers, rather than inherently different 3D velocity structures. The slightly smaller stellar velocities are expected, since rotating gas quickly collapses to a disk, whereas the stars are likely in a more spheroidal distribution, and hence have a larger fraction of their kinetic energy locked up in dispersion, rather than rotation.
We therefore conclude that the eastern nucleus, like the western nucleus, is a kinematically coherent structure. This, in conjunction with our stellar population analysis (\S\ref{sec:sfh}), leads to the obvious conclusion that most likely the nuclei are (remnants of) the progenitor galaxies' bulges.

Since mm-wavelength CO(2-1) observations are significantly less affected by extinction than K-band tracers, and given that our analysis above clearly shows that our CO(2-1) observations are probing deeply into the eastern nucleus, we can use these to assess the depth to which our K-band tracers are probing. We do this by utilising our model from \S\ref{sec:EN:kin} again, and making the simplifying assumption that the gas and stars we are observing in the K-band are also moving in a (thick) disk structure in the same gravitational potential as the cold gas. From this starting assumption, we then calculate at what distance above the disk midplane the different K-band emitters must be rotating to have velocities as observed. We find that generally they are 0.5--1 times as high above the disk midplane as their radial distance from the rotation axis -- implying that we are only probing the outer regions of the eastern nucleus in the K-band, in agreement with the large extinction we found earlier (\S\ref{sec:extinction}).

\begin{figure}
\begin{center}
\includegraphics[width=0.65\textwidth]{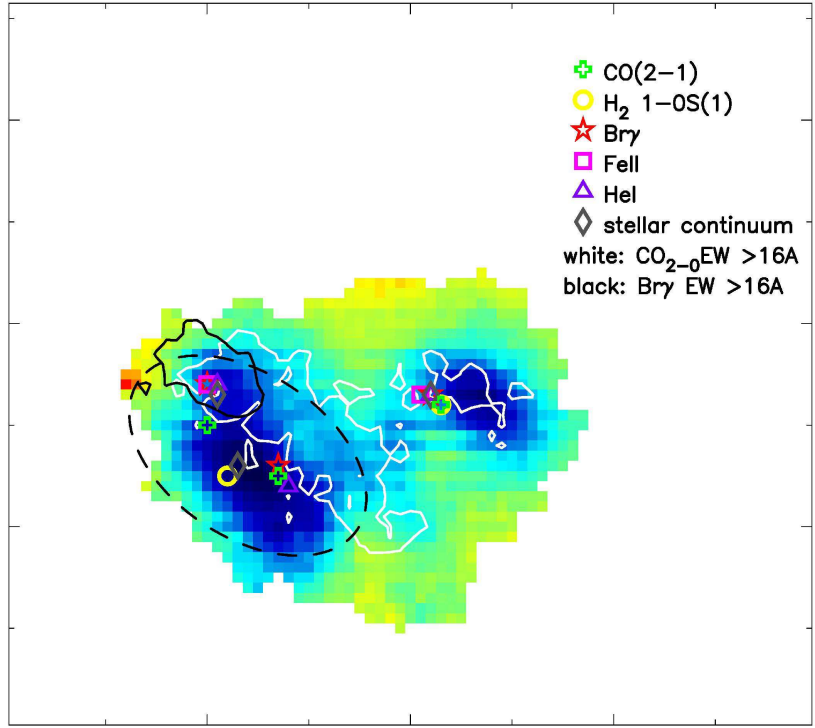} 
\caption{Extinction map (cf.~Fig.~\ref{fig:ext}), with the emission peaks of the various tracers, and the regions of high EW$_{Br\gamma}$ ($>$\,16\,\AA, black contours) and EW$_{CO}$ ($>$\,16\,\AA, white contours). The latter indicate regions with large flux contributions from recent star formation. As can be seen, both the region with lots of recent star formation and the region with high extinction are elongated along the major axis of the CO(2-1) rotation, but offset from each other, with one either side of the major axis. This suggests that we are seeing a thick disk/spheroid embedded in the larger scale gas disk, with the northwestern half above, and the southeastern half below the gas disk; leading to the offset regions of high extinction (nucleus obscured by gas disk) and high star formation (nucleus above disk plane). The black dashed line is a suggested outline of the nucleus.\label{fig:ENmod}}
\end{center}
\end{figure}

Fig.~\ref{fig:ENmod} shows the extinction map, with the emission peaks of the various tracers, and the regions of high EW$_{Br\gamma}$ ($>$\,16\,\AA, black) and EW$_{CO}$ ($>$\,16\,\AA, white) which indicate regions with large flux contributions from recent star formation. As can be seen, both the region with lots of recent star formation and the region with high extinction are elongated along the major axis of the CO(2-1) rotation, but offset from each other, with one either side of the major axis. This suggests that we are seeing a thick disk/spheroid embedded in the larger scale gas disk, with the northwestern half in front, and the southeastern half behind the gas disk; leading to the offset regions of high extinction (lower half of nucleus obscured by gas disk) and high star formation (nucleus above disk plane). We have indicated the suggested outline of the nucleus in Fig.~\ref{fig:ENmod}. We note that this is supported by Fig.~8 of \cite{sakamoto08}, which shows the location of the radio supernovae/supernova remnants \citep{parra07} in relation to the 860\,$\mu$m continuum emission.

\section{Nuclei \& Large-Scale Gas Disk}
\label{sec:setup}

As \cite{scoville97}, \cite{downes98}, and \cite{mundell01} show, the two nuclei are embedded in a larger-scale gas disk, with an estimated gas mass of $\sim$\,3\,$\times$\,10$^9M_\odot$ \citep{downes98}. Our observations are consistent with this; the H$_2$ kinematics clearly show rotational motion across the full SINFONI field of view (Fig.~\ref{fig:all1}).

As outlined in \S\ref{sec:WN:analysis} \& \S\ref{sec:EN:analysis}, the western nucleus is quite compact, whereas the eastern nucleus appears more extended. \cite{scoville98} propose that the crescent-like K-band continuum morphology of the western nucleus is due to an embedded opaque dust disk, which absorbs the stellar light from the southern half of the nucleus (their Fig.~6). This is supported by the stellar velocity map presented here (Fig.~\ref{fig:stellar}), which shows that the stellar rotation field is regular and extends to the proposedly extincted lower half. The embedded dust disk appears limited in extent, since in the \cite{scoville98} NICMOS data, emission from the southernmost tip of the western nucleus is visible.
Turning to the eastern nucleus; as we argue in \S\ref{sec:EN:nature}, the extinction pattern and EW$_{CO}$ suggest that here we are seeing a more extended spheroid, the south-eastern half of which is more strongly reddened due to being embedded in the larger-scale gas disk. While confirming the general picture put forward by \cite{mundell01} (cf.~their Fig.~7), in which the eastern nucleus is coplanar with the main CO disk, and the western nucleus lies above it, we have shown that the eastern nucleus is half embedded in the main CO disk. Furthermore, as our CO and stellar kinematics maps show, the relative velocities of the two disks are nearly identical. This, and the roughly face-on orientation of the large-scale gas disk, suggests that the merger is taking place roughly in the plane of the sky.

\section{Conclusions \& Summary}
\label{sec:summ}

We present new adaptive optics integral field spectroscopy near-infared data, and re-analyse interferometric mm CO(2-1) and continuum emission observations of the prototypical ULIRG Arp\,220. Our main conclusions are:
\begin{itemize}
\item
We show that in the central kpc, a 10\,Myr old starburst provides the majority of the K-band luminosity, with a $\gtrsim$\,1\,Gyr old stellar population accounting for the remainder of the luminosity and accounting for much of the mass. There is also a small contribution by stars $\lesssim$\,8\,Myr old, which are responsible for the Br$\gamma$ emission.
\item
The \cite{calzetti00} reddening law is found to provide the best fit to photometric data points spanning 0.45\,$\mu$m to 2.12\,$\mu$m.
\item
Estimating the system's bolometric luminosity from \textit{IRAS} fluxes, we find that we only see $\sim$\,7\% of the observed (non-reddening corrected) K-band luminosity. Since a reddening correction would only increase the observed flux by a factor $\sim$\,2--3, this implies that there is a significant amount of emitted K-band light which is completely obscured.
\item
Assuming the CO(2-1) emission and the K-band continuum peaks of the western nucleus to coincide, we find that the stellar kinematic centre is coincident with the mm continuum emission peak. This indicates that the mm continuum emission marks the centre of the gravitational potential in which the stars and gas are moving. The CO(2-1) and the stellar kinematics agree well, and all the K-band tracers peak at the same position; this indicates that the western nucleus is a compact, coherent structure.
\item
The eastern nucleus displays a markedly different morphology. Two emission peaks are seen in CO(2-1). We model the CO(2-1) data as two unresolved sources orbiting around the centre of continuum emission; this produces an excellent match. We find a clear signature for continuum absorption by the CO line-emitting gas; it is plausible that self-absorption also leads to the curious CO(2-1) morphology. Comparison of the velocities of the mm and K-band tracers shows that the CO(2-1) has a larger velocity gradient, which is expected since mm observations probe deeper into the obscured nucleus. We therefore conclude that the eastern nucleus also is a coherent structure, albeit more extended than the western nucleus. This, in conjunction with our stellar population analysis, strongly suggests that the nuclei are (remnants of) the progenitors' bulges. 
\item
This indicates that both nuclei have a similar structure in that there is a very dense and massive, but compact, structure in the central $\sim$\,50\,pc which is almost completely obscured at wavelengths shorter than far-IR; and around this is a more extended component with much lower mass surface density and lower extinction, but which is responsible for most of the observed luminosity. In general, we agree with the geometry suggested by \cite{mundell01}, and can refine it further. Comparison of the extinction map, EW$_{CO}$ and EW$_{Br\gamma}$ suggests that the lower half of the eastern nucleus is much more obscured than the upper half. This can plausibly be explained, if the eastern nucleus is coplanar with the larger-scale gas disk, and its lower half is buried in, and hence observed through, this gas disk. The very similar systemic velocities of the two nuclei, and the nearly face-on orientation of the large-scale gas disk suggest that the merger plane is approximately face-on.
\end{itemize}

\bibliographystyle{apj}

\begin{deluxetable}{lccccccc}
\tablecaption{Stellar Populations\label{tab:stellar}}
\tablewidth{0pt}
\tablehead{
\colhead{} &
\colhead{$L_K$\tablenotemark{a}} &
\colhead{EW$_{CO}$\tablenotemark{a}} &
\colhead{EW$_{Br\gamma}$\tablenotemark{a}}  &
\colhead{$L_{K,non-SB}/L_K$\tablenotemark{b}} &
\colhead{$L_{bol}/L_K$\tablenotemark{b}} &
\colhead{$L_{K,non-SB}/L_K$\tablenotemark{c}} &
\colhead{$L_{bol}/L_K$\tablenotemark{c}}
}
\startdata
SINFONI FoV & 3.25$\times10^9L_\odot$ & 14.78\AA & 7.09\AA & 0.13 & 151$\pm$25 & 0.44 & 49$\pm$9 \\
western nucleus & 9.5$\times10^8L_\odot$ & 15.72\AA & 8.69\AA & 0.08 & 111$\pm$12 & 0.26 & 51$\pm$5 \\
\enddata
\tablenotetext{a}{measured}
\tablenotetext{b}{assuming the non-starburst population to be $\gtrsim$\,8\,Myr old with EW$_{CO}$=\,0\,\AA}
\tablenotetext{c}{assuming the non-starburst population to be $\lesssim$\,20\,Myr old with EW$_{CO}$=\,12\,\AA}
\tablecomments{Stellar population analysis. We assume the starburst population to have EW$_{CO}$=\,17\,\AA, as measured between the two nuclei, and calculate the luminosity contribution of a younger (EW$_{CO}$=\,0\,\AA) or older (EW$_{CO}$=\,12\,\AA) stellar population required to dilute the CO absorption bandheads to the measured values. We furthermore calculate the expected $L_{bol}/L_K$ for each case. The EW$_{Br\gamma}$ furthermore indicates the presence of a very young ($\lesssim$\,7\,Myr) stellar population, the luminosity contribution of which however, at less than a few percent of the total, is negligible.}
\end{deluxetable}


\end{document}